\documentclass[revtex4]{emulateapj}
\usepackage{hyperref}
\usepackage{subfigure}
\usepackage{natbib}
\begin{document}

\title{An HST Survey for 100-1000 AU Companions around Young Stellar Objects in the Orion Molecular Clouds: Evidence for Environmentally Dependent Multiplicity}

\author{M. Kounkel\altaffilmark{1}, S. T. Megeath\altaffilmark{2}, C. A. Poteet\altaffilmark{3,4}, W. J. Fischer\altaffilmark{5}, L. Hartmann\altaffilmark{1}}

\altaffiltext{1}{Department of Astronomy, University of Michigan, Ann Arbor, MI 48109, USA}

\altaffiltext{2}{Ritter Astrophysical Research Center, Department of Physics and Astronomy, University of Toledo, Toledo, OH 43606}

\altaffiltext{3}{New York Center for Astrobiology, Rensselaer Polytechnic Institute, Troy, NY, USA}

\altaffiltext{4}{Currently at Space Telescope Science Institute, 3700 San Martin Drive, Baltimore, MD, 21218}

\altaffiltext{5}{NASA Postdoctoral Program Fellow, Goddard Space Flight Center, Greenbelt, MD, USA}

\begin{abstract}
We present a near-IR survey for the visual multiples in the Orion molecular clouds region at separations between 100 and 1000 AU. These data were acquired at 1.6~$\mu$m with the NICMOS and WFC3 cameras on the Hubble Space Telescope. Additional photometry was obtained for some of the sources at 2.05~$\mu$m with NICMOS and in the $L'$-band with NSFCAM2 on the IRTF. Towards 129 protostars and 197 pre-main sequence stars with disks observed with WFC3, we detect 21 and 28 candidate companions between the projected separations of 100---1000 AU, of which less than 5 and 8, respectively, are chance line of sight coincidences. The resulting companion fraction ($CF$) after the correction for the line of sight contamination is 14.4$^{+1.1}_{-1.3}$\% for protostars and 12.5$^{+1.2}_{-0.8}$\% for the pre-main sequence stars. These values are similar to those found for main sequence stars, suggesting that there is little variation in the $CF$ with evolution, although several observational biases may mask a decrease in the $CF$ from protostars to the main sequence stars. After segregating the sample into two populations based on the surrounding surface density of YSOs, we find that the $CF$ in the high stellar density regions ($\Sigma_{YSO} > 45$~pc$^{-2}$) is approximately 50\% higher than that found in the low stellar density regions ($\Sigma_{YSO} < 45$~pc$^{-2}$). We interpret this as evidence for the elevated formation of companions at 100 to 1000 AU in the denser environments of Orion. We discuss possible reasons for this elevated formation. 
\end{abstract}


\section{Introduction}

Multiplicity is a common property of stars that influences both the future evolution of a star and its potential to host habitable worlds \citep[e.g.][]{2009PASP..121..316D, 2014ARA&A..52..487S, 2014ApJ...791..111W}. Roughly one third of all field stars are multiple systems instead of single stars, with the fraction of multiple systems increasing with stellar mass, from 26\% for 0.1---0.5~M$_{\odot}$ stars, to 44\% for 0.7---1.3~M$_{\odot}$ stars, and finally to 60\% for 8---16~M$_{\odot}$ stars \citep{2006ApJ...640L..63L, 2013ARA&A..51..269D}.

Spatially resolved images of pre-main sequence stars and protostars show that most stars probably form in multiple systems \citep{1993A&A...278...81R, 2004AJ....127.1747H, 2008AJ....135.2496C, 2011ApJ...731....8K}. The number of companions over the total number of primaries and single stars (referred to as the companion fraction or $CF$) is observed to be high for protostars. The $CF$ ranges from 0.91 for Class 0 protostars to 0.5 for the more evolved Class I protostars, as measured over a separation range of 50 to 5000 AU \citep{2008AJ....135.2526C, 2013ApJ...768..110C, 2014prpl.conf..267R}. After the protostellar phase, pre-main sequence stars can still show a high incidence of multiplicity. For example, the $CF$ of pre-main sequence stars associated with the Taurus Molecular Cloud is 0.47 over a range of 18 to 1820 AU. This is 1.9 times higher than that for solar-type field stars in that separation range \citep{1993A&A...278..129L}. The implied decline in the observed companion fraction over this separation range with evolution may come from changes in the orbital parameters and potential ejection of companions. These changes could result from the formation of the companions in non-hierarchical systems with three or more stars \citep{2014prpl.conf..267R}, or from changes in the binding mass due to the ejection of gas in protostellar envelopes by outflows \citep{2008AJ....135.2526C}.

Since many stars form in young clusters, interactions with other stars in these clusters are also thought to play a significant role in determining both the fraction of multiple systems and the companion fraction. Observations of young stars in both low stellar density associations, such as that in the Taurus Molecular Cloud, and in the dense young clusters, such as the Orion Nebula Cluster (ONC), have shown that stars in the low density associations have 0.2 companions per logarithmic interval of separation, while stars in the dense clusters have 0.08 companions per logarithmic interval \citep{1993A&A...278..129L, 1994ApJ...421..517P, 1998ApJ...500..825P, 2007AJ....134.2272R, 2013ARA&A..51..269D}. This substantial difference has been attributed to the interactions between stars in clusters, which can change the orbital parameters or dissolve systems \citep{2001MNRAS.321..699K, 2012MNRAS.421.2025K}. Direct evidence of the dissolution of multiple systems by such interactions have been found in the ONC, where stars in the dense inner region of the cluster are deficient in wide companion systems compared to those in the lower density outskirts of the cluster \citep{2007AJ....134.2272R}. 

Currently, there is no consensus on whether the mass dependent fraction of multiple systems observed in the field is the result of the initial star formation process, the subsequent dynamical interactions in the clustered environment, or a combination of the two. \citet{1972MNRAS.156..437L} and later \citet{2012A&A...543A...8M} proposed that all stars form in multiple systems following an initially universal, mass invariant distribution of semi-major axes. They demonstrated that interactions in clusters with a range of stellar densities can reproduce the currently observed $CF$s and the distribution of the semi-major axes of the multiple systems. This approach has merit in explaining the higher fraction of multiple systems in the lower stellar density regions such as Taurus. Furthermore, \citet{2007AJ....134.2272R} present strong evidence for the dissolution of wider binaries in dense clusters, as required by \citet{2012A&A...543A...8M}. Although, \citet{2012A&A...543A...8M} only consider the mass fraction and semi-major axis distribution of solar-type stars, \citet{2014MNRAS.442.3722P} show that these interactions can reproduce the observed mass dependent multiplicity fraction and an increasing mean semi-major axis with mass. However, \citet{2014MNRAS.442.3722P} also show that the observed $CF$s and semi-major axis distributions can arise if the multiplicity is initially mass dependent with multiplicity fractions much less than unity \citep[e.g.][]{2013ARA&A..51..269D}. This would require the formation of some multiple systems via capture in expanding clusters \citep{2010MNRAS.404.1835K, 2010MNRAS.404..721M}. 

Ultimately, in order to disentangle the relative importance of the primordial variations in the $CF$ and semi-major axis distribution and their subsequent evolution in clustered environments, observations of the multiplicity in star forming regions (before they can be profoundly affected by interactions) are needed. Wide binaries may be most strongly impacted by the external interactions within a cluster environment, hence, observations in this regime will provide strong constraints on the theoretical models. Furthermore, low mass star formation occurs in a wide range of environments, from dense clusters to relative isolation. Within these environments the kinetic temperature, gas density, and degree of turbulence can vary greatly \citep[e.g.][]{1999ApJ...525..343W}. Detecting variations in the primordial multiplicity with the environment would be a step towards understanding how these physical factors influence the formation of the multiple systems. 

Studies of the effect of both the birth environment and early ($< 2$~Myr) evolution on multiplicity require relatively large samples of young stellar objects (YSOs) in the protostellar and pre-main sequence phase. Here we present one of the largest near-IR surveys of visual binaries around YSOs to date: a survey of multiplicity towards 375 protostars and pre-main sequence stars in the Orion molecular clouds made with the NICMOS and WFC3 cameras on the Hubble Space Telescope (HST)\footnote{Based on observations made with the NASA/ESA Hubble Space Telescope, obtained  at the Space Telescope Science Institute, which is operated by the Association of Universities for Research in Astronomy, Inc., under NASA contract NAS 5-26555. These observations are associated with program 11548.} .  The main focus of this survey is the extended Orion complex and not the ONC. This survey was executed as part of the {\it Herschel} Orion Protostar Survey (HOPS), a multi-observatory survey of protostars identified in the \textit{Spitzer} Space Telescope survey of the Orion molecular clouds \citep{2012AJ....144..192M}. The HOPS program combines 2MASS photometery, {\it Spitzer} photometry from the {\it Spitzer} Orion Survey, low resolution {\it Spitzer}/IRS 5-40~$\mu$m spectra, 70, 100 and 160~$\mu$m photometry from {\it Herschel}/PACS, and 350 and 870~$\mu$m photometry from APEX to construct 1.6-870~$\mu$m spectral energy distributions (SEDs) of the Orion protostars \citep[Furlan et al. submitted.; Ali et al. in prep.;][]{2013ApJ...763...83M,2013ApJ...767...36S,2013AN....334...53F}. To complement the spectral energy distribution, which is constructed from the data with $2-19''$ angular resolution, we targeted 280 protostars with either the NICMOS or WFC3 cameras on the HST. 

The YSOs observed in this survey include both the protostars targeted by the HOPS program, as well as other dusty YSOs\footnote{{\footnotesize The term "dusty YSOs" was used by \citet{2012AJ....144..192M} to designate all YSOs identified by the IR emission from dust in their circumstellar environments, either in infalling envelopes or in circumstellar disks. The term includes protostars (Class 0, I and flat-spectrum sources), as well as pre-main sequence stars with disks (Class II), but does not include young pre-main sequence stars without gas rich disks (Class III).}} that serendipitously fell into our targeted fields. High angular resolution near-IR imaging is a powerful means for studying the protostellar phase of the stellar evolution. Although protostars emit most of their luminosity at the mid to far-IR wavelengths, they are often detected in the near-IR, where the high angular resolution observations with space and ground-based facilities can achieve $\le 100$~AU resolution in the nearest molecular clouds \citep[e.g.][]{1999AJ....117.1490P}. The 1-2~$\mu$m emission is dominated by the central protostar and the inner ($< 1$~AU) regions of the protostellar disk. This emission can be directly detected in protostars with lower density envelopes or in high density envelopes where the orientation provides a line-of-sight through a cavity in the envelope cleared by the outflows. Alternatively, in the case of protostars with denser envelopes observed at intermediate or edge-on inclinations, the emission is scattered from dust grains in the envelope; in this case, the scattered light often delineates cavities carved into the envelopes by the outflows \citep{2014ApJ...781..123F}. In cases where unresolved emission from the central protostar/inner disk is detected, near-IR observations can also resolve and separate companions and thereby determine the incidence of binarity at separations of $\sim 100$~AU or greater. This capability can be extended to the more evolved pre-main sequence stars, where the lack of an infalling envelope eliminates biases due to the extinction by the envelope. 

We also present 3.8~$\mu$m imaging made with the NSFCAM2 camera on NASA's InfraRed Telescope Facility (IRTF). These images have a lower angular resolution due to the effect of the atmosphere, but they allow us to observe protostars at longer wavelengths where they are brighter. The primary purpose of these data is to obtain colors of some of the companions detected in the HST survey and thus determine whether they are consistent with protostars, pre-main sequence stars with disks, or perhaps unrelated field stars. In addition, five protostars and one pre-main sequence star imaged with the IRTF were not imaged with HST, and the IRTF data is our only information on multiplicity within these systems. 

Although the distance towards the Orion complex can range between 380--450 pc \citep[Kounkel et al. in prep]{2007ApJ...667.1161S, 2007A&A...474..515M, 2007PASJ...59..897H}, we assume a uniform distance to the complex of 420 pc. When discussing projected separations between binaries we convert angular separations in AU to provide a sense of physical scale, but they could vary by as much as 10\% due to the spread of distances found within the Orion cloud complex.

In the next section, we provide an overview of the properties of the target sample, and then discuss the observations and the data reduction. A catalog of multiple systems and their properties is then presented in Sec~\ref{sec:psc}. We use these data to study the $CF$ as a function of the evolution and the environment in Sec.~\ref{sec:csf}. This is followed in Sec.~\ref{sec:prop} by an analysis of the colors and magnitudes of the primaries and companions. Finally, Sec.~\ref{sec:discuss} discusses the implications of our results, and in particular, the role of the environment in the formation of multiple systems. 

\section{Observation and Data Reduction}
\label{sec:obs}

This paper is based on a program of observations obtained with NICMOS and WFC3 on the HST and NSFCAM3 on the IRTF. The program was initiated with NICMOS on HST in a program to obtain 1.6~$\mu$m and 2.05~$\mu$m images of {\it Spitzer} identified protostars. After the failure of NICMOS, the program was moved to WFC3 which obtained very sensitive 1.6~$\mu$m observations, but could not observe at 2.05~$\mu$m. Due to the larger field of view of WFC3, young stars with disks identified by {\it Spitzer} were also imaged serendipitously, allowing us to extend the survey to pre-main sequence stars. Finally, $L'$-band imaging of the protostars was obtained with NSCAM2 on the IRTF to obtain longer wavelength data.

\subsection{The Herschel Orion Protostar Survey Sample}

The HST survey targeted 280 protostars in the Orion A and B clouds that were identified with 1.2-24~$\mu$m photometry from the {\it Spitzer} Orion Survey \citep{2012AJ....144...31K,2012AJ....144..192M}. This sample of protostars was selected to be the entire sample of {\it Spitzer} identified protostars with predicted 70~$\mu$m fluxes in excess of 42~mJy; this flux cutoff was used to ensure they would be detectable with {\it Herschel}/PACS as part of the {\it Herschel} Orion Protostar Survey, or HOPS, an open time key project selected for the {\it Herschel} mission (Megeath PI). The HST survey input catalog is not a complete sample of all {\it Spitzer} identified protostars satisfying the 70~$\mu$m flux cutoff for two reasons. First, the methodology for identifying protostars in the {\it Spitzer} data was still evolving at the time the input catalog was defined; hence, several protostars in Orion were missed. Second, the flux cutoff was based on an extrapolation from the {\it Spitzer} 3.6-24~$\mu$m photometry, and in many cases the measured flux was substantially different. Furthermore, there is a small degree of contamination from reddened pre-main sequence stars and galaxies. Hence, only 248 out of the original 280 sources are still classified as protostars. The subsequent IRTF protostar survey targeted 66 {\it Spitzer} identified objects with the $m_{3.6\mu m} < 10.4$ mag.

\subsection{The Hubble Space Telescope Observations}

As part of program GO 11548, 87 orbits of observations were successfully exectuted with the NICMOS camera in August and September 2008 before the failure of the cryocooler. With the deployment of the WFC3 in June 2009, the program was redesigned for the near-IR channel of that instrument. The two primary changes were the elimination of the F205W band observations due to the wavelength cutoff of WFC3 at 1.75~$\mu$m and the inclusion of multiple protostars in a single pointing with the larger field of view (FOV) of WFC3. The remaining protostars in the target catalog were observed in 126 orbits obtained between August 2009 and December 2010. 

\subsubsection{The NICMOS Observations}

A total of 87 protostars were observed with the NIC2 camera, which has a $19.2'' \times 19.2''$ FOV and a pixel size of $0.075''$. Due to the obscuration of the molecular clouds, some of the guide stars were of marginal brightness; consequently, the guiding failed for nine protostars. Each protostar was imaged in a single orbit split between the F160W and F205W bands. The NIC-SPIRAL-DITH patten was used with 4 points and a $1''$ spacing; this pattern moved the source to the four corners of a square with $1"$ sides in the focal plane. The MULTIACCUM mode was used to maximize the dynamic range of the observations. The predefined SAMP-SEQ=STEP16 sequence was selected for the MULTIACCUM; this mode samples up the ramp starting with reads spaced by 0.303 seconds and then increases the interval up to a maximum of 16 seconds. A sequence of 25 steps were chosen for the F160W band and 16 steps for the F205W band. This resulted in total integration times of 1215.4~s and 767.6~s in the F160W and F205W bands, respectively. 

The initial calibrated frames were created with the \textit{calnica} command from the IRAF NICMOS data reduction tools provided by the Space Telescope Science Institute (STScI); this reconstructed the MULTIACCUM data into units of DN/S and performed the flat fielding and dark subtraction to the data. The resulting frames showed a spatially-varying, residual background. This background was most obvious at 2.05~$\mu$m and is due in part to thermal emission from the telescope. To reduce this artifact, which appeared to change very little over time, all images that had no detections of a protostar or only faint detections were used to generate a residual "sky" frame which was not contaminated by stars or nebulosity. The residual "sky" was then subtracted form all the data. This procedure was done separately for the F205W and F160W frames.

Even after the subtraction of the residual "sky" frame, the four individual quadrants of the NICMOS array showed varying offset levels. The offsets levels were particularly pronounced when a very bright source was present in the image. To match the offset, every image was individually adjusted by adding a constant value to each of the four quadrants with a custom Interactive Data Language (hereafter: IDL) program. These values were varied until a discontinuity was no longer apparent upon visual inspection of the image. 

A generic bad pixel mask was downloaded from the STScI, and then was supplemented by finding additional bad pixels through visual inspection. The final masks included several dead pixels, the gap between the four detector quadrants, and the circular region of the array blocked by the NICMOS coronagraph. Finally, the four dithered images were combined with the \textit{calnicb} command.
 
\begin{figure}
\centering
\epsscale{0.4}
\plotone{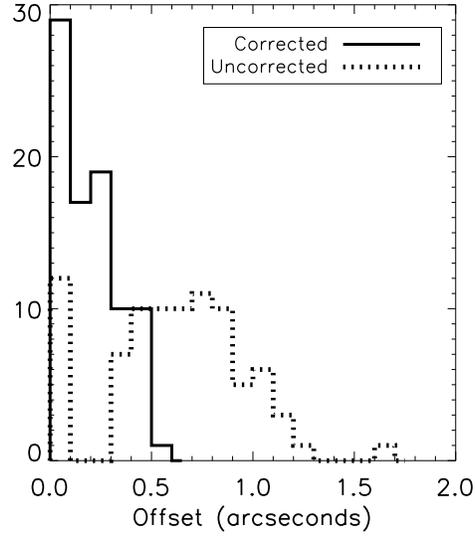} 
\caption{Total offset distance between the absolute coordinates of point sources in the NICMOS images with the positions given from the {\it Spitzer} Orion Survey. Dashed line shows the offsets before correction; solid line shows the offset after a simple correction was made to the absolute pointing.}
\label{fig:offset}
\end{figure}

While the relative astrometry within a NICMOS image is precisely calibrated, the absolute pointing can vary by as
much as $1.5''$. In Sec.~\ref{sec:psc}, we describe the detection of point sources in the NICMOS frames and the measurement of their positions and magnitudes. To correct the absolute pointing, we compared the positions of the point sources in the NICMOS fields to analogous sources in the {\it Spitzer} Orion Survey. For every NICMOS source for which there was a reliable {\it Spitzer} counterpart, we calculated the offsets between the NICMOS and the {\it Spitzer} coordinates from \citet[][Figure \ref{fig:offset}]{2012AJ....144..192M}. The {\it Spitzer} positions, which have been refined through a comparison to the 2MASS point source catalog, have positional errors $< 0.3''$\footnote{{\footnotesize \url{http://irsa.ipac.caltech.edu/data/SPITZER/docs/irac/iracinstrumenthandbook/}}}. In some of the NICMOS images there were no point sources detected; the protostars in these images were either too faint or surrounded by bright nebulosity. To facilitate the position refinement of these data, and to average out uncertainties in the {\it Spitzer} positions, we applied a global position correction to the images in bulk, as opposed to correcting each objects individually. We used the median offsets between the NICMOS and IRAC positions; these were $\delta {\rm R.A.} = 0.161''$, $\delta {\rm dec} = -0.376''$ before August 13 2008, and $\delta {\rm R.A.} =0.298''$, $\delta {\rm dec.} =-0.628''$ after that date. Although these offsets improved the pointing significantly (Figure \ref{fig:offset}), there were six outliers where the total offset was greater than $1''$ or the position offset had a different sign as compared to the average. In these cases, individual offsets were applied so that the NICMOS coordinates corresponded to those of the {\it Spitzer} counterparts.

\subsubsection{The WFC3 Observations}

One hundred and twenty six fields were imaged with the WFC3 camera between August of 2009 to December of 2010. In two of those fields, the tracking failed, resulting in poor imaging, and they were not included in the subsequent analysis. The near-IR channel of the camera has a $123'' \times 136''$ FOV with a pixel size of $0.13''$. Five frames were obtained per orbit using the DITHER-LINE pattern with a $2.106''$ spacing and a $45^{\circ}$ pattern orientation. To maximize the dynamic range, the MULTIACCUM mode was used with the SAMP-SEQ=STEP50 and 15 samples. This sequence samples up the ramp, starting with readouts spaced by a 2.932~sec interval and continuing with a logarithmically increasing intervals for the first 50~s, this is followed by linearly spaced intervals up to a total exposure time of 2496.17~sec. 

The initial frames were calibrated with {\it calwf3}. To mosaic the resulting five frames, we used the PyRAF tool {\it MultiDrizzle} \citep{2009hmdi.book.....F}. We applied the recommended parameters for all the steps of the reduction with the exception of setting the drop size for drizzling to be 0.75, and setting the pixel scale to be $0.065''$. The sampling and drop sized were chosen to take full advantage of the angular resolution resolution of the instrument, which is approximately $0.18''$, or 74 AU at the distance of Orion.

To refine the absolute pointing of the data, we again compared the WFC3 positions of the detected stars to those in the {\it Spitzer} Orion survey. Unlike the NICMOS data, there are numerous point sources in each WFC3 field. We adjusted the pointing of the WFC3 data so that the average differences in the positions of point sources detected by {\it Spitzer} and WFC3 were zero.

\subsection{The IRTF/NSFCAM2 Observations}

A total of five nights of observations were obtained at the NASA IRTF with the {NSFCAM2} camera, which houses a $2048 \times 2048$ detector with a FOV of $80'' \times 80''$ and a pixel scale of $0.04''$ per pixel. Our observations were obtained in the $L'$-band, centered at 3.76~$\mu$m, which was selected as a compromise between maximizing the flux from the observed protostars, all of which exhibit rising SEDs with increasing wavelength, and maximizing the sensitivity of the observations, which declines with wavelength due to the rising thermal background of the Earth's atmosphere. The typical seeing was $0.75''$, which corresponds to a spatial resolution of $\sim300$~AU at 420 pc. We imaged 36 fields in January 2010 and 24 fields in December 2010, all of them containing at least one confirmed protostar. A five point dithering pattern that put the target at the center of the frame and in each of the quadrants. Depending on the brightness of the central source, the dither pattern was repeated three to five times resulting in 15 or 25 dithered frames per targeted field. Each frame had an exposure time of 30 seconds using 1 second integrations and 30 coadds. In total, we obtained images of 98 \textit{Spitzer} identified YSOs.

The NSFCAM2 data were reduced using custom IDL routines. The flat fields were constructed using frames targeting different sources that were imaged close in time, but had differences in their airmasses of $\sim 0.4$. The frames were organized into pairs, and the low airmass frame of the pair was subtracted from the high airmass frame. The resulting images were then normalized and stacked, and a sky frame was constructed by taking the median value at each pixel in the stack. A flat field was created for all five nights of the observations; these were consistent to 2\% from night to night within a given observing run.

To subtract out the sky contribution, all the flat fielded frames toward a given source were then grouped into sets of five consecutive frames; these correspond to the five spatially offset fields of our five point dither pattern. A constant offset was added to each image so that the median value of the pixels in each of the five frames was equal to the median pixel value of all five frames. We then stacked the images and produced a sky frame by calculating the median of the five values at each pixel location. This approach provided enough redundancy to eliminate stars and compact nebulosity as well as artifacts such as cosmic rays, but ensured that the intensity of the sky did not change substantially during the sky measurement. Each of the five frames were then subtracted by the sky frame. 

NSFCAM2 at the time of the observations utilized an engineering grade detector that possessed many low responsivity or dead pixels. A bad pixel mask was created by removing the pixels where flat field showed more than 10\% deviation from unity. The sky subtracted, flat fielded, and masked frames were then mosaicked together using a custom IDL routine provided by R. Gutermuth. This routine determined offsets between overlapping frames by comparing the positions of sources in common between the frames. The images were then interpolated onto a mosaic grid using a bilinear interpolation. This was done iteratively with the mosaic being assembled one image at a time. During each iteration, the signal level of the mosaic and the overlapping sky frame was compared, and a constant value was subtracted off the new image so that it had the same signal level as the mosaic. 

During the first observing run, two different standard stars were observed: HD40335 and SAO112626. In the second run, we expanded to four standard stars: HD40335, SAO112626, GL105.5, HD44612 and HD1160. During both runs, one standard, HD 40335, was observed throughout the night at different air masses. 
The data were reduced in an identical manner to the science targets, but the photometry was measured for the
individual frames instead of using the mosaicked data. We adopted the tabulated $L'$-band magnitudes from the UKIRT faint infrared standard catalog \footnote{{\footnotesize \url{http://www.ukirt.hawaii.edu/astronomy/calib/phot_cal/faint_stds.html} \citep{2003MNRAS.345..144L}}}, with an exception of GL105.5, photometry for which came from the UKIRT bright infrared standard catalog \footnote{{\footnotesize \url{http://www.ukirt.hawaii.edu/astronomy/calib/phot_cal/bright_stds.html}}}. We determined the zero point magnitudes as a function of airmass for each night.

The images were registered using absolute coordinates from the {\it Spitzer} Orion Survey \citep{2012AJ....144..192M}. After identifying counterpart point sources in the {\it Spitzer} data, a linear transformation between the pixel values of the NSFCAM2 sources and the absolute coordinates were determined for each of the NSFCAM2 mosaic using least square fits. By adopting a linear transformation, we disregarded any non-linear distortion in the IRTF data; this was necessary given the small number of sources in common. The adopted linear coordinate transformation is only accurate to half an arcsecond around the edges of the mosaic; this level of precision is adequate for the analysis in this paper. 

\section{Source Detection and Point Source Catalog}
\label{sec:psc}

For each of the reduced NICMOS, WFC3 and NSFCAM2 mosaics, we identified all the point sources in the images and measured their magnitudes. This section describes the procedures and software used to extract photometry and compile the results into the final point source catalog.

\subsection{The WFC3 Photometry}

Point sources were identified with PhotVis, a tool for IDL that facilitates both automated and interactive source finding and photometry \citep{2008ApJ...674..336G}. The sources were found using an automated search with a 20$\sigma$ detection limit. After the automated source finding, each field was visually examined, previous unidentified sources were added, and false detections of noise and galaxies were eliminated. We adopted a zero point of 24.51~mag. To estimate the uncertainty, we adopted the gain for the the exposure time of 2496.17~s as given by the HST documentation. We limited our catalog to point sources and excluded extended or nebulous sources. 

Point spread function (PSF) fitting photometry was then performed using the IDL implementation of DAOPHOT. We used a modified version of DAOPHOT designed to mask out saturated pixels in the images \citep{2012AJ....144...31K}. This allowed us to remove the contribution of bright stars, find faint companions, and perform simultaneous PSF fitting of the companions using a 2nd iteration of DAOPHOT. In addition, the PSF fitting recovered the fluxes for several sources that were saturated. Fitting the data to the WFC3 PSF also allowed us to distinguish contamination from slightly extended galaxies initially identified as point sources by the automatic detection routine.

\begin{figure}
\centering
\subfigure{\includegraphics[width=0.33\textwidth]{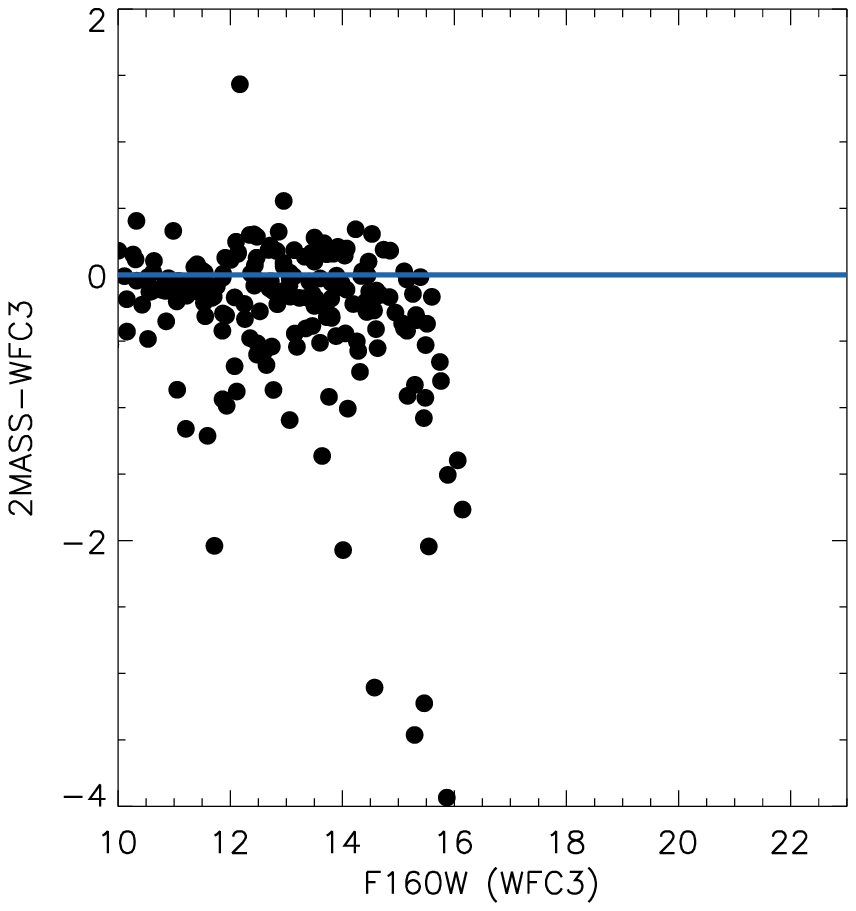}}
\subfigure{\includegraphics[width=0.33\textwidth]{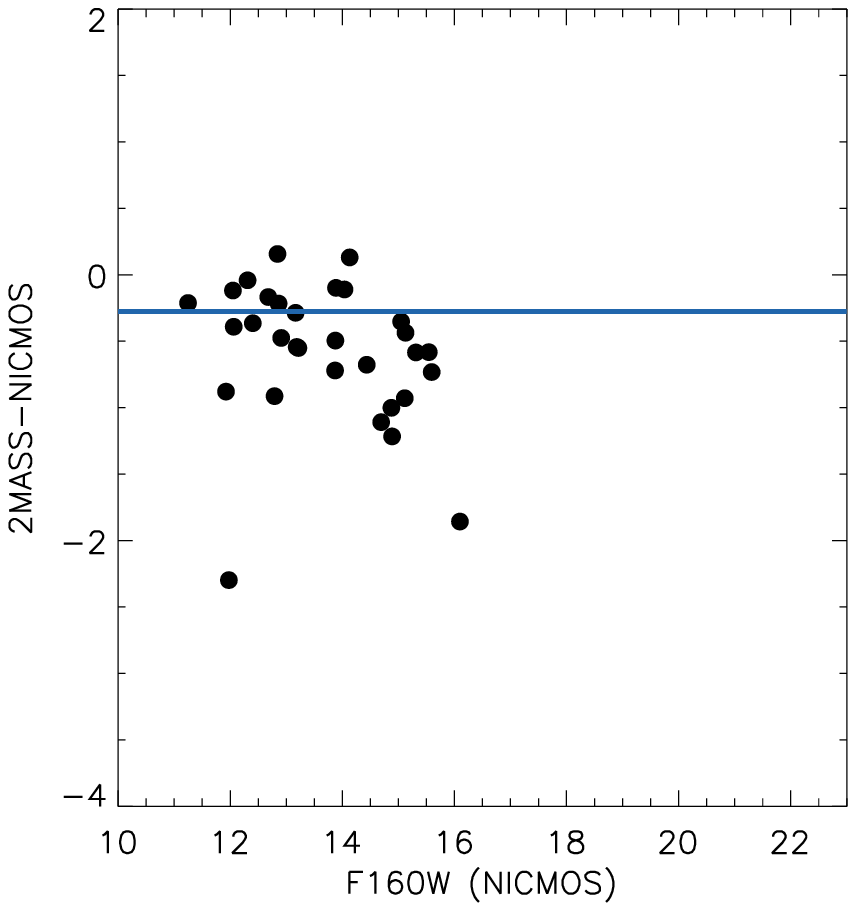}}
\subfigure{\includegraphics[width=0.33\textwidth]{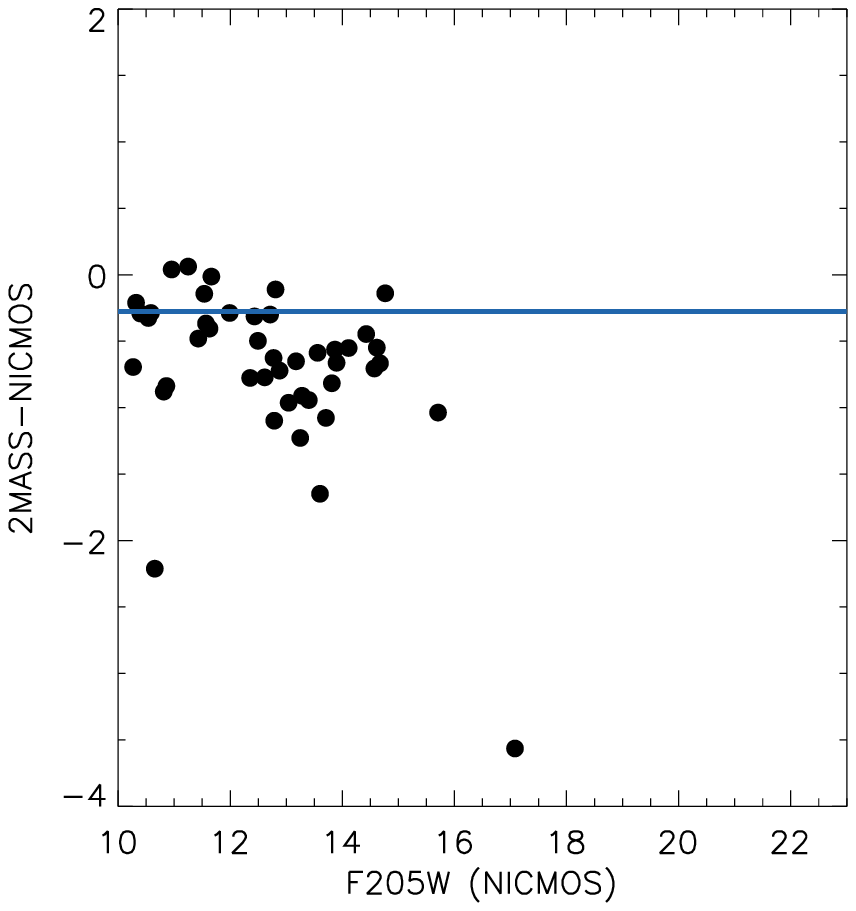}}
\subfigure{\includegraphics[width=0.33\textwidth]{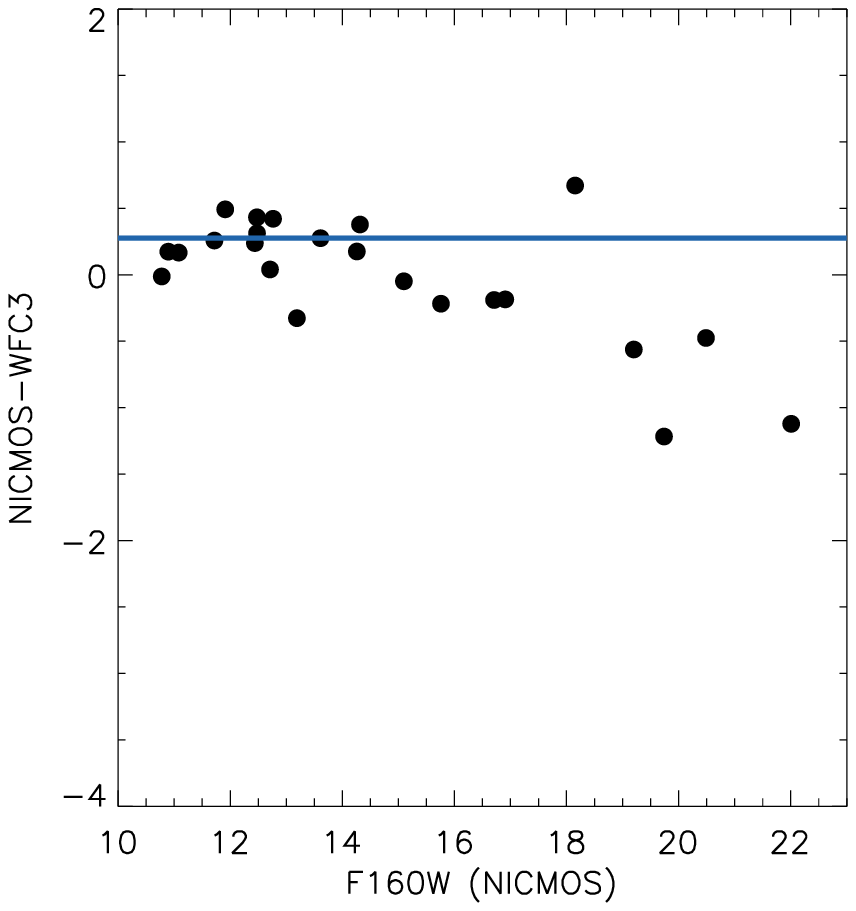}}
\caption{The difference in magnitude for sources with multiple epochs of WFC3, NICMOS and/or 2MASS PSC photometry after applying the standard calibration to the NICMOS and WFC3 data. The WFC3 camera 1.60~$\mu$m photometry is consistent with the 2MASS $H$-band photometry; the differences are due to the lower angular resolution and sensitivity of 2MASS. Due to the small size of the NICMOS FOV, there are fewer stars in the overlap region, and these stars are known YSOs which typically exhibit variability. Nevertheless, for the bright sources common to the NICMOS fields, there is a systematic offset between the NICMOS data and the WFC3/2MASS data which motivates a correction to the zero point which is applied to both the NICMOS F160W and F025W data. The offset used in our zero point correction is shown with the blue line. Fluxes presented in the Table \ref{tab:list} are corrected for this offset. \label{fig:compare}}
\end{figure}

The WFC3 photometry is consistent with the 2MASS $H$-band photometry (Figure \ref{fig:compare}). In the cases where the 2MASS and WFC3 photometry disagreed, the WFC3 magnitudes are fainter with the differences highest near the detection limit of the 2MASS data. This is expected since the lower angular resolution 2MASS point source photometry can be contaminated by extended emission, and the faintest 2MASS sources are most affected by contamination. 

\subsection{The NICMOS Photometry}

In contrast to the WFC3 data, the much narrower NICMOS FOV typical contained only a few sources. This allowed us to identify point sources through visual inspection with the ATV package in IDL \citep{2001ASPC..238..385B}. The photometry was measured at the positions identified by ATV using the procedure "aper" from the IDL astronomy library \citep{1993ASPC...52..246L}. We adopted an aperture of 6.5 pix ($0.4875''$) with a sky annulus from 10 pix ($0.75''$) to 20 pix ($1.5''$). The zero point magnitude and aperture corrections are from the NICMOS documentation. However, a comparison between the small number of sources that were observed with both WFC3 and NICMOS showed a constant offset between the F160W magnitudes. Thus, we corrected the NICMOS zero point by -0.276, which was the median difference in photometry between stars detected both in the WFC3 and NICMOS datasets with $12<m_{F160W}<15$. We applied the same offset for both F160W and F205W NICMOS photometry. Comparisons of the uncorrected NICMOS photometry to the WFC3 and 2MASS photometry are shown in Figure~\ref{fig:compare}.

We subtracted the PSFs of the primary sources to search for faint objects hidden by the wings of the PSFs. Unlike the WFC3 images, the NICMOS images do not contain enough point sources to construct a PSF directly from the image; for this reason we did not use DAOPHOT, which relies on a PSF generated from the data itself. Instead, we created a PSF model with TinyTim for both the F160W and F205W bands \citep{1998nicm.rept...18S}. A custom IDL PSF fitting routine was used to remove the contribution of the point source from the data; this was used to search for close companions. For sources where a companion was detected, simultaneous fits were used to determine the magnitudes for both objects. We adopted the values from the pre-fitted aperture photometry for the uncertainties of those sources. For the five secondary sources that were not detected prior to fitting the adopted uncertainties, we assigned them the uncertainty of a source found by NICMOS with a similar magnitude. 

\subsection{The NSFCAM2 Photometry}

The PhotVis routine in IDL was used to identify points sources in the NSFCAM2 mosaics and measure their photometry. We searched for sources manually, as the automatic finding routine identified too many spurious sources due to noise spikes and artifacts in the data, particularly toward the edges of the frames. PSF fitting to the NSFCAM2 data was not performed due to the variations of the PSF with both position on the detector and between observations due to changes in the seeing. A 50 pixel ($2''$) aperture, with a sky annulus from 50 to 73 ($2.92''$) pixels was typically used; however, for close binaries, we reduced the aperture to less than half the separation of the sources. In these cases, we maintained the same size for the sky annulus. We then calculated an aperture correction for these photometry using several isolated stars in the same images. The uncertainty was calculated from the variation in the aperture correction between different isolated stars in the image. 
 
\subsection{The Final Point Source Catalog}

Individual point source catalogs were generated for all of the three cameras; these were then merged with each other and with the catalog of all point sources in the {\it Spitzer} Orion survey \citep{2012AJ....144..192M}. For the NICMOS observations, there were 78 fields with usable data. Due to the high levels of extinction through the Orion Molecular cloud, particularly in regions of high gas column density that typically surround protostars, and the small field of view of NICMOS, six of the fields did not have detections. A total of 72 frames observed in F205W and a total of 63 in fields in F160W had at least one detected object. The NICMOS catalog contains F205W photometry for 123 point sources and F160W photometry for 98 point sources. The WFC3 catalog contains 4984 point sources including 128 HOPS objects, of which 123 are protostars and five are young stars with disks (Furlan et al. in submitted) In addition to this, the larger FOV encompassed six additional protostars and 192 young stars with disks that were not part of the HOPS catalog. The final point source catalog from NSFCAM2 has 337 YSOs.

A total of 178 protostars have point source photometry in the F205W and/or F160W bands. An additional 23 were detected with NSFCAM2, these are either to red to detect with HST or in the case of 5 objects, they were not observed with HST. The remaining HOPS protostars that are not part of our final catalog were either too faint to observe at our targeted wavelengths or were extended and did not contain a point source. Since every source was visually identified or found using an algorithm that identified sources with peaks 20$\sigma$ over the surrounding noise level, we did not filter the data based on their uncertainties or magnitudes. For the WFC3 catalog, we find that 99\% of the sources have $m_{160} \le 23.3$~mag and $\sigma_{160} \le 0.14$~mag. The 95\% limit for the NICMOS sources are $m_{160} \le 21.3$~mag and $m_{205} \le 20.8$~mag (we use the 95\% since there are only about 100 sources in the NICMOS data). For the IRTF $L'$-band, 99\% of the sources are brighter than $14.1$~mag and have uncertainties $\le 0.16$~mag. The photometry of dusty YSOs and their companions extracted from the final point source catalog are presented in Tables \ref{tab:list}, \ref{tab:single}.  We note that several of the sources imaged by both NICMOS and WFC3 show significant differences in their $m_{160}$ mags; these will be studied in future papers to search for variability among the observed YSOs and their companions.   

\section{The Incidence of Companions and its Dependence on Evolution and Environment}
\label{sec:csf}

Companions to the known protostars and pre-main sequence stars can be found by identifying nearby point sources in the point source catalog. The primary challenge is distinguishing bonafide companions from chance line of sight coincidences. The lack of data on the motions of the stars and the very limited wavelength coverage in our survey precludes distinguishing companions from line of sight coincidences on the basis of their observed properties. Instead, companions are identified by a statistical excess in the number of sources surrounding the targeted YSOs. In this section, we estimate the number of companions and then examine how the fraction of stars with companions depends on the evolutionary classes of the primaries and the environment in which they are found. We adopt the evolutionary classification of the primaries by \citet{2012AJ....144..192M}, and for the protostars observed by the HOPS program, we use the revised classification of the protostars based on the 1.6 to 870~$\mu$m spectral energy distributions compiled from 2MASS, {\it Spitzer}, {\it Herschel} and APEX data (Furlan et al. submitted). 

\subsection{The Identification of Companions}
\label{sec:identification}

To search for an excess density of sources, we utilize the mean surface density of companions \citep{1995MNRAS.272..213L}, which is related to the two point correlation function \citep{1998MNRAS.297.1163B}. In this approach, the density of objects is determined in concentric annuli centered on each {\it Spitzer}-identified dusty YSO within the NICMOS, WFC3 and NSFCAM2 fields. The density is then averaged over all the dusty YSOs using the equation

\begin{equation}
\Sigma_{comp}(r_i) = \frac{1}{N_{YSO}} \sum^{N_{YSO}}_{j=1} \frac{N_{comp}(j,r_i - \Delta i/2. < r \le r_i+\Delta i/2)}{\pi [(r_i+\Delta i/2)^2-(r_i-\Delta i/2)^2]},
\end{equation}

\noindent
where the {\it Spitzer} identified YSOs at the center of the annuli (hereafter: primaries) are numbered $j= 1$ to $N_{YSO}$, $\Delta i$ is the width of the $i^{th}$ annuli that extends from $r_i-\Delta i/2$ to $r_i+\Delta i/2$, and $N_{comp}$ is the number of NICMOS, WFC3 and/or NSFCAM2 sources within the $i^{th}$ annulus centered on the $j^{th}$ {\it Spitzer} YSO. 

To determine $N_{comp}$ within the annuli, we first identified a sample of primary sources by selecting the {\it Spitzer}-identified YSOs which appear as point sources in the HST data. This excludes sources that are too reddened to be detected in the F160W, F205W, and $L'$-band images and extended nebulous sources in which the central protostar is not detected as a point source. To find the number of sources as a function of radii from these {\it primary} sources, we use the point source catalog extracted from the NICMOS, WFC3, and NSFCAM2 data.

\begin{figure}
 \centering
			\subfigure{\includegraphics[width=0.45\textwidth]{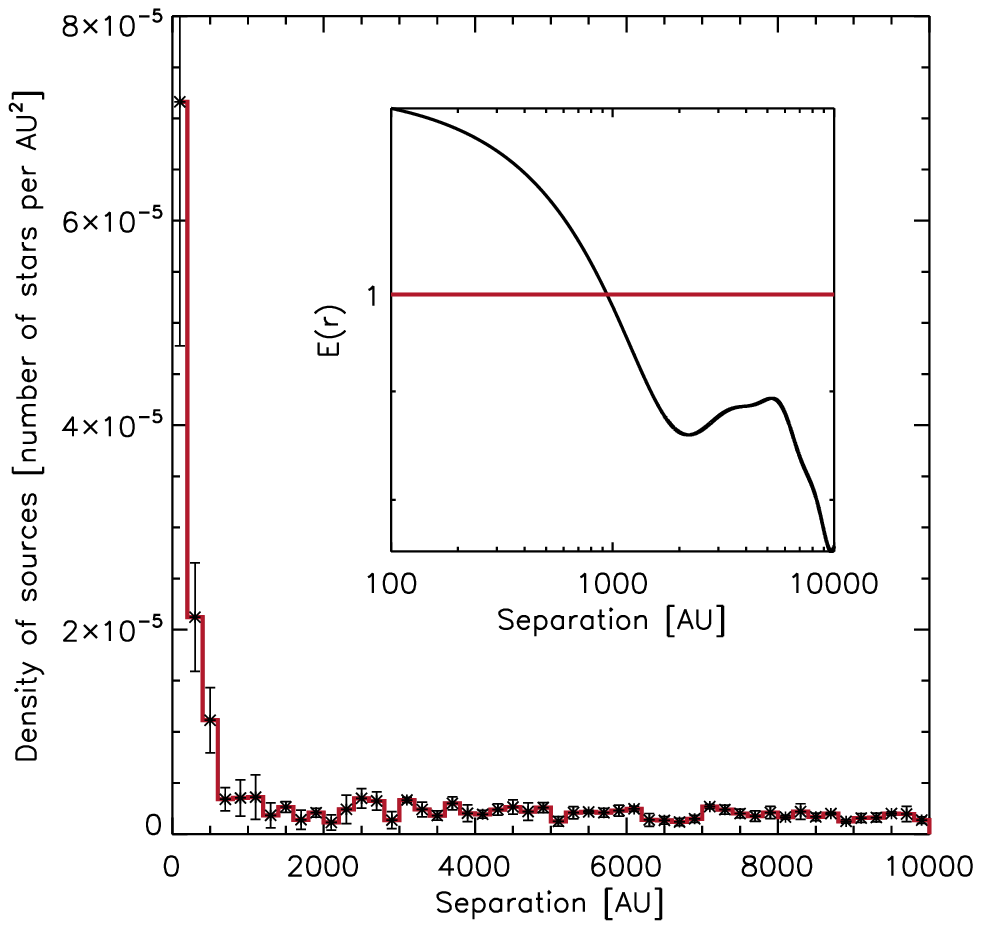}}
 			\subfigure{\includegraphics[width=0.45\textwidth]{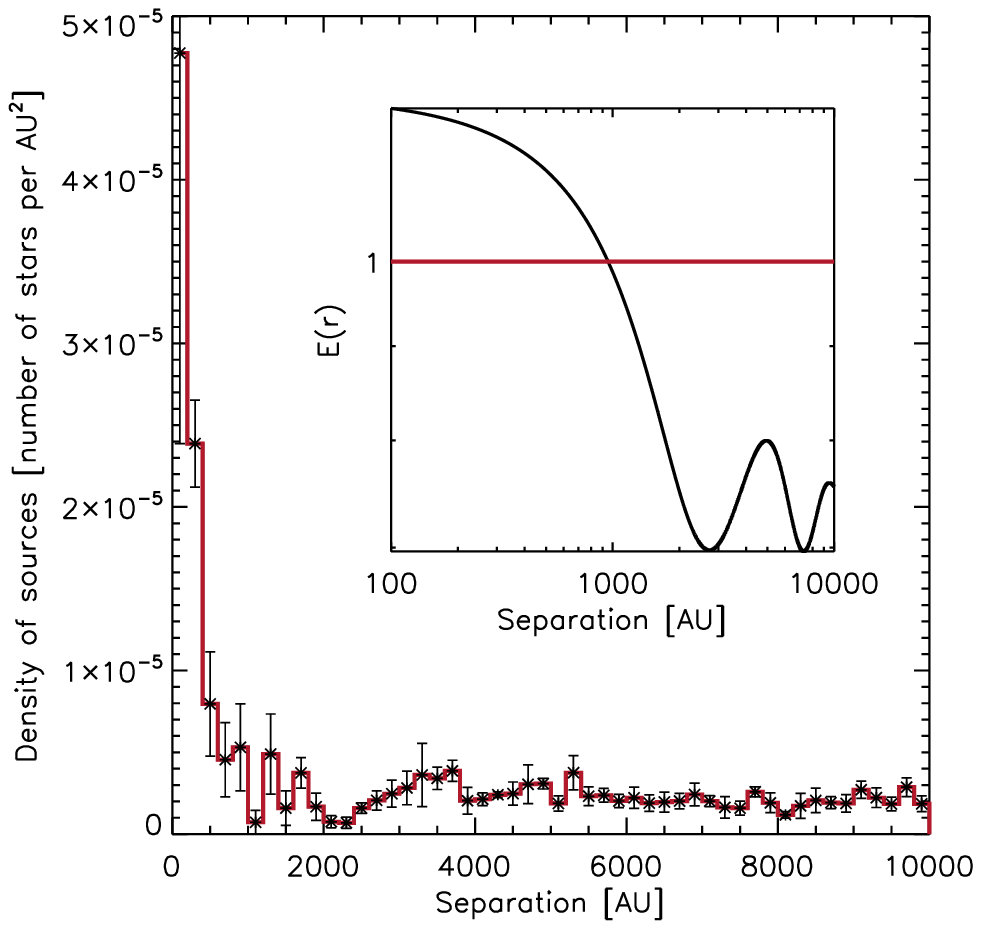}} 
 \caption{The mean surface density of companions, $\Sigma_{comp}$, as a function of projected separation. The bin size is 250 AU, except for the first bin, which contains companions with separations of 80--250 AU. The error bars were determined by dividing the sample randomly into four equal parts and finding the maximum variation between the $\Sigma_{comp}$ in those four samples. \textbf{Left:} $\Sigma_{comp}$ around protostars. \textbf{Right:} $\Sigma_{comp}$ around pre-main sequence stars with disks. Inserts show the values of Emark, a diagnostic for spatial dependence  \citep{baddeley2003,RSSB:RSSB433},  between all point sources and YSOs, normalized to be independent at $E(r)=1$, and evaluated up to $r$ of 40,000 AU (or half of FOV of WFC3). It shows that the sources with a neighboring source within 1000 AU are more likely to be YSOs.}\label{fig:mean1}
\end{figure}

In Figure~\ref{fig:mean1}, we plot the $\Sigma_{comp}(r_i)$ for the protostars and pre-main sequence star primaries separately.  The pre-main sequence primaries are only those that can be identified by their infrared excess and thus have disks. At larger separations, the plot shows a constant surface density of sources due to chance alignments with other YSOs and background stars. It also shows a clear spike in the density of sources within 1000 AU of the primaries; we identify this spike as the signature of a population of companions. We note that there may also be companions at larger distances; however, the surface density of those companions is not high enough to rise above the surface density of the line of sight contamination. Sources at separations $< 80$~AU cannot be resolved, therefore we are only able to identify companions between 80 and 1000~AU.  

To further confirm this result, we also show the results of the Emark routine from the R package \citep{baddeley2003, RSSB:RSSB433}.  Emark performs an analysis of a marked point process where each dusty YSO identified in the Spitzer catalog is assigned a mark of 1 and the remaining sources a mark of 0.  The routine determines the expectation value of the marks given the presence of an additional source at a specified projected separation.  The increase of the expectation value at small separations indicates that sources with a neighboring source within 1000 AU are more likely to be YSOs, once again demonstrating that there is an enhanced density of sources around the YSOs.

Between 80 and 1000 AU, we find that there are 29 candidate binary and one candidate ternary systems around the protostellar primaries, and 27 candidate binary and one candidate ternary systems around the pre-main sequence primaries. The primaries and companions are displayed in Table~\ref{tab:list} with their positions, separations and HST or IRTF magnitudes. Three of the companions around protostars are at separations less than 100~AU (HOPS 281, 268 and 24 at separations of 80, 96, and 99.9~AU, see Table~\ref{tab:list}) and one of the binaries around pre-main sequence stars has a separation less than 100~AU (MGM~3376, separation 95~AU). All the detected YSOs that do not have a companion are listed in Table~\ref{tab:single}.

\begin{figure}
\epsscale{0.9}
\plotone{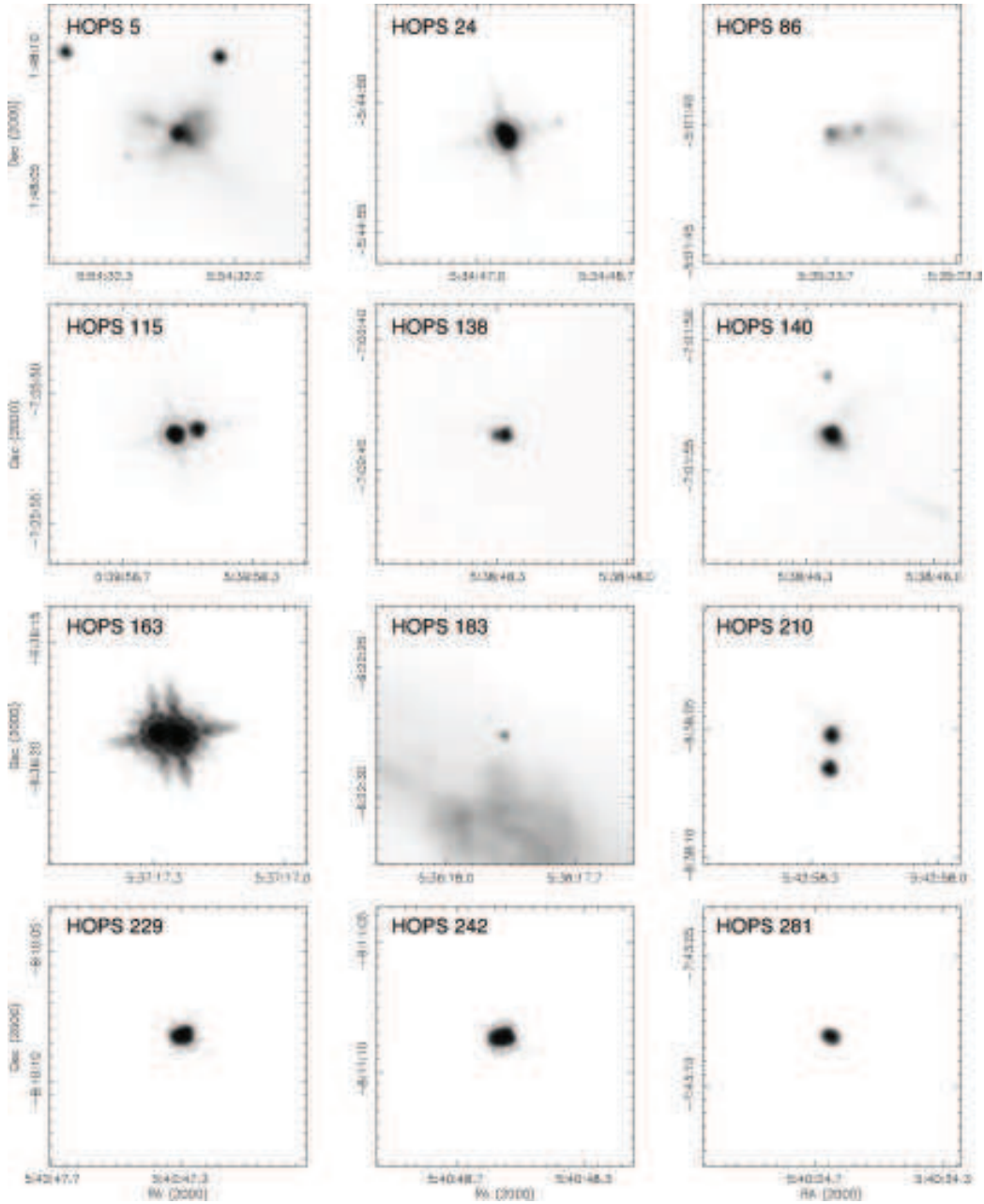} 
\caption{F160W images of the binary systems imaged solely with the WFC3 camera. Each image covers a 0.6 x 0.6$''$ region.\label{fig:a00}}
\label{fig:figs1}
\end{figure}

\begin{figure}
\epsscale{0.9}
\plotone{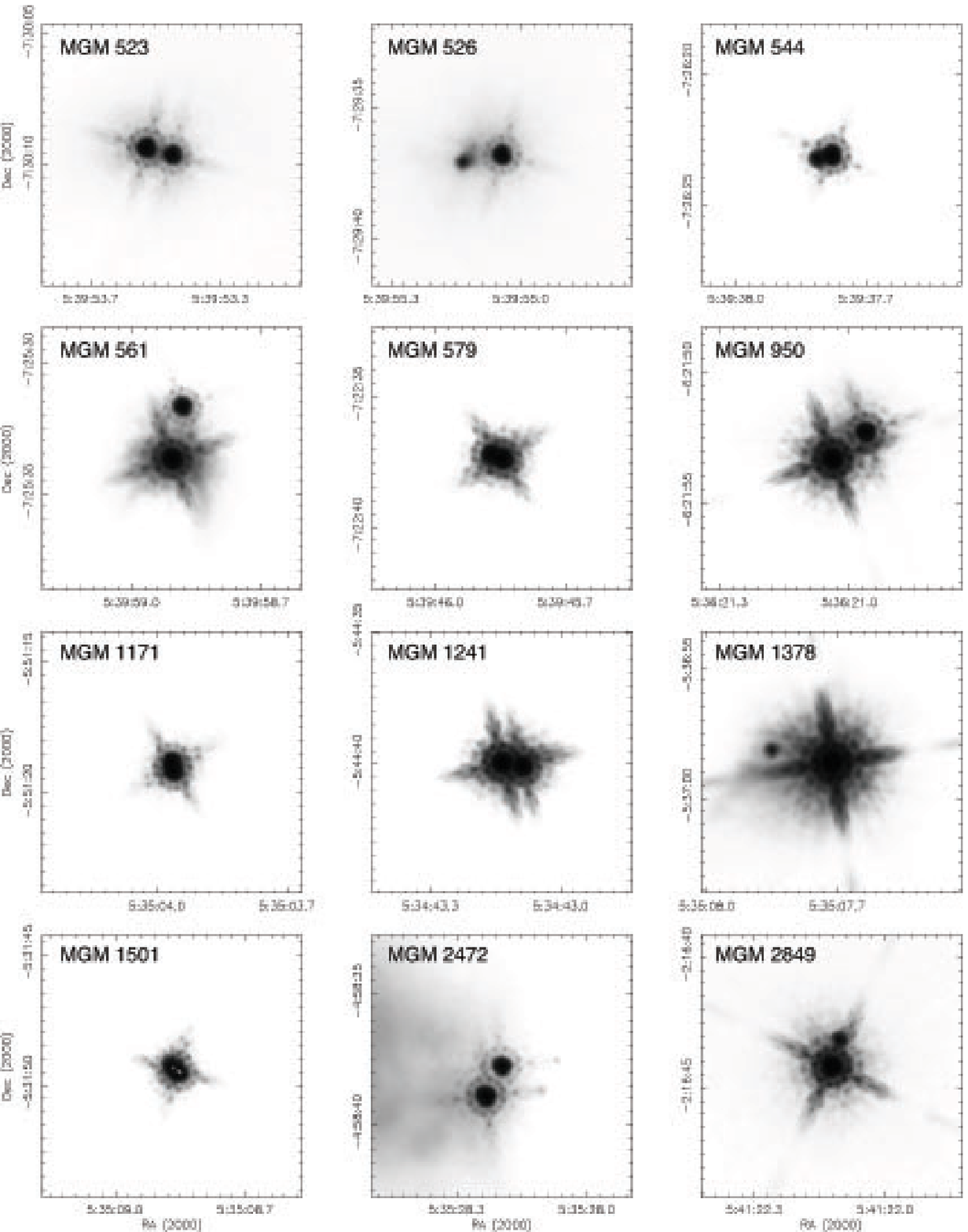} 
\caption{ F160W images of the binary systems imaged only with the WFC3 camera, continued.\label{fig:a01}}
\end{figure}

\begin{figure}
\epsscale{0.9}
\plotone{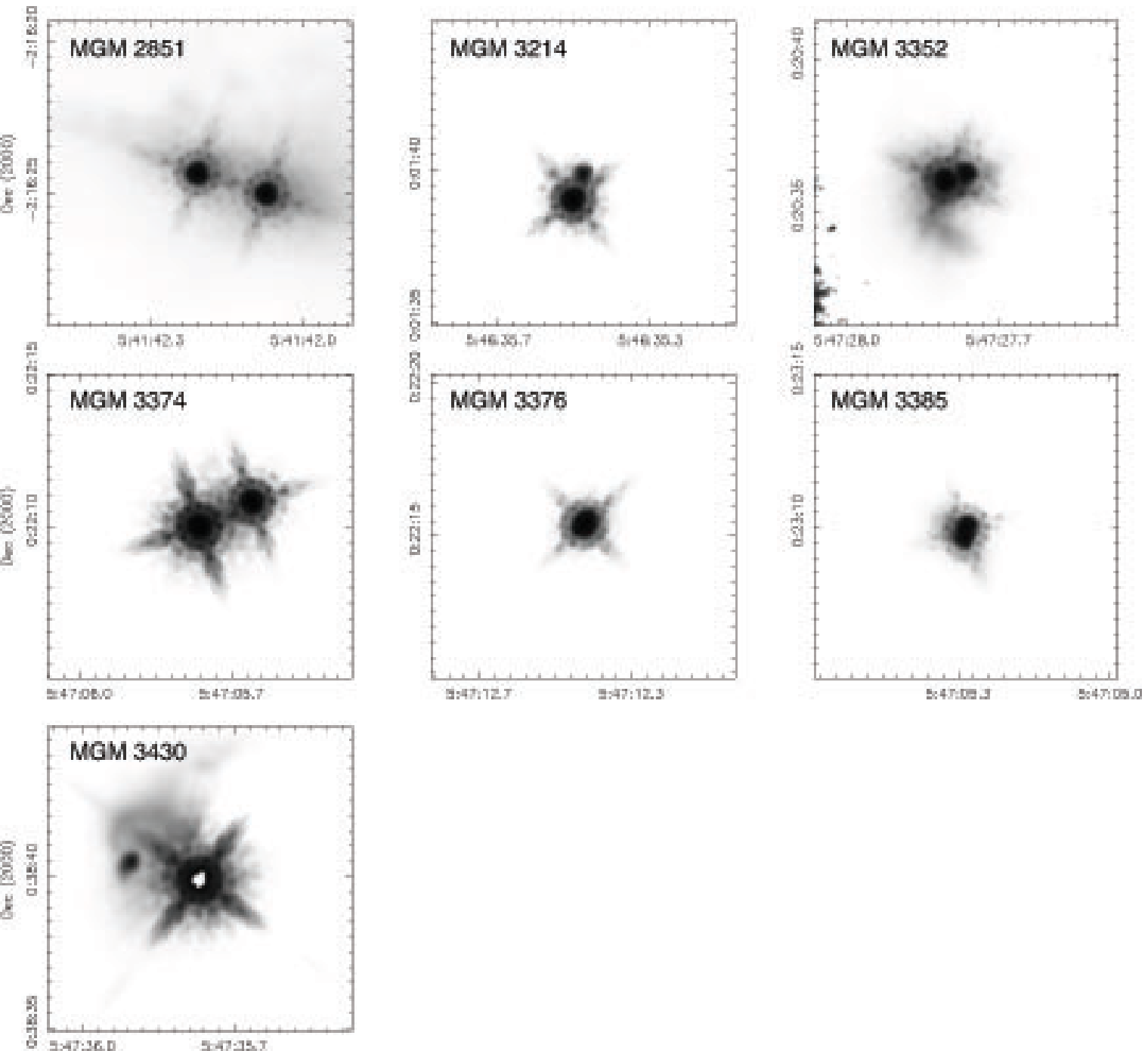} 
\caption{F160W images of the binary systems imaged only with the WFC3 camera, continued.\label{fig:a02}}
\end{figure}

\begin{figure}
\epsscale{0.9}
\plotone{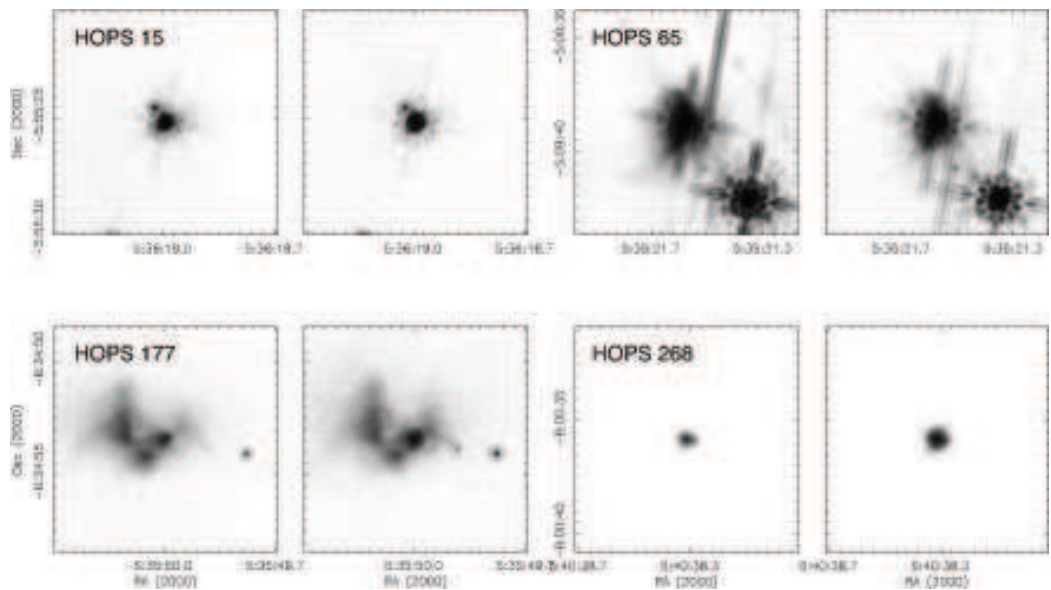} 
\caption{F160W and F205W images of the binary systems imaged only with the NICMOS camera.\label{fig:a1}}
\end{figure}

\begin{figure}
\epsscale{0.9}
\plotone{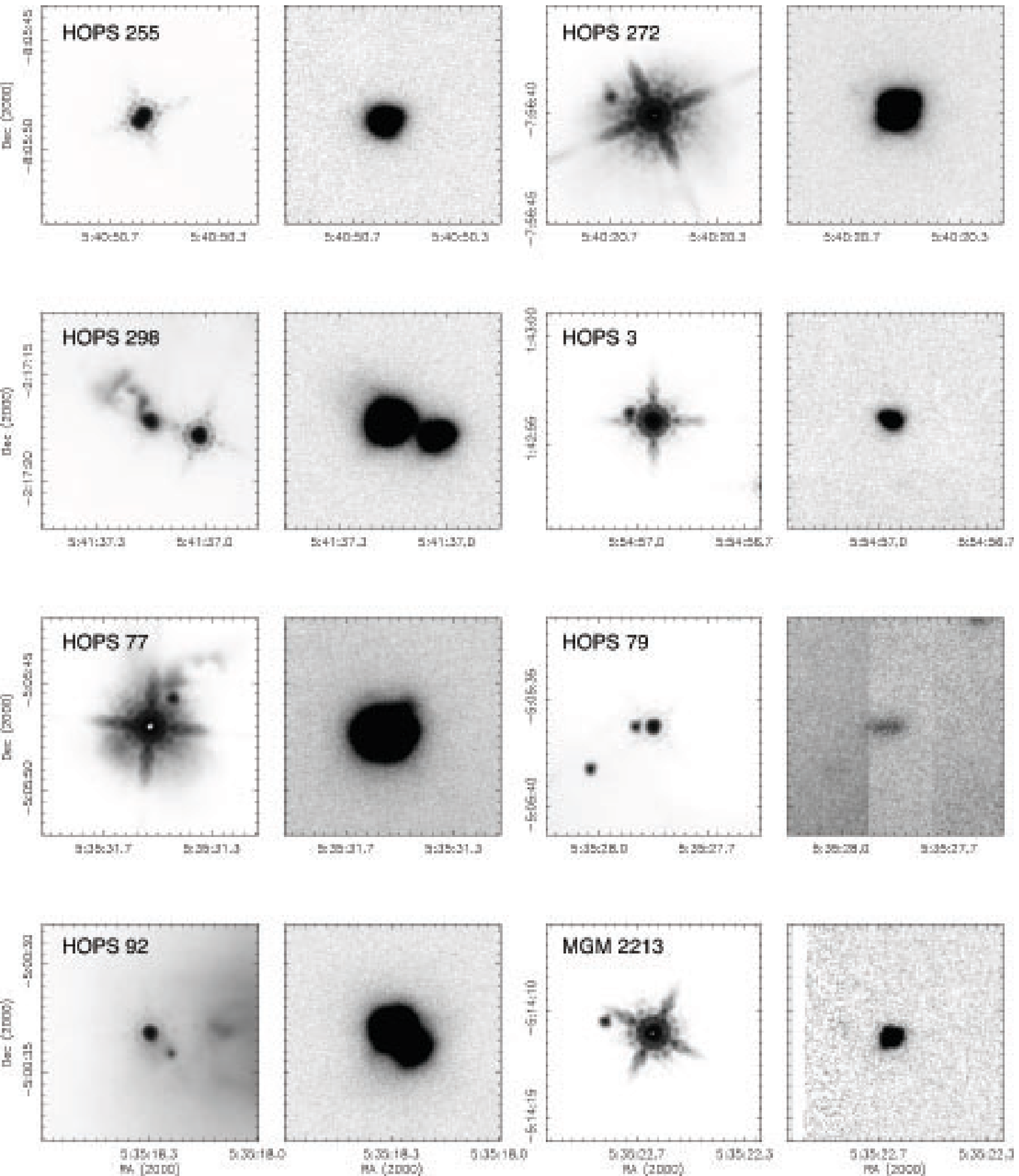} 
\caption{F160W and $L'$-band images of binary systems imaged with WFC3 and NSFCAM2.\label{fig:a20}}
\end{figure}

\begin{figure}
\epsscale{0.9}
\plotone{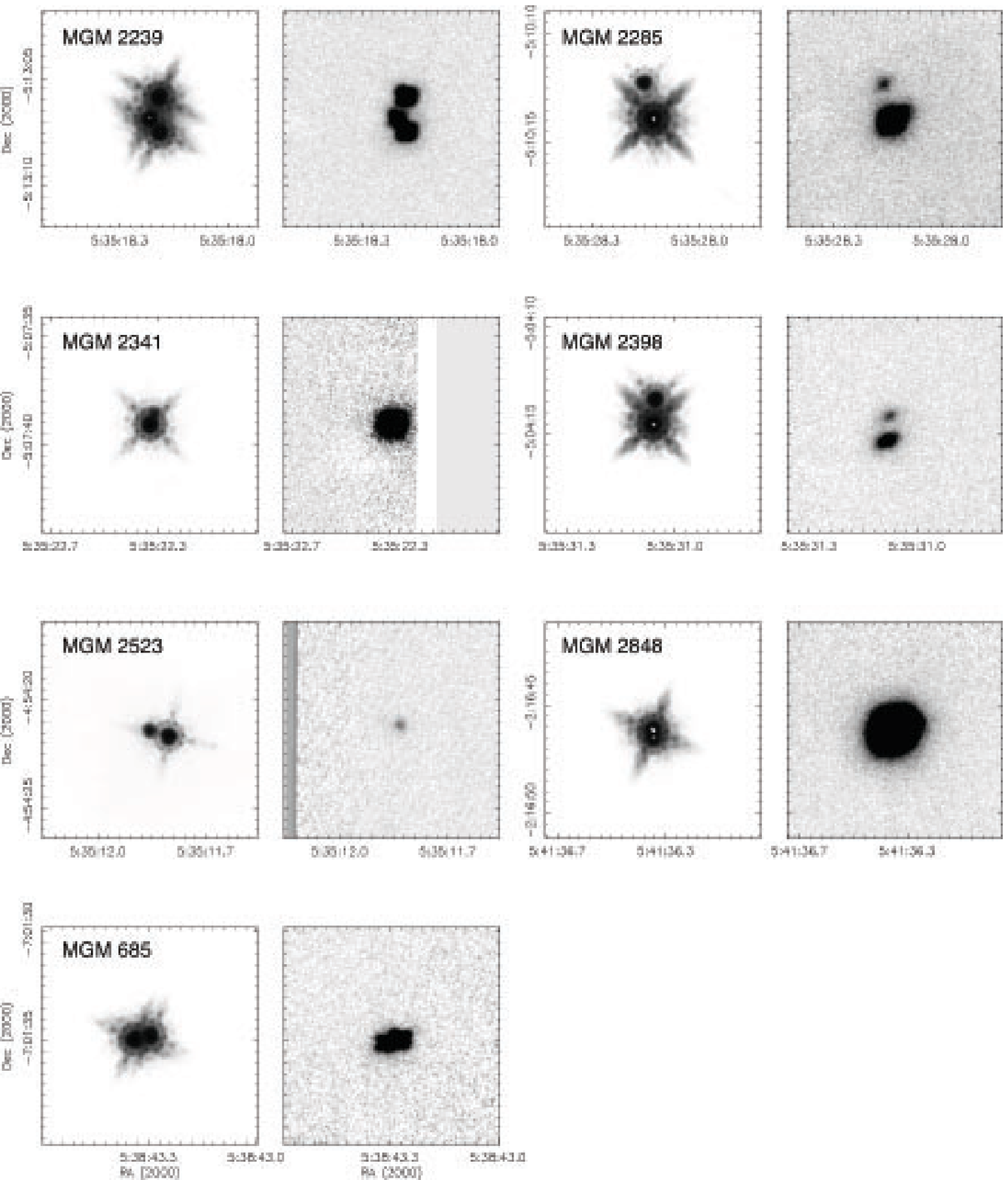} 
\caption{F160W and $L'$-band images of binary systems imaged with WFC3 and NSFCAM2, continued.\label{fig:a21}}
\end{figure}

\begin{figure}
\epsscale{0.9}
\plotone{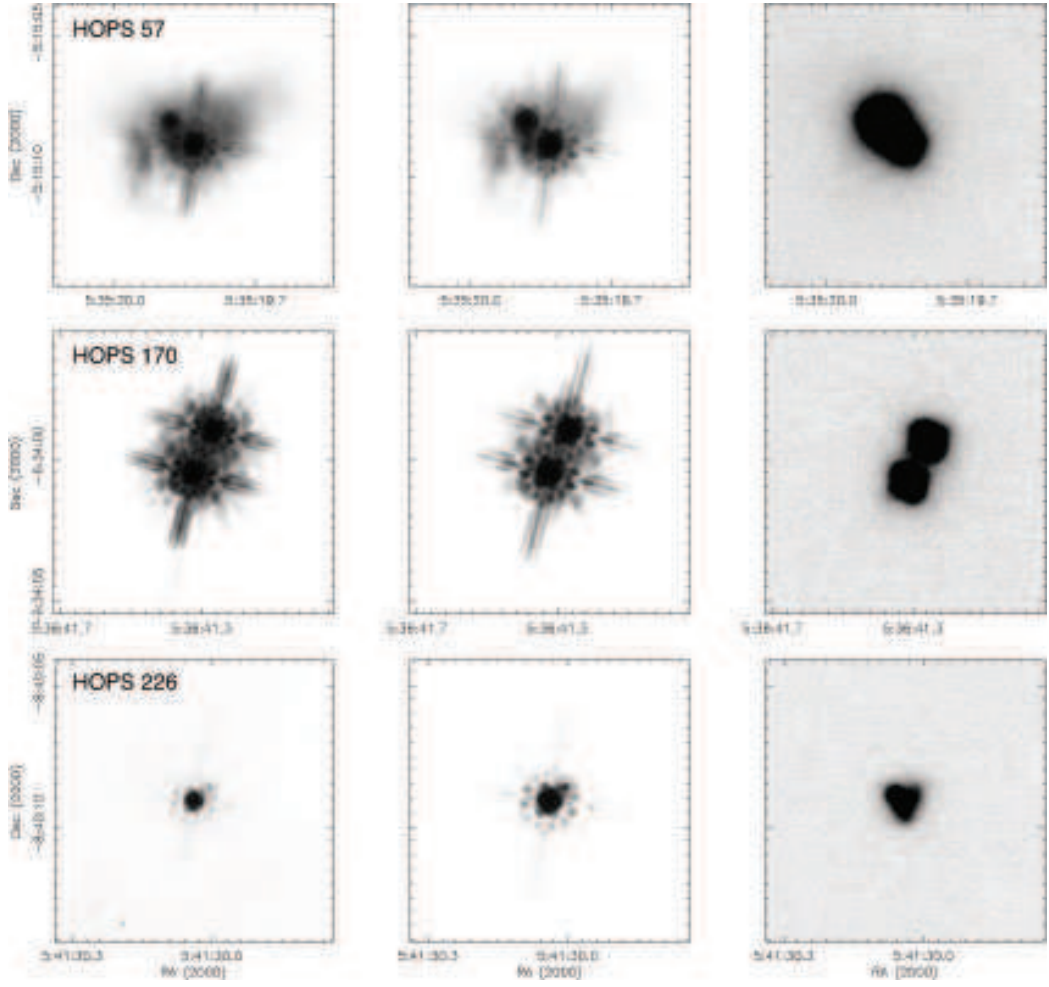} 
\caption{F160W, F205W and $L'$-band images of binary systems imaged with NICMOS and NSFCAM2.\label{fig:a3}}
\end{figure}

\begin{figure}
\epsscale{0.9}
\plotone{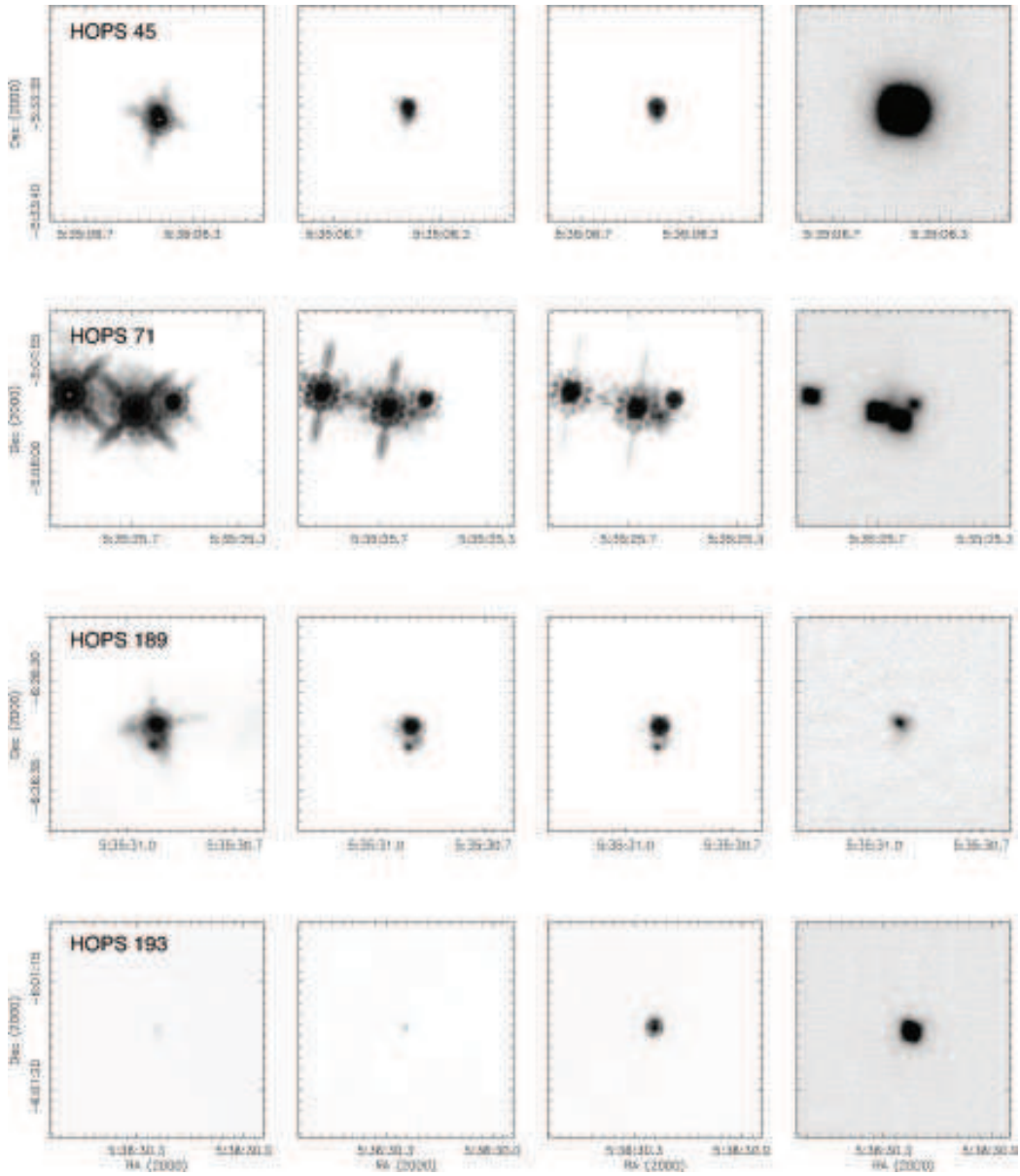} 
\caption{F160W, F205W and $L'$-band images of binary systems imaged with all three cameras.\label{fig:a4}}
\label{fig:figs2}
\end{figure}

We present the numbers of primaries and candidate companions between 100 and 1000 AU in Table~\ref{tab:comp_combined}; we will further motivate these limits in the following sub-section. The numbers are determined for three different permutations of our combined IRTF, NICMOS and WFC3 data set: the combined set from all three cameras, the NICMOS and WFC3 data, and the WFC3 data only. While the combined data from all three cameras (hereafter the {\it combined} sample) gives us the most companions, the IRTF angular resolution is lower than that of the WFC3 and NICMOS data, motivating a combined look at the data from the two HST cameras alone (hereafter the {\it HST} sample). Finally, the WFC3 data has the highest sensitivity and it has the largest FOV, and hence it is worthwhile looking at the data from this camera alone (hereafter the {\it WFC3} sample). All but one of the pre-main sequence stars with disks are found only in the WFC3 sample; the remaining one is in the IRTF sample. The images of the candidate multiples obtained with the three cameras are shown in Figures~\ref{fig:figs1} through \ref{fig:figs2}.

\subsection{The Discovery Space of Companions}

The detection of faint companions can be limited by confusion with the structure in the wings of the primary's PSF. To determine the detection limits for companions as a function of projected separation, we conducted a fake star analysis on the WFC3 data. We created a PSF using the IDL implementation of the DAOPHOT package utilizing a densely populated field in the WFC3 data. The PSF was then added to the vicinity of four isolated stars in different WFC3 fields. The PSFs were added at different projected separations, ranging from 45 to 1000 AU with a 5 AU increment. At each separation, the PSF was added to 12 positions. This process was repeated with fake stars that were 0 to 8 magnitudes fainter than the four stars, in steps of 0.5 magnitudes. The fake stars were then recovered with PhotVis using the same search parameters as the WFC3 data. The fraction of sources identified for each combination of projected separations and $\Delta m_{160}$ was then determined, where $\Delta m_{160}$ is the fake star magnitude minus the primary magnitude. The results at each separation were averaged over concentric annuli with a uniform spacing in separation. Since the magnitudes of the faintest companions we can detect depends on the brightness of the primary, the detection fractions are also averaged over uniform intervals of $\Delta m_{160}$. 

The curves of the projected separations vs $\Delta m_{160}$ determined from this analysis are shown in Figure~\ref{fig:wfc3_comp}. Overlaid are the separations and $\Delta m_{160}$ for the the primaries and all the candidate companions. In addition, double stars found around the remaining stars in the observed fields are shown: these may be visual binaries around pre-main sequence stars without disks or they may be optical doubles due to random coincidences between stars in the line of sight. In either case, they demonstrate that the discovery space delimited by the fakestar test is consistent with the observed visual binaries and optical double stars. 

The 99\% completeness limits for the WFC3 data range from $\Delta m_{160} \sim 2.5$~mag at 200 AU, to 5~mag at 500 AU, and to almost 7~mag at 1000~AU. The resulting mass limits depend on the brightness of the primary, the amount of extinction, and the age of the source. Given that the last two cannot be ascertained directly from this data, we cannot uniquely determine the mass ratio probed. However, a $\Delta m_{160}$ range of 2.5~mag translates into a mass range of approximately a factor of 3 to 5, a range of 5~mag implies a mass range covering approximately a factor of 15, and 7~mag entails a mass range in excess of a factor of 50 \citep[using the pre-main sequence tracks of][]{1998A&A...337..403B}. 

To determine if the completeness values are comparable for NICMOS, we repeated the analysis on the NICMOS data for a smaller number of separations and mass ratios. Unlike the WFC3 data, the sources in the NICMOS data were identified by visual inspection. The analysis showed that the completeness of NICMOS was similar to that of the WFC3. A more detailed analysis will be done as part of a future paper of the separation and $\Delta$mag distributions. In contrast, the magnitude range and separations probed by the IRTF data are much more limited; these data were primarily obtained to provide color information for the companions. For this reason, we do not include the sources only imaged by the IRTF in the following analysis. We also limit our analysis to the candidate companions with projected separation of 100 to 1000 AU. Since the density of sources at larger radii are dominated by either field stars or other YSOs that happen to be found in projection near the targeted source, we cannot look for companions at wider separations with these data. For separations less than 100~AU, we are only sensitive to mass ratios close to unity and thus highly incomplete. 

\subsection{Correcting for Line of Sight Contamination}
\label{sec:contamination}

The number of companions must be corrected for chance coincidences with stars in the line of sight. To do this, we estimate the surface density of the contaminants by counting the number of sources detected in the annulus between 2000 and 5000~AU and summing over all the primaries. At these separations, the mean surface density of companions is flat and does not increase with decreasing separation, indicating that we no longer detect signifiant numbers of companions (Figure~\ref{fig:mean1}). The density is measured near enough to the primary that it provides an accurate assessment of the surface density contaminants. We then calculate the number of companions from the equation 

\begin{equation}
N_{comp} = N_{cand}-N_{cont} \left(\frac{\Omega_{comp}}{\Omega_{cont}}\right),
\end{equation}

\noindent
where $N_{cand}$ is the number of candidate companions with projected separations from their primaries between 100 and 1000 AU, and $N_{cont}$ is the number of sources in the region between 2000 and 5000~AU). 

We assume that $N_{cont}$ gives the number of contaminating stars, either foreground and background stars or other Orion YSOs in the line of sight, which we scale by the ratios of the solid angles of the companion and contamination regions to estimate the number of contaminating objects in the companion region. If we ignore the effect of incompleteness, the ratio of solid angles becomes the area of the annulus extending from 100 to 1000 AU for the companions over the area of the annulus extending from 2000 to 5000 AU for the contaminants:

\begin{equation}
\frac{\Omega_{comp}}{\Omega_{cont}} = \left(\frac{1000^2-100^2}{5000^2-2000^2}\right) = 0.047.
\end{equation}

\noindent 
The number of contaminants in the 2000-5000 AU annulus and the corrected number of companions are given in Table~\ref{tab:comp_combined} for each of the three samples. The contamination is the total number of contaminants for all of the primaries. The number of contaminants is small due to the high extinction towards the YSOs: 0.04 companions per primary between 100 and 1000 AU. Given the small number of companions, however, as many as 26\% of the companions may be contaminants. Thus, the primary uncertainty in the number of companions is due to the contamination subtraction. In Appendix A, we discuss how we measure the uncertainty in the contamination subtraction using the Poisson statistics of the line of sight contamination.
 
Since the reduction of completeness near the primary can also limit the detection of contaminants, the number of contaminants at 100 to 1000 AU can be overestimated. Due to confusion with the primary's PSF, fainter stars can be detected in the 2000 to 5000 AU annulus, where we measure the surface density of contaminants, than can be detected within 1000 AU of the primary. This will result in an over-subtraction in the number of contaminants, and therefore the number of companions in Table~\ref{tab:comp_combined} are lower limits to the actual number of companions. The over-subtraction will be partially mitigated by the uniform spatial distribution of contaminants, which implies that most of the contamination will be at the largest separations that subtends the largest angle on the sky. 

\begin{figure}
\epsscale{0.5}
\plotone{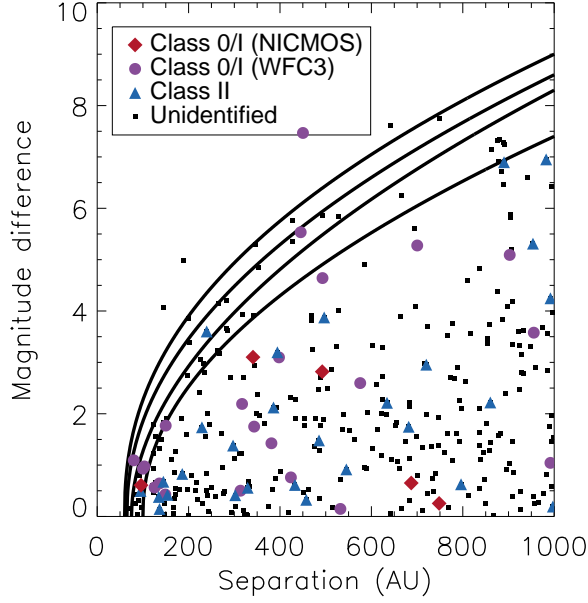} 
\caption{Completeness as a function of separation and $\Delta m_{F160W}$ performed by adding fakestars to the WFC3 data. The contours give the boundaries region of separation and $\Delta m_{F160W}$ space where 99\%, 75\%, 50\% and 25\% of the fakestars are detected. Overplotted are all the apparent systems with projected separation at the distance of Orion of less than 1000 AU. Red diamonds represent identified binary systems found in WFC3 data around protostars whereas blue triangles are systems around pre-main sequence stars with disks. Black dots are visual binaries for sources which are not dusty YSOs, many of these may be optical binaries. Binary systems found only in NICMOS data are plotted on the same scale in purple for comparison.}
\label{fig:wfc3_comp}
\end{figure}

To correct for this bias in the WFC3 sample, we repeat our estimates of the number of companions and the $CF$s. We use the 25\% and 99\% completeness curves for the WFC3 data to determine the inner radii for calculating $\Omega_{comp}$ (Figure~\ref{fig:wfc3_comp}). The correction is calculated by determining $\Delta m_{160}$ between each object in the 2000 to 5000~AU annulus and the targeted primary star in the center of the annulus, and then adopting the criteria that contaminating objects in the line of sight can only be detected outside a limiting radius where the completeness is $\ge 99\%$ for that value of $\Delta m_{160}$. We determine the ratio with the equation:

\begin{equation}
\frac{\Omega_{comp}}{\Omega_{cont}} = \frac{1}{N_{cont}} \sum^{N_{cont}}_{i=1} \left(\frac{1000^2-r_i^2}{5000^2-2000^2}\right),
\end{equation}

\noindent
where $r_{i}$ is set to $r_{99\%}$ for the i$^{th}$ object in the 2000-5000 AU annulus, down to a minimum value of 100~AU. The number of contaminates is reduced by a factor of 0.46 for protostars and a factor of 0.42 for pre-main sequence stars. This analysis was repeated using the $\Delta m_{160}$ vs radius curve for a 25\% level of completeness by adopting $r_{25\%}$ for the inner radius (Figure~\ref{fig:wfc3_comp}). This reduces the number of contaminants by a factor of 0.6 for protostars and 0.56 for pre-main sequence stars. By reducing the estimated number of line of sight contaminants, the number of companions to protostars increases from 15.9 to 18 for $r_{25\%}$ and 18.6 for $r_{99\%}$, and the number of companions to pre main sequence stars increases from 20.1 to 23.6 and 24.7 for $r_{25\%}$ and $r_{99\%}$, respectively. This an increase of 1.35$\sigma$ for the protostars and 1.5$\sigma$ for the pre-main sequence stars. The number of candidate companions and the level contamination for the different choices of $r_{inner}$ are given for the WFC3 sample
in Table~\ref{tab:comp_wfc3}. Again, the determination of the uncertainties is described in Appendix~A. 

\subsection{The Companion Fraction}
\label{sec:CF}

The incidence of multiplicity among a population of stars can be quantified by the companion fraction, or $CF$, which is defined by the ratio of the number of companions to the total number of single stars and systems,

\begin{equation}
CF = \frac{B + 2T + 3Q}{S+B+T+Q},
\end{equation}

\noindent
where $S$ is the number of single star systems, $B$ the number of binary systems, $T$ the number of ternary systems, and $Q$ the number of quaternary systems \citep[e.g.][]{1997ApJ...490..353G}. (This equation can be expanded to include larger systems; however, there are only binary and ternary systems in our sample.) The $CF$ can be calculated by finding the total number of candidate companions with separations between 100 and 1000 AU and then subtracting the estimated number of line of sight contaminants. Alternatively, we can think of the $CF$ as the integral of the mean surface density of companions minus the average density of line of sight contaminants, over an integration range of 100 to 1000 AU (Figure~\ref{fig:mean1}). Multiplicity fraction (MF) is also frequently used in the literature; this is given by the fraction of stars hosting multiple systems and counts each system only once. Since the $CF$ is derived from the surface density of point sources from which we have identified the Orion companions, it is the natural statistic to use in our analysis. The $CF$ is then

\begin{equation} 
CF = \frac{N_{comp}}{N_{prim}}-\frac{N_{cont}}{N_{prim}}\left(\frac{\Omega_{comp}}{\Omega_{cont}}\right),
\label{eqn:cf}
\end{equation}

\noindent
where $N_{prim}$ is the number of primaries. The calculation of the uncertainties for the $CF$ is described in Appendix~B.

The $CF$s are given in Table~\ref{tab:csf_all} for an $r_{inner} = 100$~AU. The $CF$ of all the dusty YSOs is $10.3^{+1.0}_{-1.0}\%$ for the combined sample, $10.8^{+1.0}_{-0.9}\%$ for the HST sample, and $11^{+1.2}_{-1.2}\%$ for the WFC3 sample. Thus, the $CF$ does not change significantly between the different samples. For the HST sample, the $CF$s are $11.4^{+1.5}_{-1.3}\%$ for protostars and $10.2^{+1.5}_{-1.6}\%$ for the pre-main sequence stars. Although the $CF$ for pre-main sequence stars is slightly lower, the $CF$s are consistent within the uncertainties. The $CF$s for different choices of $r_{inner}$ are given for the WFC3 sample in Table~5. $CF$ increases from $12.3^{+1.7}_{-1.4}\%$ to $13.9^{+1.6}_{-1.5}\%$ and $14.4^{+1.1}_{-1.3}\%$ for the protostars, using $r_{25\%}$ and $r_{99\%}$, respectively (Table~\ref{tab:csf_wfc3}). For the pre-main sequence stars, the $CF$ increases from $10.2^{+1.5}_{-1.6}\%$ to $12^{+1.2}_{-1.3}\%$ and $12.5^{+1.2}_{-0.8}\%$ for $r_{25\%}$ and $r_{99\%}$, respectively (Table~\ref{tab:csf_wfc3}). The $CF$s for the pre-main sequence stars and protostars both increase and remain consistent, and the absence of a significant difference between the two evolutionary stages persists. 

\subsection{The Companion Fraction as a Function of Environment}
\label{sec:csf_env}
 
A comparison of the $CF$ in regions of high and low stellar density is motivated by the observed variations in the protostellar luminosity function with stellar density discovered by \citet{2012AJ....144...31K}. Although the changes in the protostellar luminosity function may not be directly related to changes in the multiplicity, the variations in the luminosity function indicate that the low mass star formation process is altered by the environments found in regions of low and high stellar density. To compare the incidence of multiplicity in regions of low and high stellar density, we divide our sample of YSOs by adopting an analysis similar to that of \citet{2012AJ....144...31K}. Specifically, we used the {\it Spitzer} Orion Survey point source catalog of \citet{2012AJ....144..192M} to determine the projected distance to the 4th nearest {\it Spitzer} YSO, hereafter NN5, for each primary or single star in the HST survey\footnote{We adopt here the nomenclature of \citet{2009ApJS..184...18G} who use NN5 for the 4th nearest neighbor}. The fourth nearest neighbor was chosen since it is not biased by binary, ternary, or quaternary systems, and yet still measures the local density of YSOs surrounding a particular object. When computing NN5, we do not include any of the sources that were identified as likely companions since none of them have been resolved in the {\it Spitzer} sample. We then use the NN5 to divide the YSOs into high and low density regions by whether or not their nearest neighbor density, as given by $3/(\pi r_{NN5}^2)$, exceeds the threshold surface density (see Megeath et al. 2016 for the definition of the nearest neighbor density adopted in this paper). 

\begin{figure}
\epsscale{1.}
\plotone{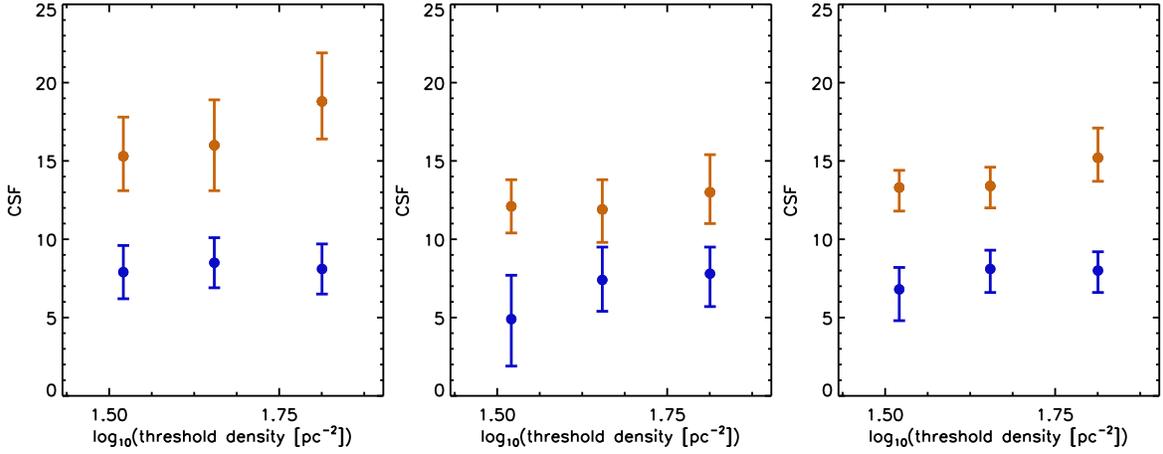}
\caption{The $CF$ for the HST sample as a function of the threshold surface density used to divide the sample into those found in high YSO surface density regions (orange) and those in low YSO density regions (blue). This is shown for the sample of protostars (left panel), pre-main sequence stars with disks (middle panel), and the merged sample of all dusty YSOs (right panel). The error bars show the 1 $\sigma$ uncertainty in the subtraction of the line of sight contamination (Appendix~B).}
\label{fig:csf_env}
\end{figure}

In Figure~\ref{fig:csf_env}, we plot the $CF$s of the HST sample for the high and low surface density regions for threshold densities corresponding to $NN5 = 25000$~AU (65~pc$^{-2}$), 30000~AU (45~pc$^{-2}$) and 35000~AU (33~pc$^{-2}$); these values bracket the NN5 where the number of objects in high and low density regions are equal. For each of these we include error bars for the line of sight contamination. In this figure, we set $r_{inner} = 100$~AU which implies there may be an over subtraction of the contamination. The subtraction of the line of sight contamination ensures that the variations in the $CF$ are not due to a higher number of chance coincidences with other YSOs in regions of higher density. We find that for the range of threshold NN5 values, the protostars and young stars with disks found in dense regions have a higher of $CF$ than the low density regions. Note that the $CF$s for the protostars, pre-main sequence stars with disks, and combined dusty YSOs in the high density regions are consistent with each other within the uncertainties, as are the $CF$s of the protostars, disks, and dusty YSOs in the low density regions. 

\begin{figure}
\epsscale{0.7}
\plotone{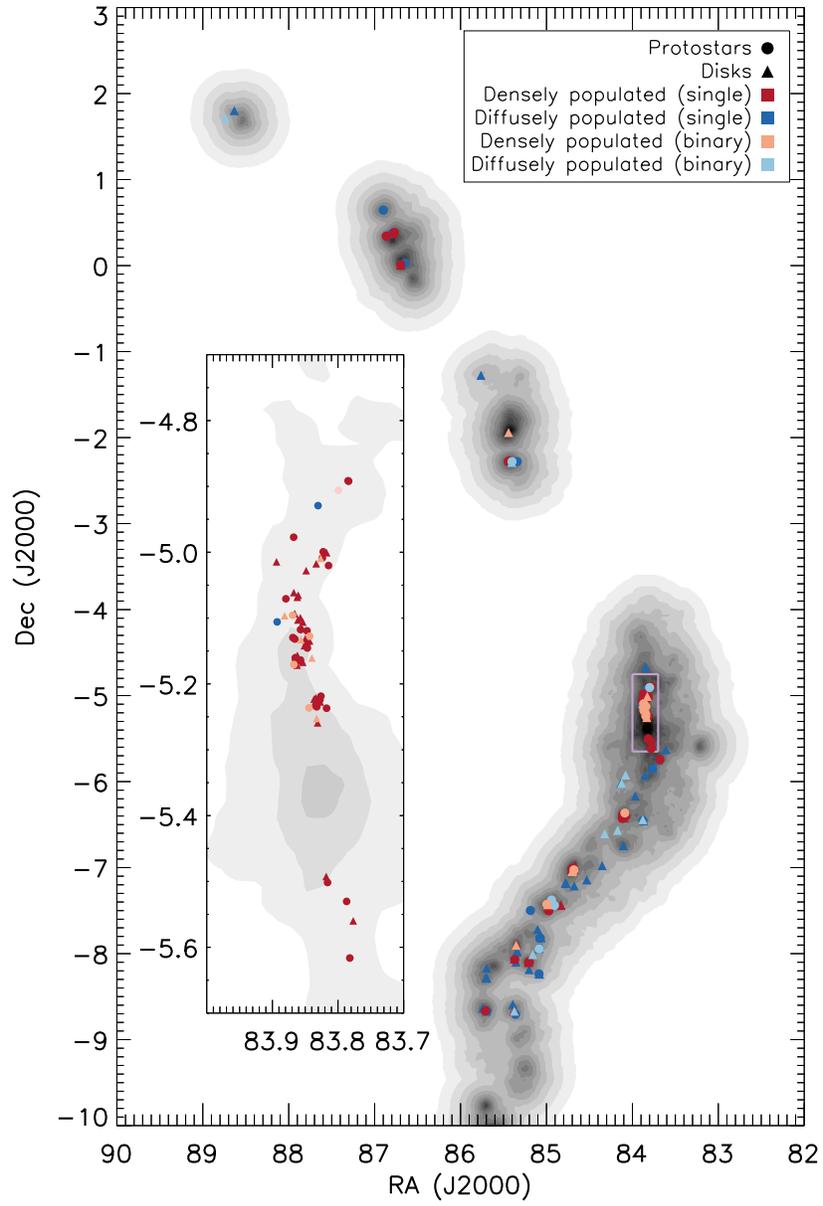}
\caption{Spatial distribution of the surveyed YSOs in the regions of high and low stellar density, distinguishing sources with and without companions. Insert shows a close-up look at the crowded ONC region, including OMC 1, 2, 3 and 4. The grey background shows the nearest neighbor surface density of dusty YSOs from Megeath et al. (2016)}
\label{fig:map}
\end{figure}

A more rigorous analysis can be done with the WFC3 data using different values of $r_{inner}$. In Table~\ref{tab:csf_wfc3}, we show the number of primaries and companions in the low and high density regions as a function of $r_{inner}$. The resulting ratios,

\begin{equation}
R = \frac{CF(\Sigma > 45~{\rm pc}^{-2})}{CF(\Sigma < 45~{\rm pc}^{-2})},
\end{equation}

\noindent
are calculated using a Bayesian parameter estimation in Appendix B and presented in Table~\ref{tab:ratio_wfc3_limit}. Since the ratio of $CF$s does not appear strongly dependent on the threshold NN5 value, we use the $45~{\rm pc}^{-2}$ density threshold only. The spatial distribution of YSOs in our sample based on this threshold is presented in the Figure \ref{fig:map}. As we adjust the value of $r_{inner}$ from 100~AU to $r_{99\%}$ , the median value of R decreases from 1.96 to 1.74 for protostars, 1.5 to 1.28 for pre-mains sequence stars, and from 1.64 to 1.45 for the merged samples. Each of these three samples consistently show an elevated $CF$ in the dense environments, although the change in $CF$ is highest for protostars. 

The primary uncertainty in the $CF$ is Poisson fluctuations in the amount of contamination (Appendix A); hence, we estimate the significance of this enhancement with respect to these fluctuations using three methods. First, our parameter estimation determines the probability distribution of $R$ (Table~\ref{tab:ratio_wfc3_limit}, Appendix B). These show that the probability for $R \le 1$ depends on both $R_{inner}$ and the sample. For protostars, the probability decreases from 0.01 to 0.001 as we change $r_{inner}$ from 100 AU to $r_{99\%}$. Similarly, for stars with disks, the probability decreases from 0.08 to 0.01 as we change $r_{inner}$. Finally, for the merged sample, the probability declines from 0.008 to 0.0004. Parameter estimation techniques, however, are not always appropriate for hypothesis testing. Thus, we also employ a Bayesian hypothesis test to determine the probability of $R \le 1$ (Appendix B). We find that for protostars the probability that $R \le 1$ ranges from 0.05 to 0.007 as we change $r_{inner}$ from 100 AU to $r_{99\%}$, for pre-main sequence stars the probability ranges from 0.22 to 0.008, and for the merged sample, the probability ranges from 0.036 to 0.004. Hence, the significance of the result remains. Finally, we employ a frequentist hypothesis test to determine the probability of the null hypothesis: that the $CF$ is the same in both environments (Appendix B). The probability that R=1 change ranges from 0.008 to 0.0007 for protostars as we change $r_{inner}$ from 100 AU to $r_{99\%}$, from 0.07 to 0.04 for disks, and from 0.009 to 0.0008 for the merged sample. We conclude that the variation in the $CF$ between high and low density regions is significant and is highly unlikely to result from fluctuations in the contamination.

\subsection{The Companion Probability as a Function of Environment}
\label{sec:CP}

The uncertainties in the number of companions and $CF$s in Tables~\ref{tab:comp_combined}, \ref{tab:comp_wfc3}, \ref{tab:csf_all} and \ref{tab:csf_wfc3} are the uncertainties from the contamination correction: no additional sources of uncertainty are included. In the literature, the number of companions typically include an additional uncertainty typically given by $\sqrt{P N_{prim}} = \sqrt{N_{comp}}$, or more correctly by $\sqrt{(1-P) P N_{prim}}$ \citep{2003ApJ...586..512B}. This additional uncertainty comes from modeling the presence of a companion as the result of a Bernoulli trial. In this model, the number of companions is described by the binomial distribution with the number of trials equal to the number of primaries,  the number of successful trials equal to the number of companions, and $P$ being the probability of having a companion. Modeling the presence of a companion as a Bernoulli trial has two possible physical interpretations. First, the formation of a companion can be approximated as a random process, which has a probability $P$ of creating a companion between 100 and 1000 AU. Alternatively, if the observed primaries are drawn from a much larger population of primaries, in which a fraction $P$ have companions between 100 and 1000 AU, then the choice of a primary with a companion could once again be treated as a Bernoulli trial. In both cases, the uncertainty in the number of companions, ignoring the uncertainty from contamination subtraction, is given by $\sqrt{(1-P) P N_{prim}}$ \citep{2003ApJ...586..512B}.

The modeling of the presence of a binary as a Bernoulli trial may not always be appropriate. In the first interpretation, the presence of a companion is assumed to be the result of a random process with probability $P$; this may not be true if the presence of a companion is strongly influenced by external, environmental factors. In the second interpretation, the observed sample is assumed to be drawn from a much larger population. This may be relevant for binary surveys of field stars; however, since we observed most of the protostars in the Orion molecular cloud outside of the ONC, the assumption that we are drawing primaries randomly from a much larger population is not appropriate for our analysis of Orion protostars. It is, however, appropriate for our sample of pre-main sequence stars which are drawn from a much larger population. 

For these reasons we have not included binomial statistics in our analysis of the $CF$ and the ratio of the $CF$s in different environment. This motivates the question whether the difference in the $CF$ would be significant if binomial statistics were included. To address this question, we define the companion star probability, or $CP$, as the probability of a star having a companion. We assume the presence of a companion can be treated as a random process similar to a Bernoulli trial, but with the additional outcomes of a ternary and quaternary system. The $CP$ can be written as 

\begin{equation}
CP = P_{binary}+2 P_{ternary} + 3 P_{quatenary},
\end{equation}

\noindent where $P_{binary}$ is the probability of a binary system, $P_{ternary}$ is the probability of a ternary system, and $P_{quaternary}$ is the probability of a quaternary system. (Additional terms can be added for larger system). The average number of companions is given by

\begin{equation}
N_{comp} = CP \times N_{prim} .
\end{equation} 

\noindent
If we adopt a particular value for the $CP$, the resulting $CF$s follow a binomial distribution with a probability equal to the $CP$.

To examine whether the $CP$ varies between the high and low density samples, we repeat the analysis applied to the $CF$ for the $CP$ (Appendix~C). First, we use parameter estimation to estimate the value of $R_{CP}$ by substituting $CP$ for $CF$ in Equation 7. In Table~\ref{tab:ratio_wfc3_limit}, we find that the median value of $R_{CP}$ is within 4\% of the value given by the $R$ determined with the $CF$s. However, the significance of the result is lower: the probability that $R_{CP} \le 1$ is 0.11-0.10 for protostars, 0.22-0.26 for disks and 0.10-0.11 for the merged sample. 

If we apply a Bayesian hypothesis test to the $CP$, we must consider the probability that $CP_{high} > CP_{low}$, that $CP_{high} = CP_{low}$ and that $CP_{high} < CP_{low}$ (in contrast, $CF_{high}$ cannot equal $CF_{low}$ since the number of contaminating stars can only be an integer and the resulting $CF$s of high and low density regions cannot be exactly the same, see Appendix B). Since the $R_{CP}=1$ hypothesis has half the number of parameters (i.e. a single value of $CP$ vs independent values of $CP$ for the high and low density regions), the resulting Ockham factor favors the $R_{CP}=1$ hypothesis: the probability that $R_{CP} = 1$ is 0.48-0.58 for protostars, 0.60-0.63 for disks and 0.58-0.61. These assume that the prior probability for $CP$ is uniformly distributed between 0 and 1. Finally, we can use our frequentist hypothesis test (Appendix C) to find that the probability that $R _{CP}= 1$. This probability is 0.10-0.12 for protostars, 0.17-0.20 for disks, and 0.07-0.09 for the merged sample. Thus, despite having one of the largest surveys for multiplicity to date, a larger sample is needed. We note that if we used the combined HST sample and $r_{inner}= 100$~AU; the frequentist hypothesis test for the merged sample gives a probability as low as 0.03 that $R_{CP} = 1$; suggesting that the differences in the $CP$ become more significant as we increase the sample size. A future investigation will employ the complete HST data, taking into account the different sensitivity levels, to re-examine variations in the $CP$ and its significance.

\section{The Properties of the Companions}
\label{sec:prop}

Our information on the properties of candidate companions is limited by the small number of bandpasses through which the multiple systems were resolved; the 1000 AU outer separation limit for our companions precludes their detection by the {\it Spitzer} space telescope or {\it Herschel} space observatory. For four companions we do not have a F160W magnitude; three of these four were detected with NICMOS in the F205W filter but they were too red to detect in the F160W filter. The other source, HOPS 304, was imaged only with NSFCAM2. For the subset of protostars observed with NICMOS, we have F160W and F205W magnitudes, and hence the additional information of the $m_{F160W}-m_{F205W}$ color. For YSOs detected by WFC3 and NSFCAM2, we have the $m_{F160W}-m_{L'}$ color, while for protostars detected with NICMOS and NSFCAM2 we have the $m_{F160W}-m_{F205W}$ and $m_{F160W}-m_{L'}$ colors. In this section, we examine the colors of stars and their companions to assess their properties.

\begin{figure}
\epsscale{0.5}
\plotone{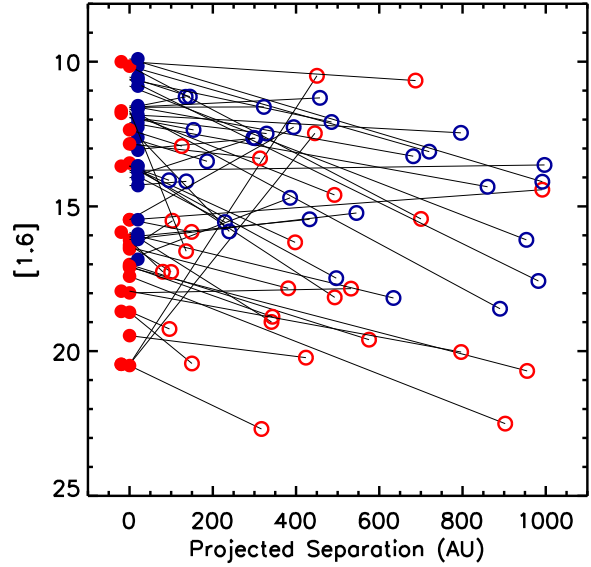}
\caption{The separation of the observed systems vs the F160W magnitudes of the primaries and companions. The red circles are the protostars and the blue circles are the pre-main sequence stars with disks. The solid circles mark the primaries, which are selected to be closest to the {\it Spitzer} position. The open circles give the companions which are connected to their primaries by the lines. The primaries are all at position $0$, but the markers of the protostars and pre-main sequence stars are offset for clarity.}
\label{fig:sep_mag}
\end{figure}

Figure~\ref{fig:sep_mag} shows the separation of the systems vs the F160W magnitudes of both the primaries and their companions. The primaries are defined as the sources closer to the coordinates given by the {\it Spitzer} observations since these sources are likely brighter in the mid-IR and, at least in the case of protostars, are expected dominate the luminosities of the systems. As expected, the secondaries are generally fainter than the primaries at F160W; however, in 6 protostellar systems and in 10 disk systems this is not the case. For the protostars, this can occur if the companions are less obscured than the primary. For the pre-main sequence stars, this typically only occurs when the companion and primary have a $\Delta m_{160} < 0.6$~mag. (although in one case $\Delta m_{160} = 2$~mag.). In these cases, it's not clear which is the companion and which is the primary; however, this does not affect any of the conclusions of this paper. 

\begin{figure}
 \centering
			\subfigure{\includegraphics[width=0.3\textwidth]{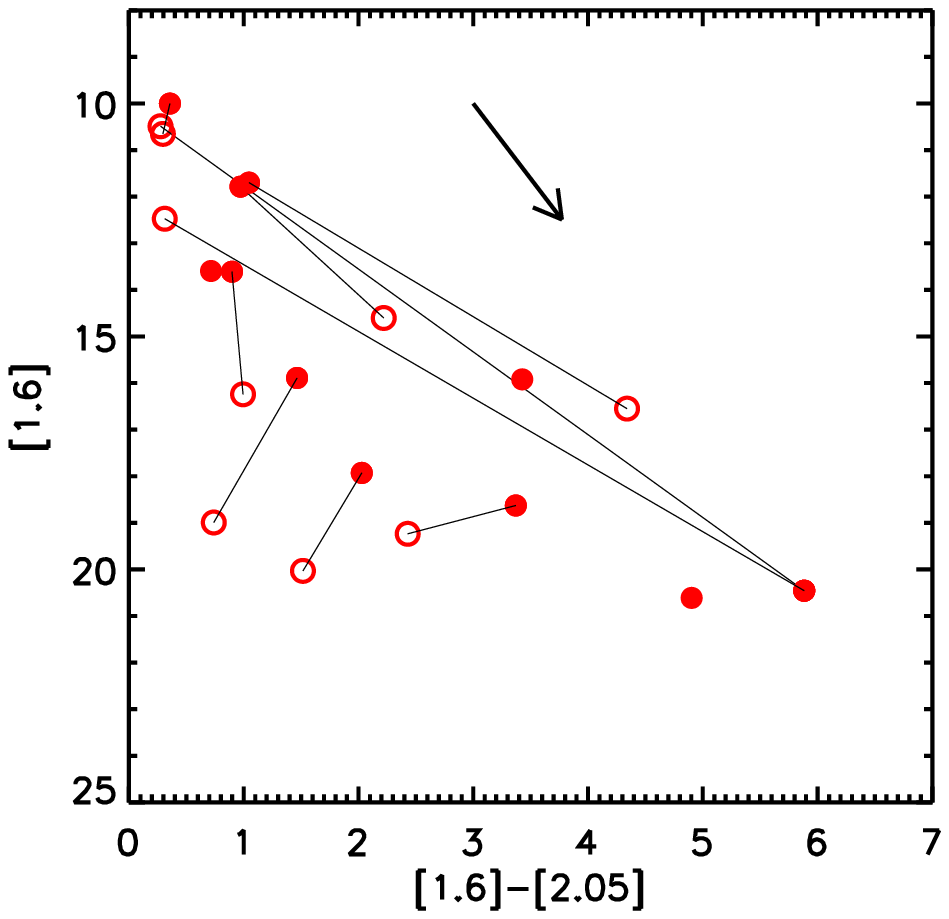}}				
			\subfigure{\includegraphics[width=0.3\textwidth]{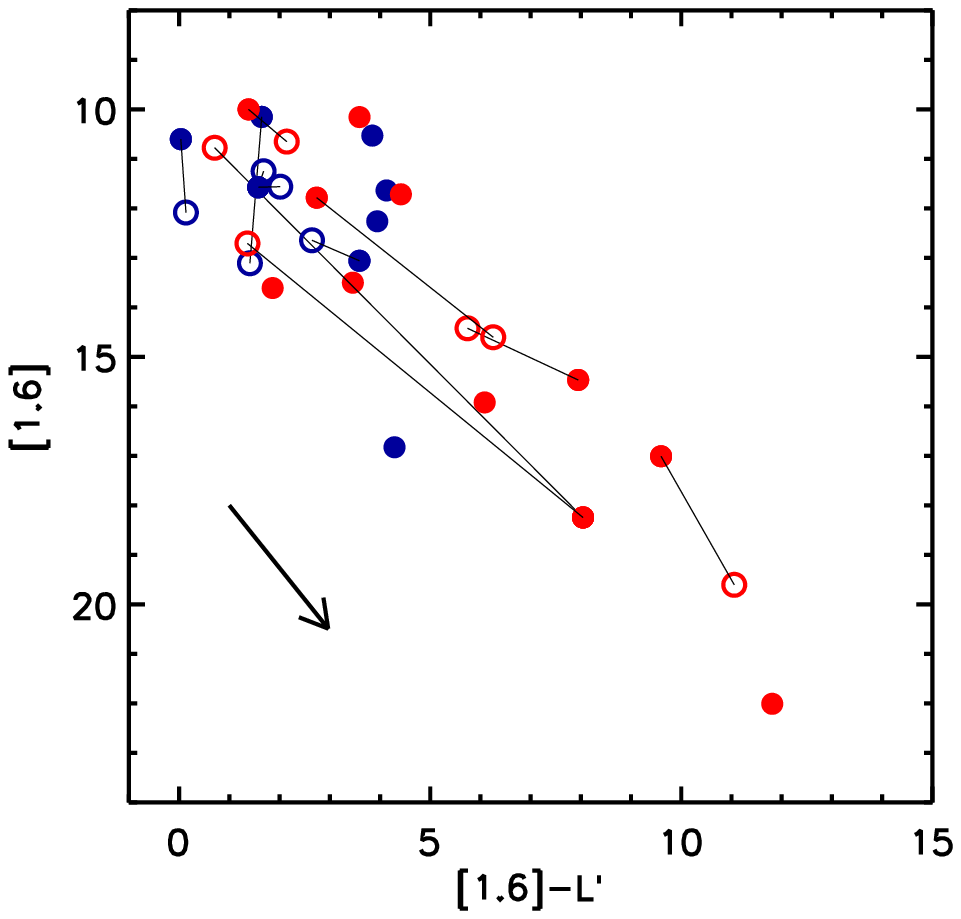}}
 	 		\subfigure{\includegraphics[width=0.3\textwidth]{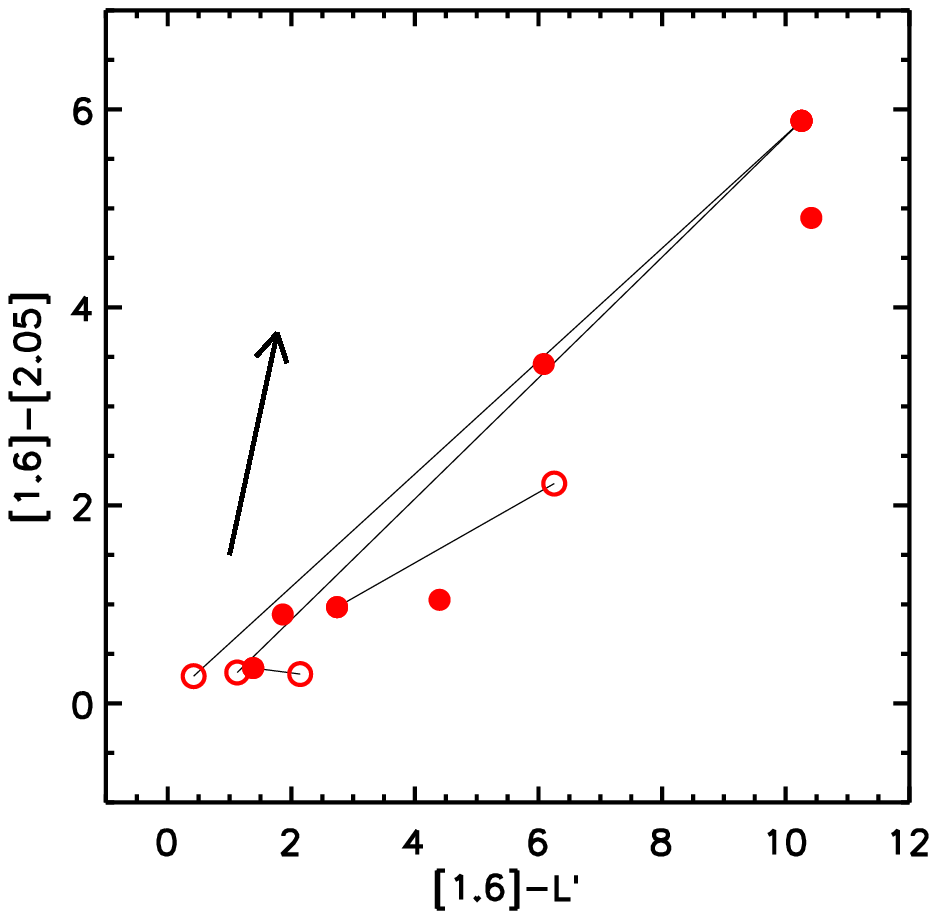}} 

\caption{Color-magnitude and color-color diagrams of primaries and companions. In all three diagrams, the solid circles mark the primaries, which are selected to be closest to the {\it Spitzer} position. The open circles give the companions which are connected to the primaries by the line. The extinction vector for an $A_K = 2$~mag is displayed in all three figures. {\bf Left:} a F160W vs F160W-F205W diagram for the protostars and companions observed with NICMOS. {\bf Middle:} a F160W vs F160W-L' diagram for the sources observed with the HST and NSFCAM2. The red circles are the protostars and the blue circles are the pre-main sequence stars with disks. {\bf Right:} Color-color diagram for protostars observed with NICMOS and NSFCAM2. In all three plots, the ternary system of HOPS 71 is distinguished as having a the reddest protostellar primary in the sample associated with two of the bluest companions.}
\label{fig:color_mag_color}
\end{figure}

For the sources observed with NICMOS, we show the $m_{F160W}-m_{F205W}$ vs $m_{F160W}$ color-magnitude diagram for 8 protostellar systems in Figure~\ref{fig:color_mag_color}. With one notable exception, the ternary system HOPS 71, they show companions fainter than the primary. The colors for the companions to the protostars can be significantly different than the primaries, with values ranging from 3.3~mag redder to 5.6~mag bluer than the primary. With the exception to the two companions to the ternary system HOPS 71, the companions have red colors consistent with protostars. Furthermore, since the colors of protostars are largely determined by extinction and scattering in their local envelopes of infalling gas and dust (e.g. Furlan et al. submitted), the significant differences observed between the companions and primaries is also consistent with both sources being protostars. Due to its small field of view, no pre-main sequence objects with disks were serendipitously observed by NICMOS.

Figure~\ref{fig:color_mag_color} shows the $m_{F160W}$ vs $m_{F160W}-m_{L'}$ diagram for the all sources found in both the WFC3 and NSFCAM2 images; this sample includes sources without companions. All five companions to pre-main sequence stars with disks appear to have similar colors to the primaries; the apparent differences are probably due to differences in the emission from the inner disks of these sources. In contrast, the protostars show a significant amount of variation which probably results from extinction, scattering and emission by dust grains in the infalling envelope surrounding each protostar. Again, with the exception of HOPS~71, the red colors of the companions to the protostars and the large differences in colors between the companions and primaries are consistent with the companions being protostars as well. However, we cannot rule out on the basis of this diagram that the companions are not pre-main sequence stars with disks or even reddened photospheres without circumstellar disks. 

Finally, Figure~\ref{fig:color_mag_color} also shows the $m_{F205W}-m_{L'}$ vs $m_{F160W}-m_{F205W}$ diagram. This diagram only includes sources observed with NICMOS and thus only contains protostars. Due to low resolution of NSFCAM2, there was sufficient data only for 3 systems. Through this analysis, the system that stands out the most is HOPS 71. The protostar in the system appears to be very reddened, with $m_{F160W}-m_{F205W}=5.9$~mag. and $m_{F205W}-m_{L'}=5.4$~mag. On the other hand, both candidate companions show photosphere-like colors and appear to be significantly brighter at F160W and F205W than the primary. Follow-up observations are needed to confirm that the HOPS 71 system is not the result of a chance alignment. If this is a bound system, it could possibly result from non-coeval evolution, the rapid dispersal of the protostellar envelopes of the two blue companions, or perhaps a viewing angle in which the two companions are seen through outflow cavities carved in the surrounding envelopes. 

\section{Discussion}
\label{sec:discuss}

This paper establishes the Orion molecular clouds as an important laboratory for studying multiplicity across a diverse range of star-forming environments. Orion contains the largest sample of protostars within 500 pc of the Sun; this paper presents a survey of 177 protostars in which the central protostar is detected at 1.6~$\mu$m. This is the largest survey of multiplicity towards protostars in a single region to date. The clouds also contain the largest sample of young stars with disks within 500 pc. Although we have only imaged 197 pre-main sequence stars with HST, a small fraction of the total population of pre-main sequence stars, the size of our sample is second only to the HST/ACS survey of the ONC by \citet{2007AJ....134.2272R}. Furthermore, our survey focuses on star formation in Orion outside the ONC and therefore encompasses a much broader range of environments and evolutionary stages than found in the ONC by \citet{2007AJ....134.2272R}. Employing an analysis of the mean surface density of companions, we have identified a population of companions between 80 and 1000 AU. The 80 AU limit is set by the resolution limit of the data. The outer limit is where the surface density of companions becomes comparable to the surface density of line of sight contaminants; at larger separations we are no longer able to reliable identify companions. We concentrate our analysis on the 100 to 1000 AU sample since the survey is sensitive to only a small range of mass ratios at distances $< 100$~AU. 

\subsection{The Mean Frequency of Companions: Comparisons to Previous Work}
\label{sec:comp_mfc}

The analysis of the mean frequency of companions for pre-main sequence stars was pioneered by \citet{1995MNRAS.272..213L} using existing catalogs of the Taurus-Auriga region. Over spatial scales of 5 AU to 5 pc, Larson found that the mean surface density of companions could be fit by two power laws that intersected at a separation of 0.04 pc. The inner power law fit the data for projected separations less than 0.04 pc and had an exponent of $-2.15$. The sources included in this part of the power law were considered to be part of multiple systems. This power law is in approximate agreement with the flat distribution of orbital periods per logarithmic interval of period for solar-type stars in the solar neighborhood \citep{1991A&A...248..485D, 1998MNRAS.297.1163B}. In contrast, the power law for separations greater than 0.04 pc was much shallower, with an exponent of $-0.62$. \citet{1995MNRAS.272..213L} proposed that this power law reflected the fractal spatial distribution of young stars in the Taurus cloud as a whole. In this interpretation, sources within 0.04~pc were true companions which were formed within one Jeans length of each other, while objects at larger separations were young stars in the Taurus-Auriga region that were not gravitationally bound companions. Using HST/NICMOS observation of the Lynds~1688 region of the Ophiuchus molecular cloud, \citet{2002ApJ...566..993A} found a similar dual power law structure for the mean surface density of companions in that cloud. \citet{1997ApJ...482L..81S} compiled data sets on three star forming regions, and fit dual power laws to the mean surface densities of the Taurus, Ophiuchus and Orion clouds. The power laws were similar to those found by \citet{1995MNRAS.272..213L}, but the break point between the power laws varied from 12000 AU in Taurus to 400 AU in Orion. \citet{1997ApJ...482L..81S} proposed that the separation was dependent on the Jeans mass and the stellar density. \citet{1998MNRAS.297.1163B} clarified the dependence on stellar density by showing that the break point occurs at the point at which the surface density of companions exceeds that of young stars in the line of sight. This break point occurs at shorter and shorter separations as the surface density of line of sight stars increases. Because we are comparing column densities, the break point depends not only on the volume density of young stars but also the depth of the observed star forming region and the corresponding depth of the column over which the surface density is measured. This length is much larger than the orbital separations in multiple star systems, and the break point can occur at a separation much smaller than the maximum orbital separations of gravitationally bound systems. 

\begin{figure}
 \centering 
			\subfigure{\includegraphics[width=0.45\textwidth]{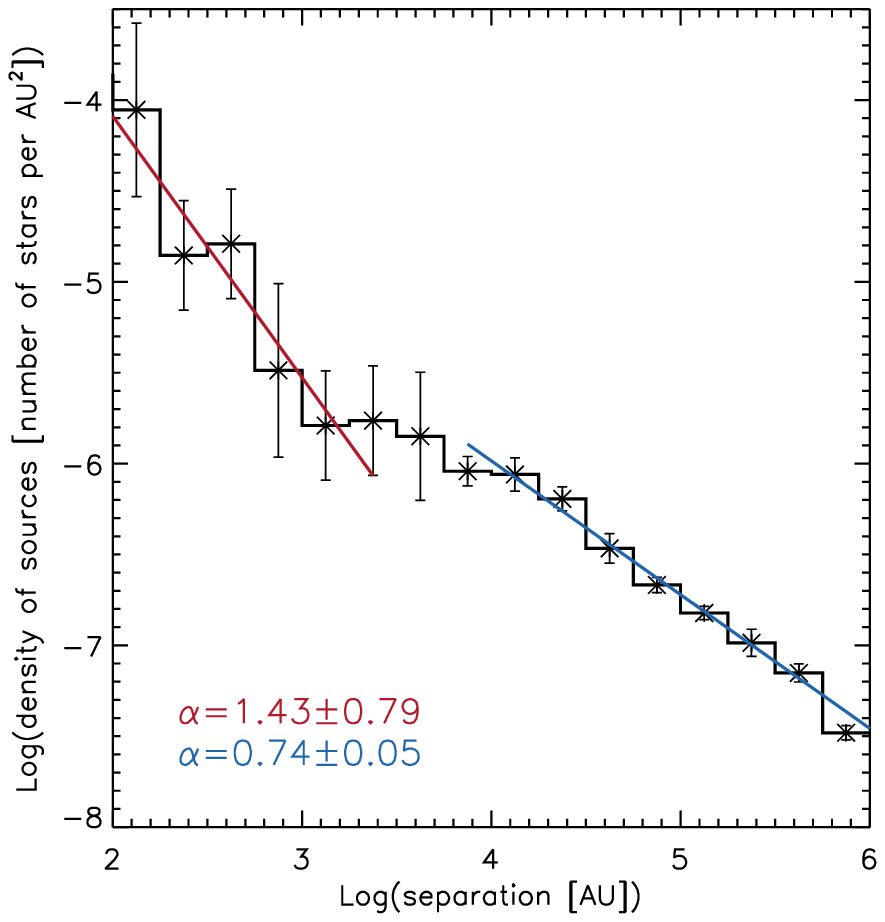}}
 			\subfigure{\includegraphics[width=0.45\textwidth]{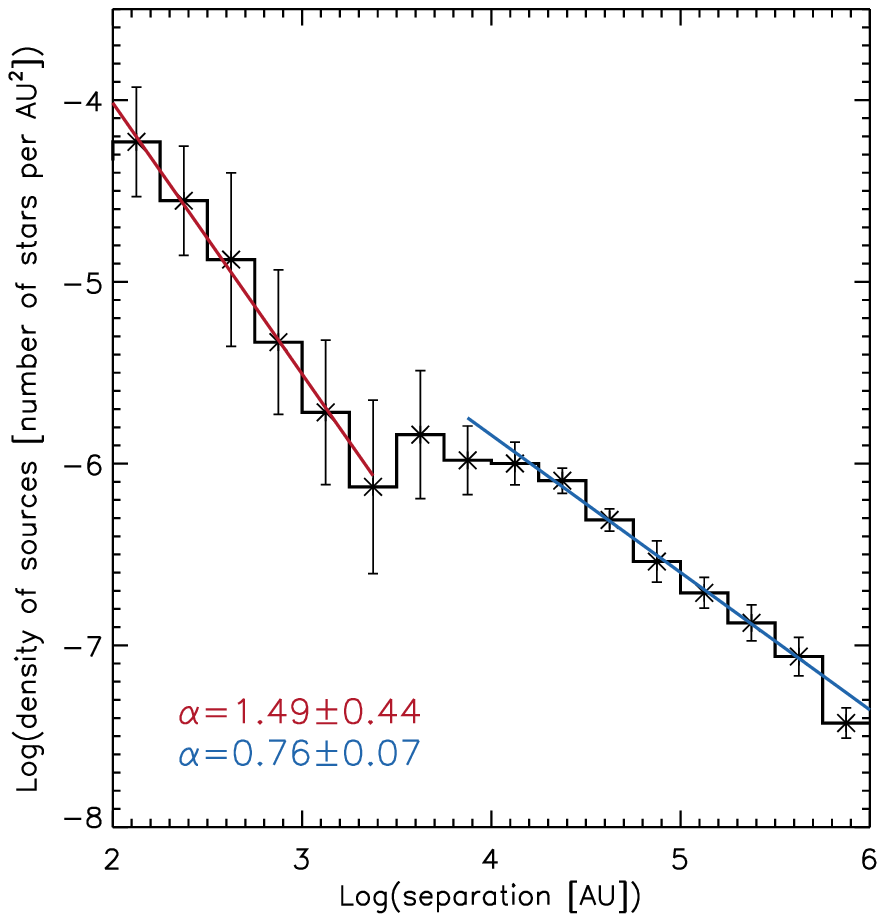}} 
 \caption{The mean surface density of companions using a logarithmic binning. The bin size is $\Delta log(r_{proj}/1~{\rm AU})= 0.25$. To extend the distribution beyond the field of the WFC3 camera, we added all the dusty YSOs from the {\it Spitzer} Orion Survey. In addition, for separations observed by WFC3 and {\it Spitzer}, WFC3 sources without IR excesses or fainter than 20th mag are removed from the sample to reduce background contamination. This cut is imposed in attempt to filter out background objects. The error bars were determined by dividing the sample randomly into four equal parts and finding the maximum variation between the $\Sigma_{comp}$ in those four samples. \textbf{Left:} $\Sigma_{comp}$ around protostars. \textbf{Right:} $\Sigma_{comp}$ around pre-main sequence stars with disks.}\label{fig:csflog}
\end{figure}

To compare our data to these previous studies, we merge our HST with the {\it Spitzer} Orion Survey of \citet{2012AJ....144..192M} to look at the mean surface density of companions between 100 AU and 5 pc. For separations $\le 1000$~AU, these sources come from the HST data. For separations $\ge 10000$ AU, these sources are the dusty YSOs from the {\it Spitzer} survey. In the in-between region of overlap, we eliminated sources which were likely faint background objects and instituted a F160W magnitude cutoff of 20~mag. The mean surface density of companions from this combined data set is shown in Figure~\ref{fig:csflog}. As found previously, the mean surface density of companions shows a steep power law at small separations and a much shallower power law at large separations. However, there are several differences. First, we find a plateau in the mean surface density that extends from 1000 to 20,000~AU (i.e. from 2.5'' to 50''). Since the extent of the plateau corresponds to approximately half the size of the WFC3 field ($135'' \times 127''$), we find it likely that the plateau is an observational artifact. This artifact arises from the inclusion of HST sources for which there is not sufficient {\it Spitzer} photometry to distinguish between a dusty YSO and a background source. Since these sources could be faint background stars and galaxies, many of the objects in the plateau are likely to be background objects detected by WFC3. Second, we find that the break point between the inner power law and the plateau is at 1000 AU, or 0.005 pc, much less than the 0.04~pc found by \citet{1995MNRAS.272..213L}. As explained by \citet{1998MNRAS.297.1163B}, this point is set by the surface density of point sources in the line of sight equals the surface density of companions. In the case of the Orion data, this occurs when the surface density of companions equals the density of sources in the plateau. Hence, the smaller distance in Orion is not due to a physical difference, but a higher level of line of sight contamination. 

A final difference is that the inner power law is less steep; the slopes of the inner power law from the HST Orion survey are $-1.43 \pm 0.79 $ and $-1.49 \pm 0.44 $ for the protostars and pre-main sequence stars, respectively, compared to $-2.15$ for Taurus \citep{1995MNRAS.272..213L}. This may be due to incompleteness at short separations; however, the uncertainties are large and the difference is of a marginal statistical significance. Finally, the outer power law is steeper than those found in other regions. Part of this may result from the plateau, which has forced us to fit the outer power law over a different range of separations than \citet{1995MNRAS.272..213L} or \citet{1997ApJ...482L..81S}. Alternatively, the difference may be due to structural differences in the distribution of young stars in Orion compared to Taurus or Ophiuchus. Nevertheless, our result is qualitatively consistent with the interpretation of previous authors, reinforcing our claim that the mean surface density of companions within 1000 AU is tracing companions and therefore can be used to characterize multiplicity in the Orion molecular clouds.

\subsection{The Change of Multiplicity with Evolution} 
\label{sec:evolution}

We find no evidence for a change in the $CF$ between the protostellar and pre-main sequence systems. Our $CF$ is also consistent with that found by the HST/ACS survey of pre-main sequence stars in the ONC: between 100-660 AU both \citet{2007AJ....134.2272R} and this study find a $CF$ of 0.09. Furthermore, the $CF$ for stars between 100-1000 AU for solar-type (G) stars is 0.13 \citep{1991A&A...248..485D}.\footnote{\citet {1991A&A...248..485D} tabulate the semi-major axes of the solar-type stars. To compare these to our projected separations, we use the empirical relationship between the projected separations $\rho$ and the semi-major axis $a$ of $\rho = 0.776 a$, derived for nearby stars by \citet{1935PASP...47..121K}.} The fact that the $CF$ for the Orion protostars and pre-main sequence stars is similar to that for solar-type, field stars may result from the observed YSOs in the Orion sample having masses lower than solar-type star, which would be expected for a standard IMF. Since main sequence stars at 0.5~M$_{\odot}$ have $CF$s lower than that for G stars \citep{2013ARA&A..51..269D}, the dusty YSO $CF$s may be larger than the field star $CF$ for the appropriate mass range, and the agreement between the dusty YSO $CF$s and that of solar-type stars may be spurious. However, given the uncertainty in the masses of the Orion YSOs, our observed $CF$s are consistent with those of field stars and we find no clear evidence for evolution. 

As mentioned in the introduction, previous studies provided evidence for a large decrease in multiplicity from the Class 0 protostellar phase to the main sequence. This change is observed at the $> 50$~AU separations that can be resolved in sub-millimeter and IR observations. In an analysis of SMA data toward 33 Class 0 protostars, \citet{2013ApJ...768..110C} find a MF and $CF$ of $0.64 \pm 0.08$ and $0.91 \pm 0.05$, respectively, in various star forming regions, with a median resolution of 2.5'' or $\sim 600$~AU for a typical distance of 240~pc. For a separation range of 100-1000 AU, the $CF$ is $0.24 \pm 0.09$; since their survey is incomplete at these separations, this $CF$ should be considered a lower limit. Near-IR surveys have found the $CF$ for more evolved Class I protostars. \citet{2004A&A...427..651D} determine the $CF$ between 110--1400 AU to be $0.23\pm 0.09$ for the Taurus Class I protostars, and $0.29\pm 0.07$ for similar protostars in Ophiuchus. Subsequent higher angular resolution observations gave $CF$s of $0.32\pm 0.06$ between 45--1400 AU in Taurus and $0.47\pm 0.08$ between 14--1400 in Ophiuchus for a combined sample of 189 protostars \citep{2007A&A...476..229D}. If we limit the separation range to 100-1000 AU, the $CF$s are $0.14 \pm 0.08$ in Taurus and $0.23 \pm 0.10$ in Ophiuchus. \citet{2008AJ....135.2526C} find a $CF$ across various star-forming regions to be $0.43$ between 100-4500 AU; this reduces to $\sim$0.24 if we limit the separations to 100--1000 AU. While some of these values are higher than those we find in Orion over a comparable separation range, given the different completeness limits for various surveys and the large uncertainties due to the relatively small sample sizes, it is difficult to draw a meaningful comparison between them.

There are several potential reasons why the Orion data do not exhibit a changing $CF$ between protostars, pre-main sequence stars and the field. First, the masses of the protostars, pre-main sequence stars, and their companions have not been determined; hence, we cannot directly compare our Orion $CF$ to the main sequence $CF$, which is a strong function of mass \citep{2006ApJ...640L..63L,2013ARA&A..51..269D}. Furthermore, when comparing protostars and pre-main sequence stars in Orion, the extinction from protostellar envelopes limits the detection of faint, very low mass companions. The faint magnitudes of the protostars and the companions in Figure~\ref{fig:sep_mag} clearly demonstrates the effect of extinction on protostellar companions, although this is compensated, at least in part, by the high sensitivity of the HST at these wavelengths. This extinction could reduce the $CF$ of the protostars relative to that for pre-main sequence stars, thereby masking a drop in the $CF$. Third, by selecting protostars that appear as a point source at 1.60~$\mu$m, our sample is biased toward objects at the later stages of protostellar evolution. \citet{2008AJ....135.2526C} find a decrease in the $CF$ with spectral index, with flat spectrum sources showing a distinctly lower $CF$. Thus, if evolution of multiplicity occurs during the protostellar phase \citep[e.g. ][]{2000AJ....120.3177R, 2010ApJ...725L..56R}, the lack of evolution apparent in our Orion sample could result from a sample biased toward protostars around which this evolution has already occurred. Fourth, the primary variation in the $CF$ may be a drop at separations less than 100~AU or greater than 1000 AU \citep{2008AJ....135.2526C}; there is currently no clear evidence in any survey of star forming regions of a drop in the $CF$s or $MF$s between 100-1000 AU. Finally, as we will discuss in the next section, differences in the $CF$ may also have an environmental component. \citet{2012MNRAS.421.2025K} find that pre-main sequence stars in Taurus have a higher $CF$ than other regions. Thus, studies of the evolution of the $CF$ between protostars and field stars must take into account such environmental variations.

Given the potential biases listed above, our data does not rule out that there is a decrease in the $CF$ with evolution. Upcoming observations with the VLA and ALMA, as well as observations at wavelengths longer than 1.6~$\mu$m with the JWST or ground-based adaptive optics, may help resolve this issue by detecting companions around protostars at even earlier stages of their development. 

\subsection{Variation in Multiplicity in Different Star Forming Environments}

Our survey of Orion shows a dependence of the incidence of 100-1000 AU multiplicity on environment as defined by the local YSO density. In Sec.~\ref{sec:csf_env}, we found that the $CF$ increases by a factor of $\sim 1.5$ between environments of low stellar density and high stellar density. This increase occurs for both the protostar and the pre-main sequence star sample. This is inconsistent with previous surveys which either found no differences between high and low stellar density regions or found that high stellar density regions had systematically lower $CF$s. \citet{2004A&A...427..651D} compared the multiplicity of Class I and flat spectrum protostars in the Taurus Molecular Cloud, which is characterized by low stellar densities, to that of protostars in the Ophiuchus cloud, which hosts dense clustered star formation. Over a separation range of 110 to 1400 AU, they found no significant difference in the $CF$s or MFs. Comparisons of the $CF$s and MFs of pre-main sequence stars have shown a different picture. \citet{2012MNRAS.421.2025K} find that the MF of the Taurus region is higher than the ONC, but that the Chameleon I region, IC348 cluster and Ophiuchus cluster are within $\sim 1\sigma$ of Orion. In their review, \citet{2013ARA&A..51..269D} highlight this difference by showing that the $CF$s per decade of separation are 0.20 and 0.08 in low density associations and dense clusters, respectively. Interestingly, a survey of protostars in the Ophiuchus, Taurus, Serpens, and the Orion clouds by \citet{2008AJ....135.2526C} showed that the $CF$ in Orion was different than that for the other clouds. Specifically, they found that Orion has a much higher $CF$ at separations between 100-200~AU than the $CF$ for the combined sample of protostars in the other clouds. The lack of consistency in these results motivates surveys within a single complex where biases due to different distances and sensitivities can be minimized. 
 
As discussed in the introduction, the observed variations between low and high density regions have been primarily explained as the effect of the cluster environment on the evolution of pre-main sequence systems after formation \citep{2012A&A...543A...8M,2012MNRAS.421.2025K}. There are several reasons why variations in the $CF$s we observed in Orion are unlikely to be the result of interactions between stars in a dense, clustered environment. First, the spatial distribution of the YSOs found in high and low spatial density regions within the Orion cloud do not show that high density sources are concentrated in large clusters; instead, we find that regions of high YSO density are located throughout the Orion complex and found both in large clusters and small groups (Figure~\ref{fig:map}). Second, if the variations are due to the interaction with neighboring stars during flybys, we would expect a lower $CF$ between 100 and 1000 AU in dense regions \citep{2012A&A...543A...8M,2014MNRAS.442.3722P}. Such a decrease is apparent in the HST/ACS observations of the ONC that shows a deficiency of wide systems in the dense inner region of Orion \citep{2007AJ....134.2272R}. This is the exact opposite of what we find. Finally, if the changes of the $CF$ resulted from interactions between YSOs in dense environments, the effect should be stronger for the older pre-main sequence stars. This is not seen: disks exhibit a smaller change in the $CF$ between high and low density environments. For these reasons, we argue that the increased $CF$ in dense regions is due to the enhanced formation of companions between 100 and 1000 AU. These companions must be formed by the time the central protostar is detectable by 1.60~$\mu$m imaging. 
 
One possible reason why YSOs in high stellar density regions have a higher incidence of multiplicity is that they are systematically more massive. Observations of field stars show a strong increase of $CF$ with mass \citep{2013ARA&A..51..269D}. In the Orion clouds, \citet{2012AJ....144...31K} found the luminosity function for protostars in regions of high stellar density was biased to higher luminosities than the luminosity function for regions of low stellar density. To assess the effect this may have on our analysis, we compare the observed magnitudes of the systems (which are usually dominated by the primaries, Figure~\ref{fig:sep_mag}) in high and low stellar density regions of the Orion clouds. We do this in two ways: by comparing the 24~$\mu$m magnitudes of the combined sample, and by comparing the dereddened J-band magnitudes of the pre-main sequence stars.

\begin{figure}
\epsscale{1.}
\plotone{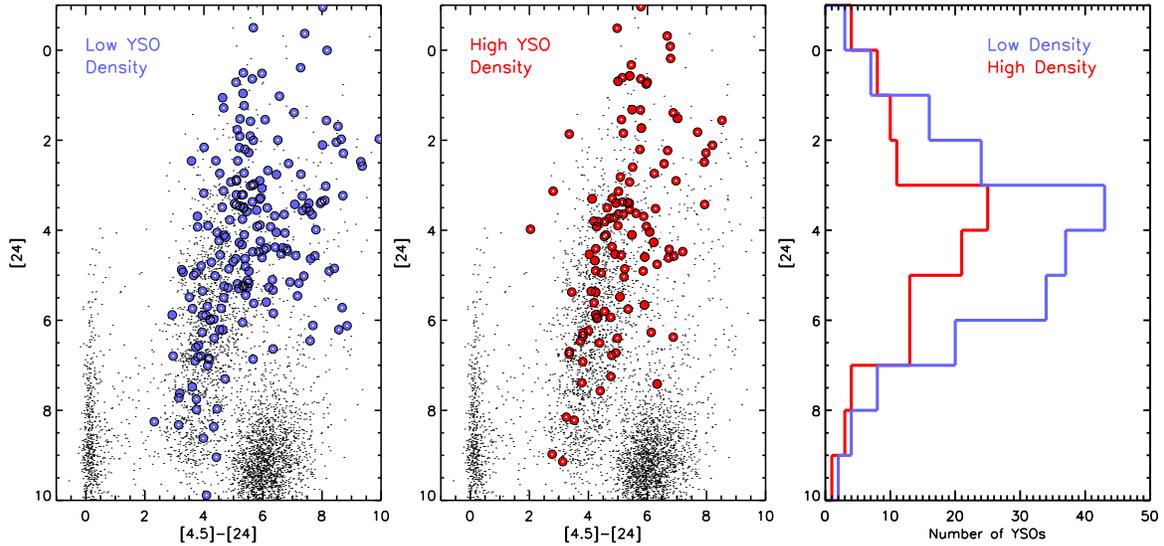}
\caption{{\bf Left panel: } the $m_{24}$ vs $[4.5]-[24]$ color magnitude diagram for YSOs found in the low stellar density regions. The black dots show the entire sample of point sources in Orion \citep{2012AJ....144..192M}. The blue circles give the locations of the YSOs for the sample of protostars and pre-main sequence stars considered in this paper; the filled circles are those with companions. {\bf Middle Panel:} the color magnitude diagram for YSOs found in high stellar density regions. The black dots are the entire sample of point sources, the red circles are the YSOs studied in this paper, and the filled circles are those with companions. {\bf Right panel:} Histograms of the $m_{24}$ for the YSOs in our sample: the blue histogram is for the sources in low stellar density regions and the red histogram is for sources in high stellar density regions. The histograms contain both single objects and those with companions. At the resolution of the {\it Spitzer} data, the companions are not resolved and the magnitudes are for the primary and secondaries combined. For the displayed plots, the separation threshold is 30,000~AU, corresponding to a density of 45~pc$^{-2}$.} 
\label{fig:cm_24}
\end{figure}

We choose the 24~$\mu$m magnitude since both pre-main sequence stars with disks and protostars show strong emission at these wavelengths. In comparison, observations of pre-main sequence stars are generally not available at longer wavelengths while shorter wavelength observations can be affected by the extinction and scattering in protostellar envelopes. Furthermore, this wavelength is usually not strongly affected by extinction. Thus, although there is not a simple conversion between mass and the 24~$\mu$m magnitude, it provides the means to compare the two samples to look for systematic differences in their observable properties. As shown in Figure~\ref{fig:cm_24}, we do not see a systematic difference in the $m_{24}$ magnitude distributions of the high and low density regions. We determine the $m_{24}$ distributions of the low and high stellar density samples using the three different critical separations in Sec.~\ref{sec:csf_env}; KS tests comparing the two distributions yield probabilities of 0.46, 0.60 and 0.22 that the two are drawn from the same parent distribution for critical NN5 thresholds of 25000~AU, 30000~AU and 35000~AU, respectively. 

\begin{figure}
\epsscale{1.}
\plotone{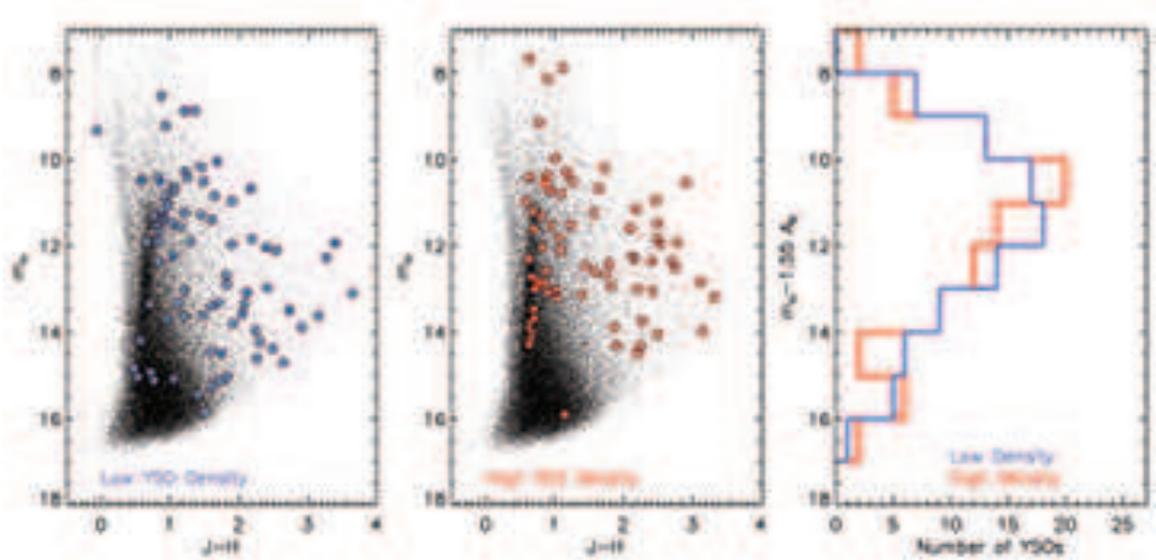}
\caption{{\bf Left panel: } the $m_J$ vs $J-H$ color magnitude diagram for pre-main sequence stars found in the low stellar density regions. The black dots show the entire sample of point sources in Orion \citep{2012AJ....144..192M}. The blue circles give the locations of the pre-main sequence stars with disks considered in this paper; the filled circles are those with companions. {\bf Middle Panel:} the color magnitude diagram for pre-main sequence stars found in high stellar density regions. The black dots are the entire sample of point sources, the red circles are the YSOs studied in this paper, and the filled circles are those with companions. {\bf Right panel:} Histograms of the {\it dereddened} $m_J$ for the pre-main sequence stars in our sample: the blue histogram is for the sources in low stellar density regions and the red histogram is for sources in high stellar density regions. The histograms contain single objects and those with companions. At the resolution of 2MASS, the systems are not resolved and thus the companion magnitudes are for the combined system. In the displayed plots, we use a density threshold of 45~pc$^{-2}$. } 
\label{fig:cm_nir}
\end{figure}

A second approach is to compare the de-reddened $J$-band magnitudes of the pre-main sequence stars. The $J$-band magnitudes of pre-main sequence stars are dominated by their photospheric emission and are strongly dependent on the masses of the stars. We use the 2MASS near-IR colors to deredden the sources following the approach of \citet{2008ApJ...674..336G} and using the extinction law of \citet{2007ApJ...663.1069F} as modified for Orion by \citet{2012AJ....144..192M}. This gives us the photospheric J-band magnitudes of the pre-main sequence stars. This magnitude depends on both the mass and age of the sources as they descend on pre-main sequence tracks. We expect that different distributions of stellar mass would be reflected in different de-reddened $m_J$ distributions, except in the unlikely case that differences in the luminosity due to different mass distribution are cancelled out by different distributions of ages. Again, as shown in Figure~\ref{fig:cm_nir}, the distributions are very similar. We compare the de-reddened J-band magnitude distributions for sources in high and low stellar density regions using the three different NN5 thresholds in Sec.~\ref{sec:csf_env}. In this case, KS tests give probabilities of 0.997, 0.96 and 0.61 that they are drawn from the same parent distribution for thresholds of 25000~AU, 30000~AU and 35000~AU, respectively. We conclude that the environmental variations in the $CF$s are not likely to be the result of systematically different masses in our two samples.
 
For these reasons, we find it likely that the variations in the $CF$ are primordial. It is not understood, however, why YSOs would form more companions in regions of high stellar density. Part of the reason may have to do with the correlation between the gas column density and the stellar surface density \citep{2011ApJ...739...84G, 2013A&A...559A..90L, 2013ApJ...778..133L}. Regions with a higher density of YSOs also tend to have a higher column density of molecular gas. This may also imply a higher volume density of gas leading to shorter Jeans lengths, smaller cores, and more prompt or turbulent fragmentation on short physical scales that lead to multiple systems \citep[see also][]{2002ARA&A..40..349T, 2010ApJ...725.1485O, 2008AJ....135.2526C}. Alternatively, protostars in regions of high gas density may be subject to higher infall rates; this may lead to the formation of larger number of companions in the outer, gravitationally unstable, regions of protostellar disks fed by infall \citep{2010ApJ...708.1585K, 2010ApJ...710.1375K}. Although disks tend to have radii of 200 AU and less \citep{2014ApJ...784...82M}, interactions in the disks may eject these objects into orbits extending beyond the disks, thereby populating the separations probed in this survey \citep{2012ApJ...750...30B}. However, neither of these suggestions can explain the high $CF$ measured in Taurus, which shows stellar surface densities and gas column densities similar to the low stellar density regions in Orion. This suggests that other physical factors may come into play. A detailed comparison of the physical properties of the molecular gas in Taurus with those in Orion is needed to identify what factors can lead to a high $CF$ in Taurus and a low $CF$ in the low stellar density regions in Orion. Finally, we have not ruled out random fluctuations as being the cause for the difference. Although the $CP$ varies in a manner similar to the $CF$, the null hypothesis of a constant $CP$ for all protostars has not been excluded. If the presence of a companion around a protostar is a random process with a probability equal to $CP$, we cannot rule out that the observed environmental variations in the $CP$ for both protostars and disks are due to random fluctuations at better than the four to ten percent level, depending on the test applied (see Section~\ref{sec:CP}). In this interpretation, the enhanced formation of companions in dense regions would be fortuitous and not the result of environmental conditions. Larger samples are needed to reliably distinguish between random fluctuations and deterministic, environmentally driven variations in the $CF$. 

The $CF$ between 100 and 1000~AU may also be influenced by complex dynamics. Simulations of nonhierarchical triples predict rapid declines in the $CF$ between 100-1000 AU during the first 100,000 years of the protostellar phase as one of the companions is ejected or launched into a wider orbit \citep{2000AJ....120.3177R, 2010ApJ...725L..56R}. This has several implications for our interpretation. First, such chaotic processes may effectively be modeled as a random process, and therefore provide a mechanism for the random fluctuation interpretation. Alternatively, changes in the initial separation due to environment can affect the subsequent evolution of nonhierarchical ternary or higher order systems, resulting in different $CF$s. For example, simulations show that the timescale for the dynamical evolution of nonhierarchical systems decreases when the initial mean separation of the system is reduced \citep{2010ApJ...725L..56R}. Finally, for our sample of more evolved, near-IR protostars and pre-main sequence stars, the rapid decrease in the $CF$ may have already occurred, as implied by the similar $CF$s that we observe for the protostars and pre-main sequence stars. Hence, if this rapid evolution of systems is present in Orion, then the $CF$ during early phases of protostellar evolution may have been higher.

The separations probed by our survey are expected to be strongly effected by dynamical effects in clusters \citep{2012A&A...543A...8M,2014MNRAS.442.3722P}. How might variations in the primordial $CF$ supported by this paper affect the interpretation of previous work on the evolution of multiplicity in the ONC and other clusters? As mentioned in the previous section, the $CF$ for separations of 100 to 660 AU for the combined high and low density is 0.085, identical to the value found in the ONC by \citet{2007AJ....134.2272R}. However, we now find that the $CF$ is higher in dense regions. If we take the $CF$ between 100 to 660 AU, it is 0.098 for dense regions and 0.071 for low stellar density regions using a threshold NN5 of 30,000 AU and the HST sample. Thus, if the initial $CF$ for the star formation in the ONC was similar to what we find in dense regions, there has been a decrease in the $CF$ for the ONC. This supports in part the conclusion of \citet{2012A&A...543A...8M} and \citet{2007AJ....134.2272R}, that the $CF$ at 100 to 1000 AU decreases due to interactions between stars in the cluster environment. However, we find that the initial multiplicity of the young stars is not invariant, as assumed by \citet{2012A&A...543A...8M}. Thus, these results imply a more complicated path to the field $CF$, where the distribution of environments affects both the initial, primordial $CF$ created by the star formation process and the subsequent evolution of the $CF$ in clustered environments. 

Future and upcoming studies promise many tests of the possible mechanisms for an environmentally varying $CF$. Ongoing measurement of the masses of the primaries and the companions in both high and low density environments can test the formation mechanism. If variations in the $CF$ result from the enhanced formation of companions in protostellar disks located in regions of high gas and star density, then the mass function should be weighted to lower mass objects \citep{2010ApJ...708.1585K, 2010ApJ...710.1375K, 2012ApJ...750...30B}. If instead, the companions are produced by prompt/turbulent fragmentation, then the companion mass function should be similar to the IMF of the primary stars. Better statistics are required, particularly to test whether random fluctuations can explain the observed variations in the $CF$. An ongoing HST/WFC3 snapshot survey of protostars within 500 pc is now being executed which will increase the sample of protostars observed by HST by more than 30\%. Furthermore, approved surveys of Orion Class 0 sources with the VLA and all Orion protostars with ALMA will allow us to examine multiplicity for very young Class 0 sources. Younger companions may also be probed with the JWST which promises higher angular resolution at longer wavelengths. Expanding the range of separations may provide new tests of environmentally dependent multiplicity and its origin. The VLA, ALMA, and {\it Spitzer} can identify wide binaries with separations greater than 1000 AU, and probe environmental differences at these separation ranges \citep[e.g.][]{2013ApJ...768..110C}. The JWST, VLA, and ALMA, as well as AO observations on $8+$ meter telescopes also promise data on companions at separations within 100~AU, providing the ability to search for environmentally dependent multiplicity at shorter separations. 

\section{Conclusions}

We conducted a near-IR survey of \textit{Spitzer} identified young stellar objects in the Orion molecular clouds acquired with NICMOS and WFC3 cameras on the HST as well as with NSFCAM2 on IRTF. In total we have observed 201 protostars and 198 pre-main sequence stars with disks. We use the mean frequency of companion to measure the density of point sources as a function of radius from these YSOs. We find a higher surface density of point sources at projected separations $< 1000$~AU which we interpret as companions. At projected separations between 80 and 1000 AU, we find 29 candidate binary and 1 candidate tertiary systems around protostars and 27 candidate binary and 1 candidate tertiary systems around pre-main sequence stars with disks. At larger separations we cannot distinguish companions from stars in the line of sight, while at closer separations we cannot resolve companions. We focus on companions at separations $\ge 100$~AU, where we have a $\ge 75$\% chance of detecting a companion one magnitude fainter than the primary. After correcting the number of 100-1000~AU companions for line of sight contamination, the resulting companion star fractions ($CF$s) between 100 to 1000 AU are $14.4^{+1.1}_{-1.3}$\% for protostars and $12.5^{+1.2}_{-0.8}$\% for the pre-main sequence stars.

The $CF$s are consistent for both the protostars and the pre-main sequence stars. Furthermore, these numbers are similar to the $CF$ of solar type G-stars in the field. Hence, in contrast to previous studies, we find that there is no clear evidence for evolution of the $CF$ between 100 to 1000 AU from the protostellar to main sequence phase. We identify a number of biases and uncertainties in our data that could mask such a change, such as extinction from protostellar envelopes, mismatches in the ranges of primary mass between samples, a lack of evolution at separations between 100-1000 AU, and the absence of evolution in the later stages of protostellar evolution traced by our survey. 

We find a dependence of the $CF$ on local surface densities of YSOs. After bifurcating our sample of YSOs into those found in high density regions and those in low density regions, both protostars and pre-main sequence stars in high stellar density regions show $CF$s $\sim 1.5$ times higher than those in low density regions. We rule out fluctuations in the line of sight contamination as the cause for the dependence. This trend is the opposite of that typically reported in the literature \citep[e.g.][]{2013ARA&A..51..269D}. The change in the $CF$ most likely results from the enhanced formation of 100-1000 AU companions in regions with higher densities of both young stars and gas. We examine possible reasons for these variations. We find it unlikely that the variations are due to systematic differences in the primary masses. Noting that regions of high stellar surface density also have a high gas column density, we suggest several possible reasons: variations due to shorter Jeans lengths in regions of high gas density, enhanced formation of companions in protostellar disks found in environments of high gas density, or random fluctuations in the formation of companions. We discuss various observational tests for these possibilities. 

\acknowledgments
We acknowledge valuable conversations with John Tobin, Fred Adams and Mike Meyer. M.K. acknowledges support from the NSF-REU program at the department of Physics and Astronomy at the University of Toledo (grant PHY-1004649). We would also like to thank the anonymous referee and the statistics editor of ApJ for valuable comments. Support for program 11548 was provided by NASA through a grant from the Space Telescope Science Institute, which is operated by the Association of Universities for Research in Astronomy, Inc., under NASA contract NAS 5-26555. We also used the {\it Spitzer} Space Telescope and the Infrared Processing and Analysis Center (IPAC) Infrared Science Archive, which are operated by JPL/Caltech under a contract with NASA. We also made use of the Infrared Telescope Facility (IRTF), which is operated by the University of Hawaii under Cooperative Agreement NNX-08AE38A with NASA, Science Mission Directorate, Planetary Astronomy Program. The authors wish to recognize and acknowledge the very significant cultural role and reverence that the summit of Mauna Kea has always had within the indigenous Hawaiian community. We are most fortunate to have the opportunity to conduct observations from this mountain. This paper makes use of data products from the Two Micron All Sky Survey, which is a joint project of the University of Massachusetts and IPAC/Caltech, funded by NASA and the National Science Foundation.

\appendix
\section{Appendix A: The Probability Distribution of $N_{comp}$}

In this paper, we estimate the uncertainty in the number of companions, $N_{comp}$, by assuming it is due solely to the uncertainty in the number of contaminants in the line of sight that fall within 100 to 1000 AU of the primary YSOs. Thus, the uncertainty in $CF$ is driven by the uncertainty in $N_{cont}$ in Equation~2. (The incompleteness to companions is another source of uncertainty, we will address the effect of incompleteness on the uncertainty in a future paper). The probability distribution for the number of contaminants has two components. The first component is the uncertainty in the surface density of line of sight sources, $\Sigma_{los}$. We estimate the surface density of the contaminants by counting the total number of sources detected in the 2000 and 5000~AU annuli centered on both the primaries of the candidate multiple systems and the single star systems. At these separations, the mean surface density of companions is constant with increasing separation, indicating that the number counts at these separations are dominated by background stars and other YSOs in the line of sight (Figure~\ref{fig:mean1}). The probability distribution of the number of contaminants is described by a Poisson distribution (assuming the contaminating sources are randomly distributed in the sky), and the resulting posterior probability distribution of surface densities is given by a gamma distribution (assuming a uniform prior),

\begin{equation}
P(\Sigma_{los} | N_{cont}) = \frac{(\Sigma_{los} \Omega_{cont})^{N_{cont}}}{N_{cont}!}e^{-\Sigma_{los} \Omega_{cont}},
\end{equation}

\noindent
where $N_{cont}$ is the number of point sources measured between 2000 and 5000~AU and $\Omega_{cont}$ is the solid angle of the annulus extending from 2000 to 5000 AU times the number of YSO primaries, $N_{YSO}$. 

The second component is the fluctuations in the number of contaminants in the annuli between 100 and 1000 AU for a given value of $\Sigma_{los}$. This is described by a Poisson distribution with a mean value equal to the surface density determined in the 2000 to 5000 AU annuli times the solid angle of a 100 to 1000 AU annulus times $N_{YSO}$. The number of contaminants is given by the observed number of candidates, $N_{cand}$, subtracted by the actual number of companions, $N_{comp}$, 

\begin{equation}
P(N_{comp}|N_{cand}, \Sigma_{los}) = \frac{(\Sigma_{los} \Omega_{cand})^{(N_{cand}-N_{comp})}}{(N_{cand}-N_{comp})!} e^{-\Sigma_{los} \Omega_{cand}}~{\rm where}~N_{comp} = 0, 1 ... N_{cand}.
\end{equation}

\noindent
To determine $P(N_{comp})$, we marginalize the distribution by integrating over $\Sigma_{los}$:

\begin{equation}
P(N_{comp} | N_{cand},N_{cont}) = \int_0^{\infty} P(N_{comp}|N_{cand}, \Sigma_{los}) P(\Sigma_{los} | N_{cont}) d\Sigma_{los}.
\end{equation}

\noindent
Note that $N_{comp}$ is the total number of companions for the entire sample of primaries, not the number of companions for a single primary. We show the resulting  probability distributions for protostars in high and low density regions in Figure~\ref{fig:appendix}. To generate the distributions, we used a Monte Carlo simulation of $\Sigma_{los}$ combined with the Poisson equation for $P(N_{comp}|N_{cand}, \Sigma_{los})$, and averaged over all the iterations of the simulation to get $P(N_{comp})$. To determine the 1$\sigma$ uncertainties given in Tables 3, 4, 5, 6 for $N_{comp}$ and $CF$, we determine the values $\sigma^u_n$ and $\sigma^l_x$ where the probability that $x$ (i.e. $N_{comp}$ or $CF$) is $1 \sigma$ above or below the mean value of $\bar x$ as given by $P(x > \sigma^u_x + \bar x) = P(x < \bar x - \sigma^l_x) = 0.16$. 

\section{Appendix B: The Probability Distribution of $CF$}

The probability of the determination of $CF$ can be calculated from $P(N_{comp})$ from Appendix~A. Since $N_{comp}$ must be an integer, $N_{comp} = N_{cand} - i$ where $i$ is the adopted number of contaminants, only discrete values of $CF$ are allowed:

\begin{equation}
CF_i = \frac{(N_{cand}-i)}{N_{YSO}}~{\rm where}~i=0,...,N_{cand}.
\end{equation}

\noindent
The resulting probability is then 

\begin{equation}
P(CF_i | N_{cand}, N_{cont}, N_{YSO}) = P(N_{comp} | N_{cont}, N_{cand})~{\rm where}~N_{comp} = N_{YSO} \times CF_i.
\end{equation}

\noindent
From this, we can calculate the $CF$s in high and low density regions, $CF^h$ and $CF^l$, respectively.  These probability distributions are displayed in Figure~\ref{fig:appendix}. We note that $CF^h$ and $CF^l$ are typically not equal since $N_{and}$ and $N_{YSO}$ are not equal in regions of low and high density. We can also define a ratio of the $CF$s, $R_{ij} = CF^h_i/CF^l_j$, where $CF^h$ and $CF^l$ is the $CF$ in high and low density regions, respectively. The probability of $R_{ij}$ is given by

\begin{equation}
P(R_{ij}) = P_{CF}(CF^l_i | N_{cont}^l, N_{cand}^l, N_{YSO}^l) P_{CF}(CF^h_j | N_{cont}^h, N_{cand}^h, N_{YSO}^h) 
\end{equation}

\noindent
where $N_{cont}^h$, $N_{cand}^h$, $N_{YSO}^h$ and $N_{cont}^l$, $N_{cand}^l$, $N_{YSO}^l$ are the data that constrain $CF$ in regions of high and low stellar density, respectively. The resulting cumulative probability distributions of the  $CF$ ratios are shown for the WFC3 sample in Figure~\ref{fig:ratio_csf_env_wfc3}.

In Section~\ref{sec:csf_env}, we examine the significance of the result that the $CF$ is higher in regions of higher stellar density using three different approaches. First, we use the distribution of $R_{ij}$ values to estimate $P(R_{ij} \le 1)$ (Table~\ref{tab:ratio_wfc3_limit}, Figure~\ref{fig:ratio_csf_env_wfc3}). Second, we use a Bayesian hypothesis testing approach where we compare the probability of $CF^h > CF^l$ to $CF^h < CF^l$. The probabilities are given by the equations

\begin{equation}
P(CF^h > CF^l) = \frac{1}{P_{tot}} \sum_{i=0}^{N_{cand}^l} \sum_{j=j_i}^{N_{cand}^h} \frac{P(CF_{i}^{l} | N_{cont}^l, N_{cand}^l, N_{YSO}^l)P(CF_{j}^{h} | N_{cont}^h, N_{cand}^h, N_{YSO}^h) }{(N_{cand}^l+1)(N_{cand}^h-j_i+1)},
\end{equation}

\noindent
where $j_i$ is determined for a given value of $i$ as the lowest integer $j$ such that $CF^h_j > CF^l_i$, and 

\begin{equation}
P(CF^{h} < CF^{l}) = \frac{1}{P_{tot}}\sum_{i=0}^{N_{cand}^l} \sum_{j=0}^{j_i-1}\frac{ P(CF_i^l | N_{cont}^l, N_{cand}^l, N_{YSO}^l)P(CF_j^h | N_{cont}^h, N_{cand}^h, N_{YSO}^h)}{(N_{cand}^l+1)(j_i)}.
\end{equation}

\noindent
The normalization of the distribution (i.e. the marginalized likelihood) is given by:

\begin{equation}
P_{tot} = P(CF^h < CF^l) + P(CF^h > CF^l).
\end{equation}
\noindent The resulting probabilities are discussed in Sec.~\ref{sec:csf_env}. 

Finally, we implement  frequentist hypothesis testing to determine the probability of obtaining the observed difference in $CF$ for the case that the null hypothesis is true: that the CFs are the same and the only differences are due to variations in the amount of contamination. We do this by determining the probability distribution of $\mu = CF^h-CF^l - (\overline{ CF^h} - \overline {CF^l})$, where  $\overline{ CF^h}$ and $ \overline {CF^l}$ are the population means for those two values.  The null hypothesis is that the population means are equal: $\overline{CF^h} = \overline{CF^l}$. Since
$\mu$ has discrete values dependent on the discrete values of $CF^h$ and $CF^l$, we write this as

\begin{equation}
\sigma(\mu_{ij}=\delta CF^h_j-\delta CF^l_i) = P(CF^l_i | N_{cont}^l, N_{cand}^l , N_{YSO}^l) P(CF^h_j | N_{cont}^h, N_{cand}^h , N_{YSO}^h),
\end{equation}

\noindent
where $\delta CF = CF - \overline{CF_{obs}}$ and $\overline{CF_{obs}}$ is the observed value of the $CF$ given by Eqn~\ref{eqn:cf}. The probability that $CF^h > CF^l$ is given by 

\begin{equation}
P(\mu \ge \mu_{obs}) = \sum_{i=0}^{N_{cand}^l} \sum_{j = j_{i}}^{N_{cand}^h} \sigma(\mu_{ij}),
\end{equation}

\noindent
in which $\mu_{obs} = CF^h_{obs}-CF^l_{obs} $ and $j_{i}$ is the smallest $j$ such that $CF^h_j-CF^l_i > \mu_{obs}$. The results for this test are given in Sec.~\ref{sec:csf_env}.

\section{Appendix C: Determining the Probability Distribution of the Ratio of $CP$}
 
\begin{figure}
\epsscale{1.}
\plotone{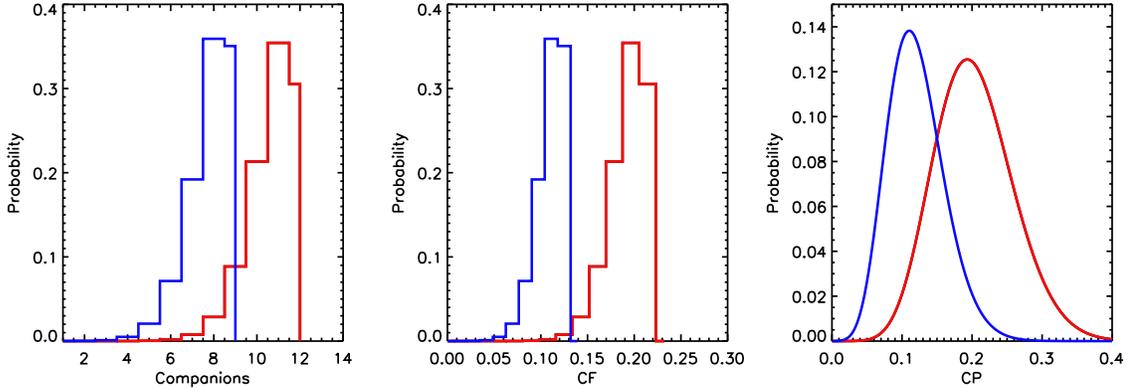}
\caption{Statistical distributions used in the parameter estimation and hypothesis testing described in Appendices~A, B and C. In all cases, the blue distributions are for the high density regions and the red distributions are for the low density regions. {\bf Left panel:} estimation of the number of candidates for the WFC3 protostar sample using the contamination estimated with $r_{inner} = r_{99\%}$. The distribution is due to uncertainties in the line of sight contamination described by a Poisson distribution (Appendix A). {\bf Middle panel:} the distributions of $CF$ values determined using the number of candidates in the left panel (Appendix B). This shows the clear difference in the $CF$s of high and low density regions. {\bf Right panel:} the distributions of $CP$ based on the distribution of $CF$ values in the middle panel and invoking an inverse beta distribution (Appendix C). Although the two distributions peak at different values of $CP$, there remains substantial overlap. Larger samples of YSOs will narrow the distributions.}
\label{fig:appendix}
\end{figure}

\begin{figure}
\epsscale{1.}
\plotone{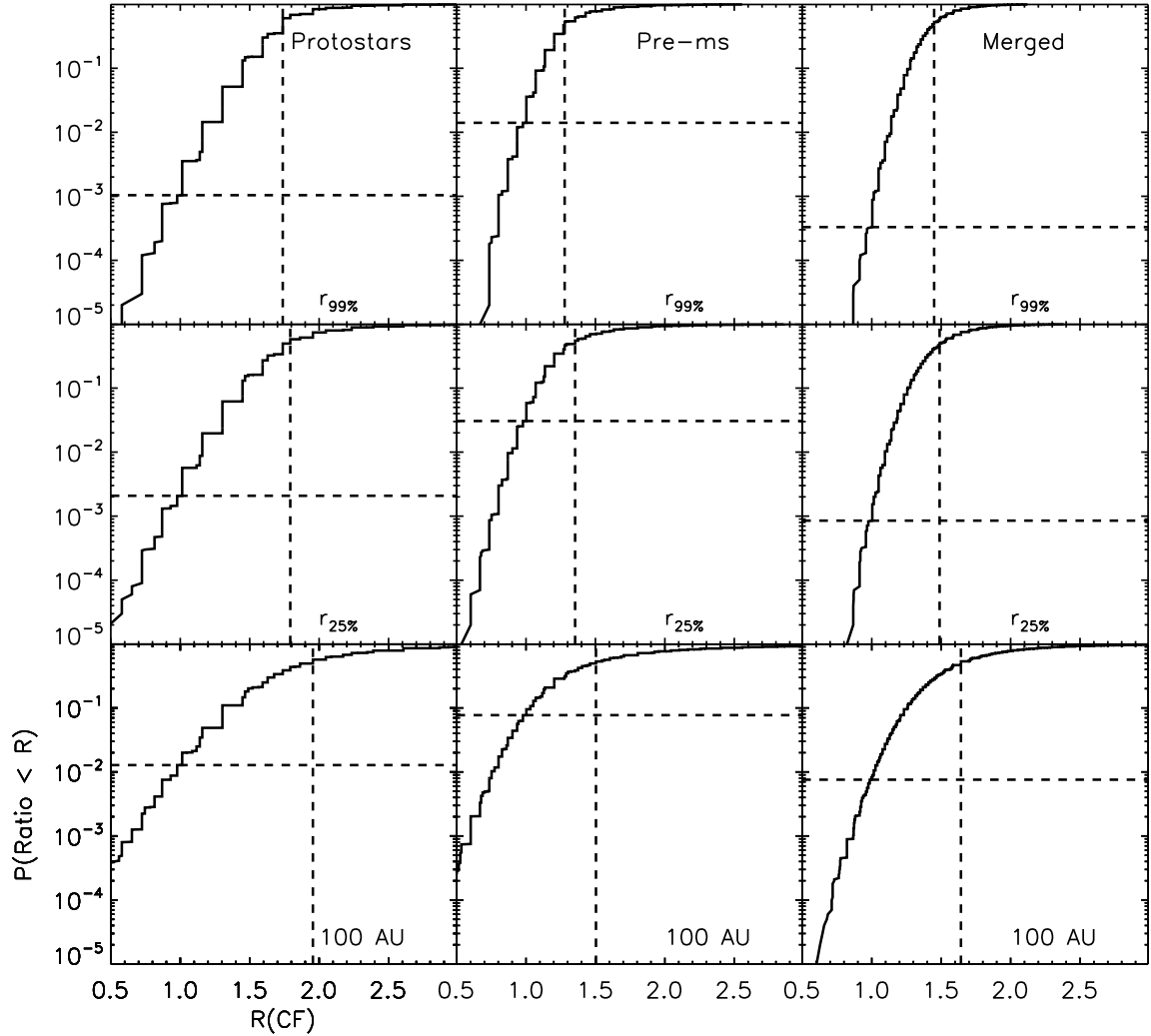}
\caption{The ratio of the high and low density $CF$s using the WFC3 sample. In this case, we have run each case for a single threshold density, $45$~pc$^{-2}$. In each row we plot the distribution using a different inner radius for the line of sight contamination based on the adopted completeness level, the inner radius is given in the panels. Without any correction for incompleteness, $r_{inner} = 100$~AU, otherwise we use $r_{25\%}$ for a 25\% completeness level and $r_{99\%}$ for a 99\% completeness level. The three columns correspond to protostars, pre-main sequence stars, and the merged sample. The vertical dashed line gives the median value of $R$, the horizontal dashed line gives the probability at which $R \le 1$.}
\label{fig:ratio_csf_env_wfc3}
\end{figure}

\begin{figure}
\epsscale{1.}
\plotone{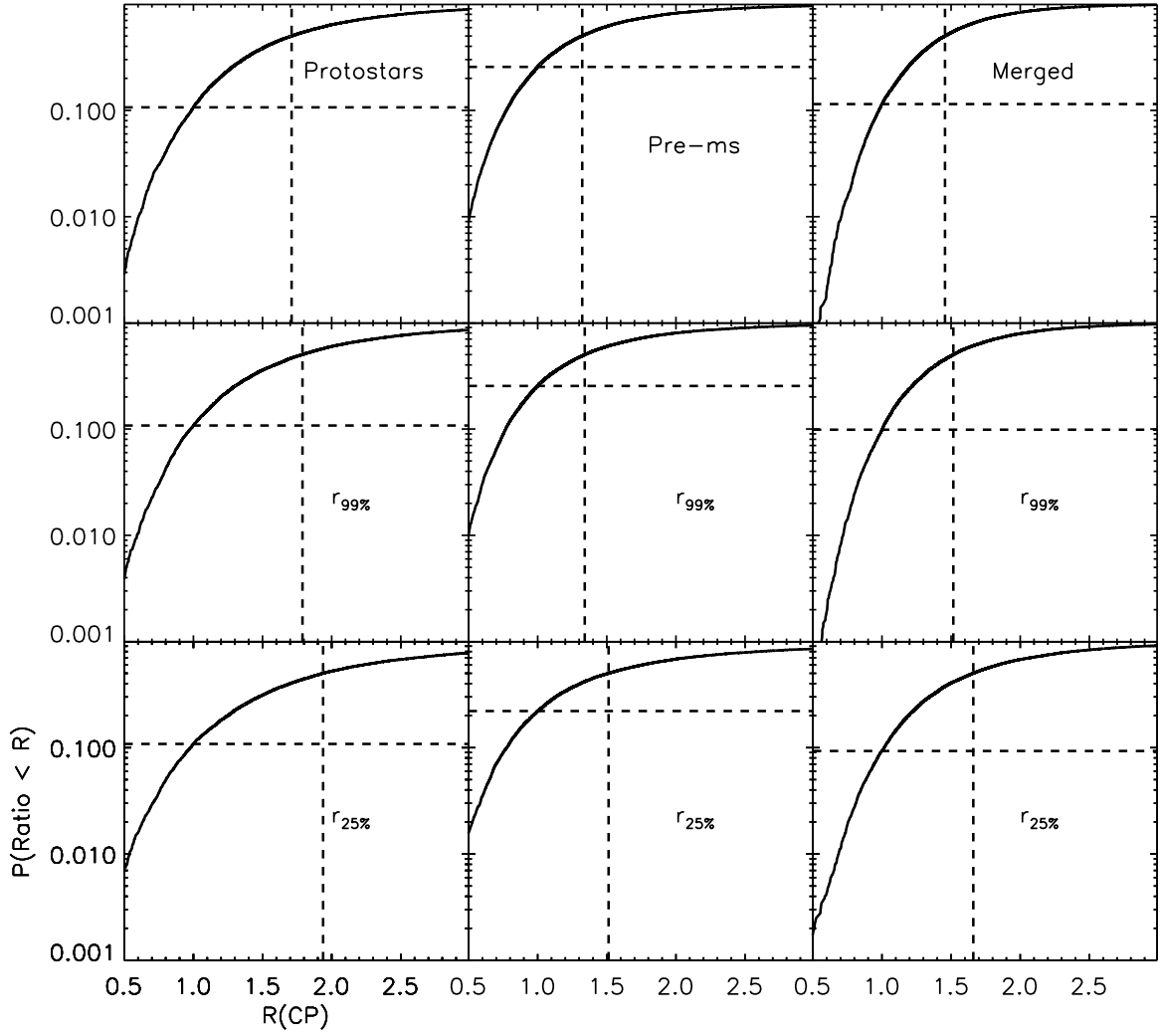}
\caption{Same as in Figure~\ref{fig:ratio_csf_env_wfc3}, but for the ratios of the $CP$.}
\label{fig:ratio_csp_env_hst}
\end{figure}

For each value of the $CF$, there is a continuous distribution of values for $CP$. In turn, for a given value of $CP$, multiple realizations of a sample of primaries would result in a distribution of $CF$s given by a binomial distribution where the probability is $CP$, the number of trials $N$ would be the number primaries, and the number of successful trials would be the number of companions. Following Bayes' theorem and assuming a uniform prior, the posterior distribution of $CP$ is described by a Beta distribution. We can then write the probability density function of $CP$ as

\begin{eqnarray}
P(CP | N_{cand}, N_{cont}, N_{YSO}) = ~~~~~~~~~~~~~~~~~~~~~~~~~~~~~~~~~~~~~~~~~~~~~~~~~~~~~~~~ \nonumber \\
\frac{1}{P_{tot}} \sum_{N_{comp}=0}^{N_{cand}} \frac{P(N_{comp}|N_{cand},N_{cont}) N_{YSO}!}{N_{comp}!(N_{YSO}-N_{comp})!}CP^{N_{comp}} (1-CP)^{(N_{YSO}-N_{comp})} 
\end{eqnarray}

\noindent
where $P_{tot}$ is the normalization of the distribution. The probability density functions for the $CP$s of protostars in regions of high and low stellar density are displayed in Figure~\ref{fig:appendix}. 
When comparing the values of $CP$ in regions of high and low stellar density, we can define a ratio similar to $R_{ij}$ in Appendix~B: $R_{CP} = CP^h/CP^l$. The determination of the probability density function of $P(R_{CP})$ requires us to marginalize over values of $CP^h$ and $CP^l$ consistent with a value of $R_{CP}$:

\begin{eqnarray}
P(R_{CP}| N_{cand}^h, N_{cont}^h,N_{YSO}^h,N_{cand}^l, N_{cont}^l,N_{YSO}^l) = ~~~~~~~~~~~~~~~~~~~~~~~~~~~~~~~~~~~ \nonumber \\
\frac{1}{P_{tot}}\int_{CP_{min}}^{CP_{max}} \! \! \! P(R_{CP} CP | N_{cand}^h, N_{cont}^h, N_{YSO}^h) P(CP | N_{cand}^l, N_{cont}^l, N_{YSO}^l)dCP.
\end{eqnarray}

\noindent
where we have set $CP = CP^l$ and $R_{CP} \times CP = CP^h$.  The cumulative probabilty density functions for the $CP$ ratios are shown in Figure~\ref{fig:ratio_csp_env_hst}.

In Section~\ref{sec:CP}, we assess the significance of the result that the $CP$ is higher in regions of higher stellar density using three approaches. First, we integrate the distribution of $R_{CP}$ values and determine $P(R_{CP} \le 1)$ (Table~\ref{tab:ratio_wfc3_limit}, Figure~\ref{fig:ratio_csp_env_hst}). Second, we use a Bayesian hypothesis testing approach where we compare the probability of $CF^h > CF^l$ to $CF^h < CF^l$. In this case, we must determine the probabilities of three hypotheses: that $CP^h = CP^l$,

\begin{eqnarray}
P(CP^l = CP^h) = ~~~~~~~~~~~~~~~~~~~~~~~~~~~~~~~~~~~~~~~~~~~~~~~~~~~~~~~~~~~~~~~~~~~~~~\nonumber \\
 \frac{1}{P_{tot}} \int_{CP_{min}}^{CP_{max}} P(CP | N_{cont}^l, N_{cand}^l, N_{YSO}^l)P(CP | N_{cont}^h, N_{cand}^h, N_{YSO}^h) \frac{dCP}{\Delta CP},
\end{eqnarray}

\noindent
where the range of possible values of $CP$ is given by $\Delta CP = CP_{max} - CP_{min}$, that $CP^h > CP^l$, 

\begin{eqnarray}
P(CP^h > CP^l) = ~~~~~~~~~~~~~~~~~~~~~~~~~~~~~~~~~~~~~~~~~~~~~~~~~~~~~~~~~~~~~~~~~~~~~~\nonumber \\
\int_{CP_{min}}^{CP_{max}}\int_{CP^l}^{CP_{max}} \frac{ P(CP^h | N_{cont}^h, N_{cand}^h,N_{YSO}^h)P(CP^l | N_{cont}^l, N_{cand}^l,N_{YSO}^l)}{P_{tot}(\Delta CP)(CP_{max} - CP^l)} dCP^h dCP^l,
\end{eqnarray}

\noindent
and that $CP^h < CP^l$

\begin{eqnarray}
P(CP^h < CP^l) = ~~~~~~~~~~~~~~~~~~~~~~~~~~~~~~~~~~~~~~~~~~~~~~~~~~~~~~~~~~~~~~~~~~~~~~\nonumber \\
\int_{CP_{min}}^{CP_{max}} \int_{CP_{min}}^{CP_l} \frac{P(CP^h | N_{cont}^h, N_{cand}^h, N_{YSO}^h )P(CP^l | N_{cont}^l, N_{cand}^l, N_{YSO}^l)} {P_{tot}(\Delta CP)(CP^l-CP_{min})}dCP^h dCP^l.
\end{eqnarray}

\noindent
The normalization of the probabilities is given by

\begin{equation}
P_{tot} = P(CP^l = CP^h) + P(CP^h > CP^l) + P(CP^h < CP^l).
\end{equation}

\noindent
We  set $CP_{max} = 1$, $CP_{min} = 0$, and $\Delta CP = 1$, and we adopt a uniform prior $P(CP) = 1$ for $0 \le CP \le 1$ and $P(CP)=0$ for $CP > 1$. 
The resulting probabilities are discussed in Sec~\ref{sec:CP}. 

Finally, we also implemented a frequentist hypothesis testing approach. In this approach, we define the probability density function for the continuous variable $\mu = CP^h-CP^l - (\overline{CP^h} - \overline{CF^l})$ where, similar to the frequentist testing in Appendix~B,  $\overline{CP^h}$ and $\overline{CF^l}$ are the population means of $CP^h$ and $CP^l$, respectively. We then assume that the null hypothesis holds and that the population means are equal: $\overline{CP^h} = \overline{CP^l}$. In contrast to our similar test for $CF$, we must integrate to marginalize over all values of $CP^h$ and $CP^l$ consistent with a given value of $\mu$:

\begin{eqnarray}
P(\mu | N_{cand}^h, N_{cont}^h,N_{YSO}^h,N_{cand}^l, N_{cont}^l,N_{YSO}^l ) = ~~~~~~~~~~~~~~~~~~~~~~~~~~~~~~~~~~~~~~~~~~~\nonumber \\
 \int_{0}^{1} P(CP^h | N_{cand}^h, N_{cont}^h,N_{YSO}^h)P(CP^l | N_{cand}^l, N_{cont}^l,N_{YSO}^l) dCP^l
\label{eqn:pmu}
\end{eqnarray}

\noindent
where $\mu = \delta CF^h - \delta CF^l$, $\delta CF = CF - \overline{CP_{obs}}$ and $\overline{CP_{obs}}$ is the observed value  given by Eqn~\ref{eqn:cf}.  Accordingly, we set  $CP^h$ in Equation~\ref{eqn:pmu} to be $CP^h = \mu +  \overline{\mu} + CP^l$, where $\overline{\mu} = \overline{CP_{obs}^h} - \overline{CP_{obs}^l}$, the difference of the actual measured values of $CF^h_{obs}$ and $CF^l_{obs}$, respectively. The probability that $\mu$ is equal to or exceeds the observed value in the case of the null hypothesis is then given by

\begin{equation}
P(\mu \ge \mu_{obs}) = \int_{\mu_{obs}}^1 P(\mu | N_{cand}^h, N_{cont}^h,N_{YSO}^h,N_{cand}^l, N_{cont}^l,N_{YSO}^l ) d\mu.
\end{equation}

\noindent
The results of this test are in Sec.~\ref{sec:CP}.

%\onecolumn

\LongTables
\begin{deluxetable}{ccccccccc}
 \tabletypesize{\tiny}
\tablewidth{0pt}
\tablecaption{List of the candidate multiple systems with projected separations less than 1000 AU. \label{tab:list}}
\tablecolumns{9} 
\tablehead{
\colhead{Name\tablenotemark{1}}&\colhead{$\alpha$}&\colhead{$\delta$}&\colhead{WFC3}&\colhead{NICMOS}&\colhead{NICMOS}&\colhead{IRTF}&\colhead{Separation}&\colhead{Common}\\
\colhead{}&\colhead{}&\colhead{}&\colhead{F160W}&\colhead{F160W}&\colhead{F205W}&\colhead{L'}&\colhead{(AU)}&\colhead{names}
}
           
\startdata
\sidehead{Protostars}
HOPS 3&5:54:56.95&1:42:56.1&13.502$\pm$0.042&---&---&10.043$\pm$0.007&---&---\\
---b&5:54:57.02&1:42:56.5&18.143$\pm$0.066&---&---&---&493&---\\
HOPS 5&5:54:32.15&1:48:07.3&17.413$\pm$0.046&---&---&---&---&---\\
---b&5:54:32.28&1:48:06.4&22.505$\pm$0.179&---&---&---&903&---\\
HOPS 15&5:36:18.99&-5:55:25.2&---&15.891$\pm$0.004&14.426$\pm$0.003&---&---&---\\
---b&5:36:19.02&-5:55:24.5&---&18.993$\pm$0.016&18.254$\pm$0.015&---&341&---\\
HOPS 24&5:34:46.93&-5:44:51.1&16.335$\pm$0.115&---&---&---&---&---\\
---b&5:34:46.92&-5:44:51.4&17.265$\pm$0.201&---&---&---&100&---\\
HOPS 45&5:35:06.44&-5:33:35.6&11.075$\pm$0.046&11.699$\pm$0.001&10.653$\pm$0.001&7.301$\pm$0.001&---&V982 Ori\\
---b&5:35:06.44&-5:33:35.3&11.717$\pm$0.058&16.549$\pm$0.01&12.209$\pm$0.001&---&136&---\\
HOPS 57&5:35:19.81&-5:15:08.9&---&11.785$\pm$0.002&10.813$\pm$0.001&9.045$\pm$0.001&---&V2359 Ori\\
---b&5:35:19.87&-5:15:08.0&---&14.604$\pm$0.011&12.384$\pm$0.004&8.350$\pm$0.010&492&---\\
HOPS 65&5:35:21.56&-5:09:38.7&---&13.595$\pm$0.002&12.880$\pm$0.002&---&---&V2377 Ori\\
---b&5:35:21.50&-5:09:40.7&---&---&19.026$\pm$0.359&---&916&---\\
HOPS 71&5:35:25.59&-5:07:57.6&18.244$\pm$0.236&20.456$\pm$0.05&14.572$\pm$0.004&10.200$\pm$0.200&---&---\\
---b&5:35:25.54&-5:07:56.8&12.710$\pm$0.030&12.474$\pm$0.002&12.162$\pm$0.002&11.350$\pm$0.003&445&---\\
---c&5:35:25.65&-5:07:57.2&10.777$\pm$0.028&10.489$\pm$0.001&10.214$\pm$0.001&10.070$\pm$0.100&450&---\\
HOPS 77&5:35:31.52&-5:05:47.0&10.157$\pm$0.055&---&---&6.565$\pm$0.000&---&V2502 Ori\\
---b&5:35:31.46&-5:05:45.7&15.433$\pm$0.082&---&---&---&700&---\\
HOPS 79&5:35:27.89&-5:05:36.3&18.813$\pm$0.030&---&---&---&---&---\\
---b&5:35:27.83&-5:05:36.3&17.064$\pm$0.026&---&---&11.661$\pm$0.040&343&---\\
HOPS 86&5:35:23.65&-5:01:40.3&19.465$\pm$0.070&---&---&---&---&---\\
---b&5:35:23.59&-5:01:40.2&20.225$\pm$0.082&---&---&---&424&---\\
HOPS 92&5:35:18.33&-5:00:33.2&17.008$\pm$0.031&---&---&7.410$\pm$0.050&---&---\\
---b&5:35:18.27&-5:00:34.1&19.605$\pm$0.044&---&---&8.550$\pm$0.150&575&---\\
HOPS 115&5:39:56.52&-7:25:51.6&16.406$\pm$0.030&---&---&---&---&---\\
---b&5:39:56.47&-7:25:51.4&17.831$\pm$0.035&---&---&---&381&---\\
HOPS 138&5:38:48.32&-7:02:43.7&18.659$\pm$0.051&---&---&---&---&---\\
---b&5:38:48.34&-7:02:43.7&20.428$\pm$0.092&---&---&---&150&---\\
HOPS 140&5:38:46.27&-7:01:53.7&17.107$\pm$0.037&---&---&---&---&---\\
---b&5:38:46.28&-7:01:51.4&20.685$\pm$0.063&---&---&---&955&---\\
HOPS 163&5:37:17.27&-6:36:18.6&12.841$\pm$0.046&---&---&---&---&---\\
---b&5:37:17.32&-6:36:18.5&13.342$\pm$0.046&---&---&---&313&---\\
HOPS 170&5:36:41.35&-6:34:00.5&---&10.002$\pm$0.001&9.645$\pm$0.001&8.620$\pm$0.010&---&V846 Ori\\
---b&5:36:41.31&-6:33:59.0&---&10.653$\pm$0.001&10.357$\pm$0.001&8.510$\pm$0.010&687&---\\
HOPS 177&5:35:49.99&-6:34:53.4&---&17.929$\pm$0.019&15.900$\pm$0.013&---&---&---\\
---b&5:35:49.87&-6:34:53.8&---&20.031$\pm$0.088&18.515$\pm$0.072&---&797&---\\
HOPS 183&5:36:17.85&-6:22:27.7&20.500$\pm$0.041&---&---&---&---&---\\
---b&5:36:17.89&-6:22:27.2&22.690$\pm$0.054&---&---&---&317&---\\
HOPS 189&5:35:30.91&-6:26:32.0&13.610$\pm$0.024&13.610$\pm$0.002&12.712$\pm$0.002&11.748$\pm$0.033&---&---\\
---b&5:35:30.92&-6:26:32.9&16.707$\pm$0.030&16.242$\pm$0.089&15.248$\pm$0.007&---&398&---\\
HOPS 193&5:36:30.27&-6:01:17.2&22.009$\pm$0.063&20.611$\pm$0.049&15.708$\pm$0.006&10.197$\pm$0.005&---&---\\
---b&5:36:30.27&-6:01:16.8&---&---&17.796$\pm$0.02&---&171&---\\
HOPS 210&5:42:58.28&-8:38:05.3&17.989$\pm$0.032&---&---&---&---&---\\
---b&5:42:58.28&-8:38:06.5&17.842$\pm$0.030&---&---&---&532&---\\
HOPS 226&5:41:30.04&-8:40:09.1&---&15.921$\pm$0.004&12.494$\pm$0.001&9.837$\pm$0.005&---&---\\
---b&5:41:30.01&-8:40:08.6&---&---&19.494$\pm$0.07&---&308&---\\
HOPS 229&5:42:47.34&-8:10:08.3&16.470$\pm$0.141&---&---&---&---&---\\
---b&5:42:47.32&-8:10:08.2&15.496$\pm$0.087&---&---&---&103&---\\
HOPS 242&5:40:48.54&-8:11:08.7&12.354$\pm$0.033&---&---&---&---&---\\
---b&5:40:48.56&-8:11:08.7&12.918$\pm$0.043&---&---&---&126&---\\
HOPS 255&5:40:50.56&-8:05:48.7&15.882$\pm$0.044&---&---&---&---&---\\
---b&5:40:50.54&-8:05:48.4&15.455$\pm$0.042&---&---&9.152$\pm$0.003&149&---\\
HOPS 268&5:40:38.34&-8:00:35.8&---&18.628$\pm$0.009&15.257$\pm$0.006&---&---&---\\
---b&5:40:38.36&-8:00:35.7&---&19.236$\pm$0.02&16.807$\pm$0.006&---&96&---\\
HOPS 281&5:40:24.62&-7:43:08.1&16.166$\pm$0.057&---&---&---&---&---\\
---b&5:40:24.63&-7:43:08.0&17.260$\pm$0.119&---&---&---&80&---\\
HOPS 298&5:41:37.17&-2:17:17.2&15.466$\pm$0.034&---&---&7.520$\pm$0.020&---&---\\
---b&5:41:37.02&-2:17:17.9&14.423$\pm$0.032&---&---&8.680$\pm$0.010&992&---\\
HOPS 304&5:41:45.94&-1:56:26.3&---&---&---&7.750$\pm$0.100&---&---\\
---b&5:41:45.90&-1:56:26.5&---&---&---&8.500$\pm$0.500&267&---\\
\sidehead{Stars with disks}\\
HOPS 272&5:40:20.53&-7:56:40.1&11.639$\pm$0.063&---&---&7.509$\pm$0.001&---&---\\
---b&5:40:20.66&-7:56:39.3&18.537$\pm$0.153&---&---&---&890&---\\
MGM 523&5:39:53.46&-7:30:09.6&16.058$\pm$0.039&---&---&---&---&---\\
---b&5:39:53.52&-7:30:09.3&15.445$\pm$0.037&---&---&---&433&---\\
MGM 526&5:39:55.05&-7:29:36.8&15.952$\pm$0.047&---&---&---&---&---\\
---b&5:39:55.15&-7:29:37.0&18.165$\pm$0.054&---&---&---&634&---\\
MGM 544&5:39:37.76&-7:26:23.1&13.807$\pm$0.026&---&---&---&---&---\\
---b&5:39:37.79&-7:26:23.3&15.542$\pm$0.033&---&---&---&229&---\\
MGM 561&5:39:58.90&-7:25:33.7&12.103$\pm$0.039&---&---&---&---&---\\
---b&5:39:58.87&-7:25:31.7&14.323$\pm$0.043&---&---&---&860&---\\
MGM 579&5:39:45.83&-7:22:37.4&12.614$\pm$0.028&---&---&---&---&---\\
---b&5:39:45.86&-7:22:37.2&13.442$\pm$0.033&---&---&---&186&---\\
MGM 685&5:38:43.33&-7:01:34.8&13.060$\pm$0.035&---&---&9.470$\pm$0.200&---&---\\
---b&5:38:43.38&-7:01:34.9&12.646$\pm$0.034&---&---&10.000$\pm$0.100&302&---\\
MGM 950&5:36:21.05&-6:21:53.3&11.520$\pm$0.028&---&---&---&---&---\\
---b&5:36:20.96&-6:21:52.3&13.267$\pm$0.030&---&---&---&682&---\\
MGM 1241&5:34:43.15&-5:44:39.9&11.926$\pm$0.029&---&---&---&---&IZ Ori\\
---b&5:34:43.10&-5:44:40.1&12.485$\pm$0.029&---&---&---&329&---\\
MGM 1171&5:35:03.96&-5:51:18.8&14.283$\pm$0.040&---&---&---&---&V2134 Ori\\
---b&5:35:03.96&-5:51:19.1&14.137$\pm$0.038&---&---&---&136&---\\
MGM 1378&5:35:07.68&-5:36:58.6&10.625$\pm$0.080&---&---&---&---&---\\
---b&5:35:07.83&-5:36:58.1&17.577$\pm$0.167&---&---&---&982&---\\
MGM 1501&5:35:08.85&-5:31:49.3&11.595$\pm$0.096&---&---&---&---&LM Ori\\
---b&5:35:08.84&-5:31:49.5&11.215$\pm$0.081&---&---&---&135&---\\
MGM 2213&5:35:22.62&-5:14:11.0&10.845$\pm$0.077&---&---&---&---&---\\
---b&5:35:22.76&-5:14:10.5&16.157$\pm$0.106&---&---&---&953&---\\
MGM 2239&5:35:18.24&-5:13:06.9&11.575$\pm$0.046&---&---&10.000$\pm$0.200&---&V2327 Ori\\
---b&5:35:18.21&-5:13:07.5&11.563$\pm$0.043&---&---&9.550$\pm$0.020&323&---\\
---c&5:35:18.21&-5:13:05.9&11.251$\pm$0.045&---&---&9.570$\pm$0.020&457&---\\
MGM 2285&5:35:28.14&-5:10:13.9&10.154$\pm$0.047&---&---&8.510$\pm$0.010&---&V419 Ori\\
---b&5:35:28.17&-5:10:12.3&13.109$\pm$0.049&---&---&11.700$\pm$0.100&720&---\\
MGM 2341&5:35:22.36&-5:07:39.1&12.262$\pm$0.031&---&---&8.317$\pm$0.004&---&---\\
---b&5:35:22.35&-5:07:38.6&15.864$\pm$0.060&---&---&---&240&---\\
MGM 2398&5:35:31.07&-5:04:14.6&10.604$\pm$0.027&---&---&10.570$\pm$0.400&---&---\\
---b&5:35:31.06&-5:04:13.4&12.085$\pm$0.028&---&---&11.950$\pm$0.100&485&---\\
MGM 2472&5:35:28.22&-4:58:37.8&16.145$\pm$0.035&---&---&---&---&V2475 Ori\\
---b&5:35:28.26&-4:58:38.9&15.228$\pm$0.033&---&---&---&545&---\\
MGM 2523&5:35:11.84&-4:54:21.4&16.827$\pm$0.048&---&---&12.538$\pm$0.063&---&---\\
---b&5:35:11.78&-4:54:21.7&14.704$\pm$0.041&---&---&---&386&---\\
MGM 2848&5:41:36.37&-2:16:46.2&10.531$\pm$0.050&---&---&6.684$\pm$0.000&---&---\\
---b&5:41:36.37&-2:16:46.5&11.208$\pm$0.062&---&---&---&145&---\\
MGM 2849&5:41:22.14&-2:16:44.2&13.604$\pm$0.045&---&---&---&---&---\\
---b&5:41:22.12&-2:16:43.0&17.480$\pm$0.040&---&---&---&497&---\\
MGM 2851&5:41:42.23&-2:16:24.4&13.761$\pm$0.043&---&---&---&---&---\\
---b&5:41:42.08&-2:16:25.0&13.569$\pm$0.042&---&---&---&997&---\\
MGM 3214&5:46:35.48&0:01:39.9&15.461$\pm$0.056&---&---&---&---&---\\
---b&5:46:35.50&0:01:39.0&12.264$\pm$0.048&---&---&---&394&---\\
MGM 3352&5:47:27.74&0:20:36.3&14.014$\pm$0.049&---&---&---&---&---\\
---b&5:47:27.78&0:20:36.0&12.633$\pm$0.047&---&---&---&297&---\\
MGM 3374&5:47:05.74&0:22:10.0&11.832$\pm$0.041&---&---&---&---&---\\
---b&5:47:05.62&0:22:10.8&12.459$\pm$0.038&---&---&---&796&---\\
MGM 3376&5:47:12.44&0:22:15.3&13.602$\pm$0.035&---&---&---&---&---\\
---b&5:47:12.43&0:22:15.5&14.097$\pm$0.045&---&---&---&95&---\\
MGM 3385&5:47:05.32&0:23:10.0&11.936$\pm$0.051&---&---&---&---&---\\
---b&5:47:05.32&0:23:09.7&12.357$\pm$0.061&---&---&---&154&---\\
MGM 3430&5:47:35.74&0:38:39.9&9.900$\pm$0.114&---&---&---&---&---\\
---b&5:47:35.90&0:38:40.5&14.148$\pm$0.056&---&---&---&992&---\\

\enddata
\tablenotetext{1} {MGM=\citet{2012AJ....144..192M}}
\end{deluxetable}

\begin{deluxetable}{cccccccc}
 \tabletypesize{\tiny}
\tablewidth{0pt}
\tablecaption{List of the YSOs without a companion. \label{tab:single}}
\tablecolumns{8} 
\tablehead{
\colhead{Name\tablenotemark{1}}&\colhead{$\alpha$}&\colhead{$\delta$}&\colhead{WFC3}&\colhead{NICMOS}&\colhead{NICMOS}&\colhead{IRTF}&\colhead{Common}\\
\colhead{}&\colhead{}&\colhead{}&\colhead{F160W}&\colhead{F160W}&\colhead{F205W}&\colhead{L'}&\colhead{name}
}
           
\startdata
\sidehead{Protostars}
HOPS 1&5:54:12.35&1:42:35.14&---&18.798$\pm$0.018&16.139$\pm$0.007&---&---\\
HOPS 2&5:54:09.12&1:42:51.98&---&15.822$\pm$0.004&13.601$\pm$0.002&---&---\\
HOPS 4&5:54:53.77&1:47:09.66&15.259$\pm$0.049&---&---&---&---\\
HOPS 6&5:54:18.41&1:49:03.69&---&19.709$\pm$0.03&17.503$\pm$0.015&---&---\\
HOPS 7&5:54:20.04&1:50:42.75&---&19.276$\pm$0.026&16.881$\pm$0.011&---&---\\
HOPS 10&5:35:09.00&-5:58:27.32&---&---&18.555$\pm$0.039&---&---\\
HOPS 11&5:35:13.41&-5:57:58.12&---&20.26$\pm$0.04&16.918$\pm$0.011&---&---\\
HOPS 13&5:35:24.55&-5:55:33.83&---&13.6$\pm$0.002&11.567$\pm$0.001&9.028$\pm$0.002&V2426 Ori\\
HOPS 16&5:35:00.81&-5:55:25.68&---&15.896$\pm$0.004&13.174$\pm$0.002&---&V2105 Ori\\
HOPS 17&5:35:07.18&-5:52:05.87&13.966$\pm$0.032&---&---&---&---\\
HOPS 18&5:35:05.53&-5:51:54.29&16.738$\pm$0.078&---&---&---&---\\
HOPS 20&5:33:30.70&-5:50:41.33&---&19.772$\pm$0.038&17.762$\pm$0.021&---&---\\
HOPS 29&5:34:49.05&-5:41:42.55&15.672$\pm$0.03&---&---&---&---\\
HOPS 30&5:34:44.06&-5:41:25.90&---&---&18.365$\pm$0.024&---&---\\
HOPS 36&5:34:26.44&-5:37:40.75&---&14.837$\pm$0.003&13.248$\pm$0.002&9.516$\pm$0.002&V1957 Ori\\
HOPS 41&5:34:29.45&-5:35:42.78&---&---&18.94$\pm$0.043&---&---\\
HOPS 42&5:35:05.05&-5:35:40.74&15.759$\pm$0.056&15.266$\pm$0.004&14.112$\pm$0.003&---&---\\
HOPS 43&5:35:04.48&-5:35:14.67&---&---&18.826$\pm$0.037&---&---\\
HOPS 44&5:35:10.57&-5:35:06.31&---&20.791$\pm$0.063&18.307$\pm$0.067&---&---\\
HOPS 52&5:35:16.32&-5:29:32.78&---&---&---&9.584$\pm$0.003&V2287 Ori\\
HOPS 56&5:35:19.46&-5:15:32.72&---&---&---&11.94$\pm$0.031&---\\
HOPS 58&5:35:18.51&-5:13:38.16&10.113$\pm$0.037&---&---&7.791$\pm$0.001&V2331 Ori\\
HOPS 59&5:35:20.13&-5:13:15.59&9.217$\pm$0.132&---&---&5.17$\pm$0.05&V2364 Ori\\
HOPS 62&5:35:24.58&-5:11:29.78&11.08$\pm$0.042&10.971$\pm$0.001&10.32$\pm$0.001&---&V2427 Ori\\
HOPS 63&5:35:24.89&-5:10:01.59&---&---&---&11.188$\pm$0.025&---\\
HOPS 64&5:35:26.98&-5:09:54.09&13.72$\pm$0.033&---&---&9.657$\pm$0.004&V2457 Ori\\
HOPS 66&5:35:26.84&-5:09:24.38&13.704$\pm$0.047&---&---&6.252$\pm$0.001&V2455 Ori\\
HOPS 68&5:35:24.30&-5:08:30.82&---&---&---&11.441$\pm$0.025&---\\
HOPS 70&5:35:22.40&-5:08:05.09&9.728$\pm$0.078&---&---&7.274$\pm$0.002&---\\
HOPS 74&5:35:24.86&-5:06:21.61&12.896$\pm$0.024&---&---&8.785$\pm$0.001&---\\
HOPS 75&5:35:26.68&-5:06:10.26&---&---&---&12.173$\pm$0.046&---\\
HOPS 76&5:35:25.75&-5:05:58.13&16.907$\pm$0.031&16.446$\pm$0.007&13.281$\pm$0.003&9.74$\pm$0.005&---\\
HOPS 80&5:35:25.17&-5:05:09.35&22.723$\pm$0.045&---&---&---&---\\
HOPS 82&5:35:19.69&-5:04:54.67&19.701$\pm$0.04&---&---&---&---\\
HOPS 84&5:35:26.54&-5:03:55.06&---&---&---&8.239$\pm$0.003&---\\
HOPS 85&5:35:28.17&-5:03:40.59&17.09$\pm$0.038&---&---&7.109$\pm$0.001&---\\
HOPS 89&5:35:19.97&-5:01:02.77&21.459$\pm$0.025&---&---&9.629$\pm$0.004&---\\
HOPS 90&5:35:34.49&-5:00:52.35&---&---&---&8.733$\pm$0.002&V2535 Ori\\
HOPS 93&5:35:15.03&-5:00:07.99&---&17.465$\pm$0.011&15.784$\pm$0.007&---&---\\
HOPS 94&5:35:16.15&-5:00:02.70&14.259$\pm$0.029&14.159$\pm$0.003&11.99$\pm$0.002&8.982$\pm$0.004&V2282 Ori\\
HOPS 95&5:35:34.19&-4:59:52.39&---&---&19.209$\pm$0.064&---&---\\
HOPS 99&5:34:29.51&-4:55:30.63&---&---&18.574$\pm$0.030&---&---\\
HOPS 105&5:35:32.29&-4:46:48.41&14.237$\pm$0.035&---&---&---&V2514 Ori\\
HOPS 107&5:35:23.34&-4:40:10.31&12.116$\pm$0.033&---&---&7.621$\pm$0.001&V2408 Ori\\
HOPS 114&5:40:01.37&-7:25:38.64&20.007$\pm$0.04&---&---&---&---\\
HOPS 116&5:39:57.89&-7:25:13.03&15.582$\pm$0.041&---&---&---&---\\
HOPS 117&5:39:55.45&-7:24:19.40&16.519$\pm$0.047&---&---&---&---\\
HOPS 118&5:39:54.59&-7:24:14.90&14.425$\pm$0.047&---&---&---&---\\
HOPS 119&5:39:50.67&-7:23:30.42&12.252$\pm$0.033&---&---&---&---\\
HOPS 120&5:39:34.30&-7:26:11.26&14.562$\pm$0.053&---&---&---&---\\
HOPS 121&5:39:33.68&-7:23:02.31&21.076$\pm$0.037&---&---&---&---\\
HOPS 124&5:39:19.98&-7:26:11.18&---&---&---&11.258$\pm$0.013&---\\
HOPS 125&5:39:19.61&-7:26:18.83&---&12.567$\pm$0.001&10.954$\pm$0.001&8.167$\pm$0.001&---\\
HOPS 127&5:39:00.96&-7:20:23.00&17.364$\pm$0.037&---&---&---&---\\
HOPS 128&5:38:52.00&-7:21:05.74&13.755$\pm$0.051&---&---&---&---\\
HOPS 129&5:39:11.84&-7:10:34.59&20.624$\pm$0.047&---&---&---&---\\
HOPS 130&5:39:02.96&-7:12:52.23&17.132$\pm$0.05&---&---&---&---\\
HOPS 131&5:39:07.55&-7:10:52.02&17.758$\pm$0.033&---&---&14.162$\pm$0.199&---\\
HOPS 132&5:39:05.35&-7:11:04.97&11.501$\pm$0.032&---&---&9.046$\pm$0.002&---\\
HOPS 133&5:39:05.81&-7:10:39.17&---&---&---&13.189$\pm$0.098&---\\
HOPS 134&5:38:42.77&-7:12:43.88&---&---&---&7.271$\pm$0.001&---\\
HOPS 139&5:38:49.63&-7:01:17.82&---&---&---&11.919$\pm$0.048&---\\
HOPS 141&5:38:48.02&-7:00:49.39&14.315$\pm$0.044&14.417$\pm$0.002&12.614$\pm$0.001&12.099$\pm$0.025&---\\
HOPS 142&5:38:47.77&-7:00:27.08&---&---&---&13.403$\pm$0.084&---\\
HOPS 143&5:38:46.18&-7:00:48.58&---&---&16.583$\pm$0.008&9.234$\pm$0.002&---\\
HOPS 144&5:38:45.01&-7:01:01.74&---&---&17.959$\pm$0.021&11.585$\pm$0.016&---\\
HOPS 145&5:38:43.84&-7:01:13.13&15.098$\pm$0.041&14.773$\pm$0.003&12.773$\pm$0.002&11.001$\pm$0.014&---\\
HOPS 148&5:38:39.51&-6:59:30.30&---&16.197$\pm$0.005&13.707$\pm$0.002&---&---\\
HOPS 149&5:38:40.46&-6:58:22.02&---&---&---&5.98$\pm$0.04&---\\
HOPS 150&5:38:07.51&-7:08:28.73&16.133$\pm$0.042&---&---&8.273$\pm$0.001&---\\
HOPS 154&5:38:20.10&-6:59:04.68&15.359$\pm$0.03&---&---&---&---\\
HOPS 156&5:38:03.40&-6:58:15.82&---&20.597$\pm$0.049&17.005$\pm$0.011&---&---\\
HOPS 157&5:37:56.55&-6:56:39.38&---&---&17.173$\pm$0.017&---&---\\
HOPS 158&5:37:24.46&-6:58:32.85&---&11.648$\pm$0.001&10.267$\pm$0.001&8.472$\pm$0.001&---\\
HOPS 159&5:37:53.73&-6:47:17.17&---&14.612$\pm$0.003&12.785$\pm$0.002&---&---\\
HOPS 160&5:37:51.03&-6:47:20.66&---&16.551$\pm$0.006&14.492$\pm$0.004&---&---\\
HOPS 165&5:36:23.51&-6:46:14.40&---&---&16.619$\pm$0.009&---&---\\
HOPS 166&5:36:25.11&-6:44:42.01&---&9.505$\pm$0.001&8.335$\pm$0.001&6.502$\pm$0.001&---\\
HOPS 167&5:36:19.78&-6:46:00.76&---&14.599$\pm$0.002&13.402$\pm$0.002&---&V2665 Ori\\
HOPS 172&5:36:19.45&-6:29:06.70&---&16.298$\pm$0.005&14.672$\pm$0.003&---&---\\
HOPS 174&5:36:25.85&-6:24:58.71&13.349$\pm$0.037&---&---&8.735$\pm$0.002&---\\
HOPS 175&5:36:24.06&-6:24:55.35&18.153$\pm$0.047&18.549$\pm$0.016&15.936$\pm$0.006&11.557$\pm$0.032&---\\
HOPS 176&5:36:23.59&-6:24:51.57&19.74$\pm$0.059&18.247$\pm$0.015&13.812$\pm$0.003&9.045$\pm$0.004&---\\
HOPS 178&5:36:24.60&-6:22:41.07&22.826$\pm$0.041&---&---&7.468$\pm$0.001&---\\
HOPS 179&5:36:21.85&-6:23:29.85&11.87$\pm$0.028&---&---&8.489$\pm$0.004&---\\
HOPS 187&5:35:50.96&-6:22:43.58&11.873$\pm$0.033&---&---&---&CG Ori\\
HOPS 188&5:35:29.83&-6:26:58.28&13.501$\pm$0.044&---&---&9.634$\pm$0.003&---\\
HOPS 190&5:35:28.50&-6:27:01.92&13.988$\pm$0.043&---&---&9.798$\pm$0.003&---\\
HOPS 191&5:36:17.26&-6:11:11.45&---&15.318$\pm$0.005&13.897$\pm$0.003&---&---\\
HOPS 192&5:36:32.46&-6:01:16.32&15.164$\pm$0.047&---&---&10.334$\pm$0.005&V2692 Ori\\
HOPS 194&5:35:52.01&-6:10:01.62&---&8.827$\pm$0.001&8.241$\pm$0.001&6.915$\pm$0.001&V1296 Ori\\
HOPS 197&5:34:15.89&-6:34:32.61&---&13.855$\pm$0.002&12.808$\pm$0.001&---&---\\
HOPS 198&5:35:22.17&-6:13:06.09&---&18.911$\pm$0.022&16.537$\pm$0.01&---&---\\
HOPS 199&5:34:39.87&-6:25:14.06&---&13.762$\pm$0.002&13.003$\pm$---&---&V2001 Ori\\
HOPS 200&5:35:33.19&-6:06:09.64&---&16.475$\pm$0.006&14.765$\pm$0.004&---&---\\
HOPS 207&5:42:38.58&-8:50:18.87&14.601$\pm$0.018&---&---&---&---\\
HOPS 209&5:42:52.89&-8:41:40.91&12.954$\pm$0.044&---&---&---&---\\
HOPS 213&5:42:48.08&-8:40:08.51&12.741$\pm$0.047&---&---&8.34$\pm$0.04&---\\
HOPS 214&5:42:47.21&-8:36:36.79&15.393$\pm$0.051&---&---&---&---\\
HOPS 215&5:43:09.58&-8:29:27.40&---&14.851$\pm$0.003&13.558$\pm$0.002&---&---\\
HOPS 216&5:42:55.55&-8:32:48.15&15.855$\pm$0.059&---&---&---&---\\
HOPS 220&5:41:29.77&-8:42:46.05&16.886$\pm$0.05&---&---&---&---\\
HOPS 221&5:42:47.10&-8:17:06.47&13.689$\pm$0.045&---&---&8.57$\pm$0.001&---\\
HOPS 223&5:42:48.52&-8:16:34.23&10.188$\pm$0.05&---&---&6.255$\pm$0.001&V2775 Ori\\
HOPS 224&5:41:32.02&-8:40:09.73&---&---&---&12.418$\pm$0.053&---\\
HOPS 225&5:41:30.33&-8:40:17.63&---&15.037$\pm$0.003&12.355$\pm$0.001&9.32$\pm$0.003&---\\
HOPS 228&5:41:34.17&-8:35:27.78&---&12.94$\pm$0.002&10.584$\pm$0.001&6.956$\pm$0.001&---\\
HOPS 232&5:41:35.46&-8:08:22.28&16.705$\pm$0.031&---&---&---&---\\
HOPS 233&5:41:52.29&-8:01:21.47&18.747$\pm$0.045&---&---&---&---\\
HOPS 235&5:41:25.32&-8:05:54.75&9.728$\pm$0.034&---&---&7.788$\pm$0.001&DL Ori\\
HOPS 236&5:41:30.20&-8:03:41.78&13.465$\pm$0.034&---&---&7.893$\pm$0.001&---\\
HOPS 237&5:41:28.96&-8:03:26.26&20.488$\pm$0.035&19.737$\pm$0.029&17.08$\pm$0.011&13.902$\pm$0.143&---\\
HOPS 238&5:41:26.64&-8:03:12.89&16.218$\pm$0.054&---&---&---&---\\
HOPS 239&5:41:27.06&-8:00:54.46&23.199$\pm$0.052&---&---&---&---\\
HOPS 240&5:41:25.96&-8:01:15.64&23.093$\pm$0.035&---&---&---&---\\
HOPS 245&5:41:22.87&-7:58:56.08&18.181$\pm$0.026&---&---&10.722$\pm$0.02&---\\
HOPS 248&5:41:22.11&-7:58:03.14&13.151$\pm$0.029&---&---&8.325$\pm$0.001&---\\
HOPS 249&5:40:52.83&-8:05:48.67&18.078$\pm$0.04&---&---&---&---\\
HOPS 250&5:40:48.84&-8:06:57.15&16.371$\pm$0.046&---&---&---&---\\
HOPS 251&5:40:53.98&-8:05:12.93&15.84$\pm$0.041&---&---&---&---\\
HOPS 252&5:40:49.88&-8:06:08.15&14.431$\pm$0.024&---&---&8.396$\pm$0.001&---\\
HOPS 253&5:41:28.76&-7:53:51.13&14.629$\pm$0.04&---&---&---&---\\
HOPS 256&5:40:45.28&-8:06:41.93&18.949$\pm$0.055&---&---&---&---\\
HOPS 257&5:41:19.88&-7:55:46.71&17.708$\pm$0.031&---&---&---&---\\
HOPS 258&5:41:24.71&-7:54:08.49&16.392$\pm$0.032&---&---&9.035$\pm$0.002&---\\
HOPS 259&5:40:20.89&-8:13:55.54&15.9$\pm$0.021&---&---&9.472$\pm$0.003&---\\
HOPS 260&5:40:19.39&-8:14:16.67&12.077$\pm$0.031&---&---&8.769$\pm$0.002&---\\
HOPS 262&5:41:23.96&-7:53:42.14&18.392$\pm$0.027&---&---&9.914$\pm$0.004&---\\
HOPS 263&5:41:23.67&-7:53:46.49&---&---&---&12.03$\pm$0.028&---\\
HOPS 265&5:41:20.33&-7:53:10.63&14.842$\pm$0.034&---&---&---&---\\
HOPS 267&5:41:19.65&-7:50:40.89&---&17.03$\pm$0.007&13.868$\pm$0.002&---&---\\
HOPS 273&5:40:20.89&-7:56:25.00&15.294$\pm$0.016&---&---&9.377$\pm$0.002&---\\
HOPS 274&5:40:20.72&-7:54:59.87&14.575$\pm$0.039&---&---&9.452$\pm$0.005&---\\
HOPS 275&5:40:36.32&-7:49:07.08&16.344$\pm$0.028&---&---&---&---\\
HOPS 279&5:40:17.78&-7:48:25.95&13.189$\pm$0.021&12.587$\pm$0.001&10.537$\pm$0.001&7.362$\pm$0.001&---\\
HOPS 280&5:40:14.92&-7:48:48.69&15.75$\pm$0.038&---&---&11.049$\pm$0.022&---\\
HOPS 284&5:38:51.48&-8:01:27.38&11.735$\pm$0.037&---&---&---&---\\
HOPS 286&5:39:58.69&-7:31:12.57&---&20.214$\pm$0.048&14.572$\pm$0.003&---&---\\
HOPS 289&5:39:56.74&-7:30:05.80&16.832$\pm$0.047&---&---&---&---\\
HOPS 291&5:39:57.94&-7:28:57.54&17.92$\pm$0.028&---&---&---&---\\
HOPS 294&5:40:51.71&-2:26:48.66&10.326$\pm$0.036&---&---&8.396$\pm$0.002&---\\
HOPS 297&5:41:23.28&-2:17:35.81&18.949$\pm$0.031&---&---&---&---\\
HOPS 299&5:41:44.59&-2:16:06.49&12.476$\pm$0.04&12.633$\pm$0.002&10.396$\pm$0.001&6.773$\pm$0.001&---\\
HOPS 300&5:41:24.21&-2:16:06.56&19.076$\pm$0.049&---&---&---&---\\
HOPS 305&5:41:45.37&-1:51:56.65&---&---&---&9.379$\pm$0.002&---\\
HOPS 311&5:43:03.05&-1:16:29.36&12.477$\pm$0.039&12.514$\pm$0.001&10.861$\pm$0.001&9.357$\pm$0.002&---\\
HOPS 315&5:46:03.64&-0:14:49.20&---&---&---&8.826$\pm$0.001&---\\
HOPS 322&5:46:46.51&0:00:16.09&---&---&---&13.363$\pm$0.148&---\\
HOPS 323&5:46:47.69&0:00:25.00&---&---&---&10.188$\pm$0.009&---\\
HOPS 342&5:47:57.10&0:35:27.19&16.231$\pm$0.033&---&---&---&---\\
HOPS 344&5:47:24.73&0:37:34.98&15.005$\pm$0.035&---&---&---&---\\
HOPS 345&5:47:38.94&0:38:36.20&14.467$\pm$0.053&---&---&---&---\\
HOPS 346&5:47:42.97&0:40:57.42&12.172$\pm$0.026&---&---&---&---\\
HOPS 363&5:46:43.13&0:00:52.55&11.053$\pm$0.045&---&---&---&---\\
HOPS 364&5:47:36.57&0:20:05.97&12.611$\pm$0.062&---&---&---&---\\
HOPS 366&5:47:04.02&0:22:10.09&16.568$\pm$0.029&---&---&---&---\\
HOPS 367&5:54:36.29&1:53:54.00&15.649$\pm$0.04&---&---&---&---\\
HOPS 369&5:35:26.96&-5:10:17.27&11.864$\pm$0.04&---&---&6.012$\pm$0.001&---\\
HOPS 370&5:35:27.63&-5:09:33.82&---&---&---&6.937$\pm$0.006&---\\
HOPS 374&5:41:25.45&-7:55:18.84&20.556$\pm$0.035&---&---&---&---\\
HOPS 377&5:38:45.51&-7:01:02.15&---&---&21.984$\pm$0.271&13.175$\pm$0.069&---\\
HOPS 385&5:46:04.77&-0:14:16.26&---&---&---&6.616$\pm$0.001&---\\
HOPS 386&5:46:08.47&-0:10:02.50&16.481$\pm$0.034&---&---&---&---\\
HOPS 389&5:46:47.03&0:00:27.73&20.076$\pm$0.041&---&---&12.888$\pm$0.098&---\\
HOPS 394&5:35:23.95&-5:07:52.86&---&---&---&13.113$\pm$0.097&---\\
MGM 695&5:38:45.63&-7:00:54.38&19.2$\pm$0.048&18.362$\pm$0.015&14.615$\pm$0.003&10.985$\pm$0.009&---\\
MGM 2236&5:35:20.77&-5:13:23.04&20.056$\pm$0.066&---&---&12.359$\pm$0.044&---\\
MGM 2351&5:35:25.43&-5:06:52.82&22.604$\pm$0.067&---&---&12.325$\pm$0.036&---\\
MGM 2405&5:35:26.91&-5:04:06.24&17.243$\pm$0.037&---&---&11.418$\pm$0.02&---\\
MGM 2853&5:41:44.21&-2:16:07.54&12.439$\pm$0.051&12.402$\pm$0.001&11.663$\pm$0.001&10.397$\pm$0.008&---\\
MGM 3366&5:47:04.79&0:21:42.59&13.953$\pm$0.057&---&---&---&---\\

\sidehead{Stars with disks}\\
HOPS 51&5:35:15.82&-5:30:05.86&---&---&---&10.736$\pm$0.013&V2275 Ori\\
HOPS 98&5:35:19.30&-4:55:45.12&13.64$\pm$0.019&---&---&7.708$\pm$0.001&V2348 Ori\\
HOPS 113&5:39:58.10&-7:26:41.17&19.265$\pm$0.057&---&---&---&---\\
HOPS 222&5:41:26.69&-8:42:24.44&12.466$\pm$0.046&---&---&9.292$\pm$0.002&---\\
HOPS 283&5:40:44.66&-7:29:54.47&11.554$\pm$0.032&---&---&9.083$\pm$0.002&---\\
MGM 209&5:41:30.48&-8:43:58.84&14.736$\pm$0.074&---&---&---&---\\
MGM 225&5:42:46.10&-8:40:01.02&10.003$\pm$0.046&---&---&8.167$\pm$0.001&---\\
MGM 227&5:42:50.51&-8:39:57.96&11.449$\pm$0.03&---&---&9.533$\pm$0.005&---\\
MGM 240&5:42:57.92&-8:38:25.42&19.21$\pm$0.043&---&---&---&---\\
MGM 245&5:42:56.37&-8:37:45.83&12.684$\pm$0.04&---&---&---&---\\
MGM 289&5:42:44.31&-8:16:45.58&16.456$\pm$0.04&---&---&---&---\\
MGM 300&5:40:20.92&-8:14:06.80&15.814$\pm$0.034&---&---&12.442$\pm$0.048&---\\
MGM 318&5:42:45.88&-8:10:40.22&15.243$\pm$0.046&---&---&---&---\\
MGM 349&5:40:44.16&-8:07:34.95&11.468$\pm$0.11&---&---&---&---\\
MGM 351&5:40:46.60&-8:07:12.80&10.428$\pm$0.032&---&---&---&---\\
MGM 355&5:40:45.03&-8:06:39.67&17.131$\pm$0.034&---&---&---&---\\
MGM 361&5:40:59.74&-8:06:03.10&12.369$\pm$0.04&---&---&---&---\\
MGM 363&5:40:49.91&-8:05:58.67&19.243$\pm$0.022&---&---&11.758$\pm$0.025&---\\
MGM 364&5:40:48.05&-8:05:58.40&9.313$\pm$0.101&---&---&8.176$\pm$0.001&---\\
MGM 366&5:40:50.00&-8:05:54.98&18.558$\pm$0.022&---&---&11.558$\pm$0.021&---\\
MGM 371&5:40:46.20&-8:05:24.08&9.503$\pm$0.046&---&---&7.579$\pm$0.001&UU Ori\\
MGM 380&5:40:53.63&-8:04:23.21&15.161$\pm$0.023&---&---&---&---\\
MGM 381&5:41:28.89&-8:04:13.47&13.808$\pm$0.048&---&---&11.2$\pm$0.018&---\\
MGM 399&5:41:49.71&-8:00:31.77&9.392$\pm$0.153&---&---&---&V1305 Ori\\
MGM 422&5:40:20.29&-7:56:25.20&12.485$\pm$0.027&---&---&9.68$\pm$0.003&---\\
MGM 427&5:41:26.44&-7:55:42.23&15.322$\pm$0.06&---&---&---&---\\
MGM 431&5:41:20.11&-7:55:23.96&12.723$\pm$0.037&---&---&---&---\\
MGM 438&5:41:29.94&-7:54:21.32&14.356$\pm$0.053&---&---&---&---\\
MGM 441&5:41:19.42&-7:53:47.73&15.067$\pm$0.046&---&---&---&---\\
MGM 446&5:41:19.07&-7:53:37.43&15.597$\pm$0.052&---&---&---&---\\
MGM 448&5:41:21.74&-7:53:16.18&14.195$\pm$0.018&---&---&---&---\\
MGM 468&5:40:17.14&-7:49:14.61&11.045$\pm$0.036&---&---&9.206$\pm$0.006&---\\
MGM 469&5:40:38.52&-7:49:09.00&14.681$\pm$0.022&---&---&---&---\\
MGM 471&5:40:18.44&-7:49:06.72&13.47$\pm$0.032&---&---&11.073$\pm$0.022&---\\
MGM 520&5:39:54.98&-7:30:20.04&18.398$\pm$0.049&---&---&---&---\\
MGM 521&5:39:56.16&-7:30:14.66&22.744$\pm$0.035&---&---&---&---\\
MGM 534&5:40:07.97&-7:27:41.32&14.566$\pm$0.027&---&---&---&---\\
MGM 535&5:40:10.34&-7:27:38.59&11.124$\pm$0.031&---&---&---&---\\
MGM 538&5:40:12.21&-7:27:02.86&13.061$\pm$0.029&---&---&---&---\\
MGM 547&5:39:35.48&-7:26:16.62&13.055$\pm$0.022&---&---&---&---\\
MGM 549&5:39:58.12&-7:26:13.24&23.125$\pm$0.071&---&---&---&---\\
MGM 559&5:40:05.15&-7:25:43.71&15.637$\pm$0.038&---&---&---&---\\
MGM 563&5:39:55.33&-7:24:39.68&16.156$\pm$0.047&---&---&---&---\\
MGM 571&5:39:48.38&-7:24:14.74&13.802$\pm$0.048&---&---&---&---\\
MGM 582&5:39:32.32&-7:22:24.66&13.397$\pm$0.033&---&---&---&V892 Ori\\
MGM 586&5:38:52.36&-7:21:09.20&9.091$\pm$0.086&---&---&---&---\\
MGM 591&5:38:50.52&-7:20:29.63&13.451$\pm$0.049&---&---&---&---\\
MGM 597&5:38:50.01&-7:20:18.37&10.331$\pm$0.053&---&---&---&---\\
MGM 635&5:38:07.52&-7:09:14.56&17.771$\pm$0.052&---&---&---&---\\
MGM 665&5:38:46.74&-7:02:49.24&13.933$\pm$0.049&---&---&---&---\\
MGM 666&5:38:44.79&-7:02:47.05&12.713$\pm$0.052&---&---&---&---\\
MGM 674&5:38:50.78&-7:02:14.06&16.374$\pm$0.026&---&---&---&---\\
MGM 676&5:38:45.71&-7:01:58.52&13.372$\pm$0.024&---&---&---&---\\
MGM 679&5:38:47.23&-7:01:53.39&15.481$\pm$0.048&---&---&---&---\\
MGM 682&5:38:50.37&-7:01:48.70&16.59$\pm$0.022&---&---&---&---\\
MGM 692&5:38:47.29&-7:00:59.98&15.315$\pm$0.042&---&---&10.773$\pm$0.008&---\\
MGM 698&5:38:48.84&-7:00:43.20&12.404$\pm$0.048&---&---&10.781$\pm$0.01&---\\
MGM 877&5:35:30.40&-6:27:07.26&14.529$\pm$0.034&---&---&14.417$\pm$0.227&---\\
MGM 887&5:35:45.92&-6:25:59.18&10.857$\pm$0.032&---&---&---&V811 Ori\\
MGM 890&5:36:27.87&-6:25:35.94&12.141$\pm$0.058&---&---&11.605$\pm$0.029&---\\
MGM 900&5:36:26.09&-6:24:51.86&13.92$\pm$0.042&---&---&13.872$\pm$0.161&---\\
MGM 911&5:36:22.44&-6:23:44.70&12.517$\pm$0.033&---&---&10.56$\pm$0.006&V2672 Ori\\
MGM 916&5:36:25.84&-6:23:31.56&17.132$\pm$0.024&---&---&---&---\\
MGM 922&5:36:20.48&-6:23:22.08&12.954$\pm$0.06&---&---&10.676$\pm$0.007&V2667 Ori\\
MGM 923&5:35:54.30&-6:23:19.82&14.853$\pm$0.06&---&---&---&---\\
MGM 924&5:36:27.69&-6:23:12.15&11.265$\pm$0.05&---&---&10.232$\pm$0.014&V2681 Ori\\
MGM 925&5:36:23.75&-6:23:11.15&12.349$\pm$0.035&---&---&8.597$\pm$0.002&V2674 Ori\\
MGM 927&5:36:21.55&-6:22:52.42&12.134$\pm$0.024&---&---&9.118$\pm$0.003&V2670 Ori\\
MGM 928&5:36:19.06&-6:22:50.39&14.078$\pm$0.049&---&---&13.043$\pm$0.103&---\\
MGM 929&5:36:17.61&-6:22:49.25&17.17$\pm$0.034&---&---&---&---\\
MGM 934&5:36:20.92&-6:22:38.60&16.808$\pm$0.046&---&---&---&---\\
MGM 937&5:36:24.46&-6:22:23.21&11.545$\pm$0.036&---&---&10.615$\pm$0.007&V2676 Ori\\
MGM 942&5:36:20.58&-6:22:15.53&15.653$\pm$0.025&---&---&---&---\\
MGM 948&5:36:27.72&-6:21:57.17&15.93$\pm$0.042&---&---&---&---\\
MGM 949&5:36:24.70&-6:21:54.14&15.78$\pm$0.025&---&---&---&---\\
MGM 954&5:35:47.64&-6:21:36.00&11.759$\pm$0.022&---&---&---&V814 Ori\\
MGM 956&5:36:22.61&-6:21:27.59&16.68$\pm$0.025&---&---&---&---\\
MGM 993&5:36:33.29&-6:15:23.97&15.151$\pm$0.057&---&---&---&---\\
MGM 1154&5:35:01.89&-5:53:01.15&15.304$\pm$0.055&---&---&---&---\\
MGM 1176&5:35:06.77&-5:51:01.27&11.419$\pm$0.044&---&---&---&---\\
MGM 1229&5:34:41.95&-5:45:22.42&13.061$\pm$0.057&---&---&---&V2009 Ori\\
MGM 1236&5:34:41.97&-5:45:00.54&12.426$\pm$0.027&---&---&---&---\\
MGM 1273&5:34:48.17&-5:42:29.19&10.311$\pm$0.034&---&---&---&KK Ori\\
MGM 1274&5:34:48.49&-5:42:28.35&12.771$\pm$0.057&---&---&---&V2038 Ori\\
MGM 1304&5:34:45.88&-5:41:09.96&11.357$\pm$0.025&---&---&---&V1447 Ori\\
MGM 1351&5:35:00.81&-5:38:07.92&10.849$\pm$0.058&---&---&---&KW Ori\\
MGM 1368&5:35:01.47&-5:37:16.46&18.543$\pm$0.037&---&---&---&---\\
MGM 1374&5:35:08.21&-5:37:04.65&10.898$\pm$0.028&---&---&---&LO Ori\\
MGM 1384&5:35:05.06&-5:36:43.73&12.079$\pm$0.039&---&---&---&V2148 Ori\\
MGM 1412&5:35:05.75&-5:35:21.97&13.838$\pm$0.049&---&---&---&---\\
MGM 1465&5:35:02.48&-5:33:10.26&11.529$\pm$0.09&---&---&---&V786 Ori\\
MGM 1483&5:35:08.01&-5:32:44.66&11.22$\pm$0.041&---&---&---&LN Ori\\
MGM 1493&5:35:02.75&-5:32:03.15&13.411$\pm$0.018&---&---&---&V2124 Ori\\
MGM 1516&5:34:37.09&-5:31:08.73&11.407$\pm$0.032&---&---&---&V1992 Ori\\
MGM 2212&5:35:16.17&-5:14:12.97&14.61$\pm$0.041&---&---&---&---\\
MGM 2216&5:35:19.78&-5:14:05.08&12.699$\pm$0.023&---&---&11.612$\pm$0.025&---\\
MGM 2219&5:35:20.23&-5:13:59.22&10.781$\pm$0.031&---&---&8.826$\pm$0.002&---\\
MGM 2227&5:35:22.92&-5:13:39.69&12.471$\pm$0.03&---&---&---&V2401 Ori\\
MGM 2232&5:35:22.59&-5:13:28.09&13.33$\pm$0.021&---&---&---&---\\
MGM 2233&5:35:18.60&-5:13:27.22&14.336$\pm$0.029&---&---&11.478$\pm$0.019&V2333 Ori\\
MGM 2234&5:35:19.64&-5:13:26.28&11.668$\pm$0.039&---&---&8.546$\pm$0.001&V2355 Ori\\
MGM 2243&5:35:19.99&-5:12:50.04&12.529$\pm$0.056&---&---&9.747$\pm$0.004&V2361 Ori\\
MGM 2252&5:35:25.32&-5:12:05.81&12.949$\pm$0.051&---&---&---&---\\
MGM 2256&5:35:24.63&-5:11:58.51&11.062$\pm$0.04&---&---&---&NO Ori\\
MGM 2265&5:35:28.15&-5:11:37.65&15.289$\pm$0.034&---&---&---&V1543 Ori\\
MGM 2274&5:35:26.86&-5:11:07.57&9.859$\pm$0.238&---&---&---&AI Ori\\
MGM 2296&5:35:25.72&-5:09:49.43&10.557$\pm$0.036&---&---&9.806$\pm$0.005&AH Ori\\
MGM 2297&5:35:27.46&-5:09:44.15&10.531$\pm$0.03&---&---&8.985$\pm$0.003&---\\
MGM 2301&5:35:27.63&-5:09:37.13&10.659$\pm$0.042&---&---&6.401$\pm$0.001&V2467 Ori\\
MGM 2307&5:35:21.26&-5:09:16.15&8.202$\pm$0.124&---&---&---&MX Ori\\
MGM 2309&5:35:25.01&-5:09:09.55&13.673$\pm$0.044&---&---&---&---\\
MGM 2310&5:35:24.07&-5:09:06.76&13.503$\pm$0.053&---&---&---&V2416 Ori\\
MGM 2312&5:35:20.66&-5:09:02.72&13.134$\pm$0.034&---&---&---&---\\
MGM 2315&5:35:23.21&-5:08:43.55&15.135$\pm$0.038&---&---&11.134$\pm$0.039&---\\
MGM 2325&5:35:23.33&-5:08:21.54&15.155$\pm$0.04&---&---&10.156$\pm$0.009&---\\
MGM 2333&5:35:27.79&-5:07:54.53&12.864$\pm$0.024&---&---&12.225$\pm$0.04&V1346 Ori\\
MGM 2338&5:35:28.50&-5:07:46.80&13.888$\pm$0.031&---&---&9.603$\pm$0.005&---\\
MGM 2345&5:35:23.32&-5:07:09.51&16.692$\pm$0.049&---&---&9.708$\pm$0.017&---\\
MGM 2348&5:35:25.69&-5:07:03.18&12.83$\pm$0.057&---&---&12.28$\pm$0.041&---\\
MGM 2356&5:35:34.26&-5:06:20.89&9.86$\pm$0.093&---&---&---&HD 37060\\
MGM 2368&5:35:28.59&-5:05:44.61&12.429$\pm$0.039&---&---&9.309$\pm$0.005&---\\
MGM 2371&5:35:32.60&-5:05:37.68&15.74$\pm$0.038&---&---&---&V2518 Ori\\
MGM 2380&5:35:31.48&-5:05:01.32&10.983$\pm$0.046&---&---&---&V422 Ori\\
MGM 2416&5:35:27.40&-5:02:41.62&13.33$\pm$0.054&---&---&---&V2465 Ori\\
MGM 2426&5:35:21.51&-5:01:53.78&15.511$\pm$0.037&---&---&---&---\\
MGM 2438&5:35:15.45&-5:01:12.76&13.517$\pm$0.037&---&---&13.974$\pm$0.482&---\\
MGM 2445&5:35:22.25&-5:00:38.93&19.56$\pm$0.038&---&---&---&---\\
MGM 2447&5:35:17.72&-5:00:31.23&13.234$\pm$0.031&---&---&10.307$\pm$0.007&---\\
MGM 2448&5:35:13.03&-5:00:26.46&18.16$\pm$0.03&---&---&---&---\\
MGM 2455&5:35:17.40&-4:59:57.38&14.947$\pm$0.039&---&---&11.299$\pm$0.032&V2305 Ori\\
MGM 2457&5:35:26.45&-4:59:52.35&14.421$\pm$0.043&---&---&---&V2450 Ori\\
MGM 2461&5:35:30.60&-4:59:36.29&11.854$\pm$0.034&---&---&---&V2493 Ori\\
MGM 2479&5:35:31.47&-4:57:47.76&15.104$\pm$0.045&---&---&---&---\\
MGM 2483&5:35:31.19&-4:57:27.06&13.068$\pm$0.038&---&---&---&V2497 Ori\\
MGM 2484&5:35:30.58&-4:57:21.45&14.042$\pm$0.03&---&---&---&---\\
MGM 2522&5:35:02.89&-4:54:30.30&16.437$\pm$0.04&---&---&---&---\\
MGM 2531&5:35:12.68&-4:54:02.89&13.896$\pm$0.031&---&---&---&V2226 Ori\\
MGM 2532&5:35:04.64&-4:54:02.68&14.451$\pm$0.021&---&---&---&---\\
MGM 2539&5:35:08.11&-4:53:31.79&14.48$\pm$0.016&---&---&11.864$\pm$0.044&---\\
MGM 2541&5:35:08.30&-4:53:29.14&14.692$\pm$0.021&---&---&11.804$\pm$0.059&---\\
MGM 2545&5:35:09.54&-4:53:17.76&13.896$\pm$0.037&---&---&---&---\\
MGM 2637&5:35:35.21&-4:47:39.69&11.502$\pm$0.042&---&---&---&V2544 Ori\\
MGM 2638&5:35:28.64&-4:47:26.58&11.798$\pm$0.046&---&---&---&V1547 Ori\\
MGM 2646&5:35:32.21&-4:46:57.33&10.638$\pm$0.022&---&---&---&V565 Ori\\
MGM 2647&5:35:34.49&-4:46:54.85&11.705$\pm$0.036&---&---&---&V423 Ori\\
MGM 2653&5:35:33.68&-4:46:23.79&11.264$\pm$0.041&---&---&---&V2528 Ori\\
MGM 2826&5:40:52.39&-2:27:12.48&14.054$\pm$0.028&---&---&11.152$\pm$0.026&---\\
MGM 2839&5:41:20.88&-2:17:52.93&13.526$\pm$0.03&---&---&---&---\\
MGM 2844&5:41:34.70&-2:17:24.07&11.36$\pm$0.036&---&---&10.429$\pm$0.011&---\\
MGM 2847&5:41:44.67&-2:16:56.54&12.007$\pm$0.021&---&---&---&---\\
MGM 2850&5:41:47.07&-2:16:37.85&11.427$\pm$0.241&---&---&6.204$\pm$0.001&---\\
MGM 2852&5:41:44.23&-2:16:16.48&11.91$\pm$0.028&12.127$\pm$0.001&11.629$\pm$0.001&10.821$\pm$0.012&---\\
MGM 3142&5:43:04.94&-1:15:46.12&19.466$\pm$0.03&---&---&---&---\\
MGM 3170&5:46:07.75&-0:09:37.45&18.589$\pm$0.038&---&---&---&---\\
MGM 3205&5:46:47.11&0:00:35.82&14.812$\pm$0.033&---&---&12.209$\pm$0.045&---\\
MGM 3212&5:46:37.06&0:01:21.94&10.267$\pm$0.034&---&---&---&---\\
MGM 3233&5:46:34.89&0:04:20.78&11.546$\pm$0.022&---&---&---&---\\
MGM 3304&5:47:02.86&0:16:52.05&13.141$\pm$0.049&---&---&---&---\\
MGM 3317&5:47:05.05&0:18:34.95&11.761$\pm$0.029&---&---&---&---\\
MGM 3318&5:47:02.97&0:18:39.06&13.723$\pm$0.034&---&---&---&---\\
MGM 3336&5:47:25.42&0:19:40.15&13.859$\pm$0.051&---&---&---&---\\
MGM 3349&5:47:26.94&0:20:31.55&17.504$\pm$0.047&---&---&---&---\\
MGM 3357&5:47:15.29&0:21:01.52&20.178$\pm$0.027&---&---&---&---\\
MGM 3358&5:47:05.68&0:21:12.12&13.802$\pm$0.035&---&---&---&---\\
MGM 3361&5:47:14.86&0:21:18.94&14.2$\pm$0.022&---&---&---&---\\
MGM 3368&5:47:17.63&0:21:55.27&21.09$\pm$0.033&---&---&---&---\\
MGM 3369&5:47:05.14&0:22:01.22&13.007$\pm$0.024&---&---&---&---\\
MGM 3371&5:47:12.89&0:22:06.67&11.865$\pm$0.03&---&---&---&---\\
MGM 3433&5:47:37.20&0:39:22.83&14.439$\pm$0.036&---&---&---&---\\
MGM 3436&5:47:36.88&0:39:47.78&14.073$\pm$0.056&---&---&---&---\\
\enddata
\tablenotetext{1} {MGM=\citet{2012AJ....144..192M}}
\end{deluxetable}

\begin{deluxetable}{lcccccccc}
\tabletypesize{\scriptsize}
\tablewidth{0pt}
\tablecaption{Number of Primaries, Companions and Contaminants vs Sample \label{tab:comp_combined}}
\tablehead{
\colhead{Sample} & \colhead{Protostars} & \colhead{Cand.\tablenotemark{1}} & \colhead{Cont.\tablenotemark{2}} & \colhead{Comp.\tablenotemark{3}} & \colhead{Pre-ms\tablenotemark{4}} & \colhead{Cand.\tablenotemark{1}} & \colhead{Cont.\tablenotemark{2}}& \colhead{Comp.\tablenotemark{3}}}
\startdata
Combined 	&	201	&	28	&	152	&	20.8$^{+3.2}_{-2.8}$  	&	198	&	28	&	168	&	20.1$^{+2.9}_{-3.1}$\\
HST		 	&	178  &      27     &      142  &       20.3$^{+2.7}_{-2.3}$ &  197 &   28 & 168 & 20.1$^{+2.9}_{-3.1}$\\
WFC3 		&	129	&	21	&	109	&	15.9$^{+2.1}_{-1.9}$	& 197	&	28	&	168	&	20.1$^{+2.9}_{-3.1}$ \\
\enddata
\tablenotetext{1} {Number of sources between projected separations of 100 and 1000~AU.}
\tablenotetext{2} {Number of sources between projected separations of  2000 and 5000~AU .}
\tablenotetext{3} {Number of companions calculated using $r_{inner} = 100$~AU}
\tablenotetext{4} {Pre-main sequence stars identified by the presence of an IR-excess due to a dusty disk. }
\end{deluxetable}

\begin{deluxetable}{lccccccccc}
\tabletypesize{\scriptsize}
\tablewidth{0pt}
\tablecaption{WFC3 sample: Number of Primaries and Companions vs.  Surface Density \label{tab:comp_wfc3}}
\tablehead{
\colhead{$\Sigma_{YSO}$} & \colhead{$r_{inner}$} &  \colhead{Protostars} & \colhead{Cand.\tablenotemark{1}} & \colhead{Cont.\tablenotemark{2}} &\colhead{Comp.\tablenotemark{3}} & \colhead{Pre-ms\tablenotemark{4}} & \colhead{Cand.\tablenotemark{1}} & \colhead{Cont.\tablenotemark{2}}& \colhead{Comp.\tablenotemark{3}}
}
\startdata
All\tablenotemark{5} & 100~AU &	129	&	21	&	109	&	15.9$^{+2.1}_{-1.9}$	& 197	&	28	&	168	&	20.1$^{+2.9}_{-3.1}$ \\
All\tablenotemark{5}  & $r_{25\%}$ & 129	&	21	&	65.1	&	17.9$^{+2.1}_{-1.9}$	& 197	&	28	&	94.1	&	23.6$^{+2.4}_{-2.6}$ \\
All\tablenotemark{5} & $r_{99\%}$ & 129	&	21	&	50.3	&	18.6$^{+1.4}_{-1.6}$	& 197	&	28	&	70.8	&	24.7$^{+1.3}_{-1.7}$ \\
$> 45$~pc$^{-2}$ & 	100~AU & 56 &	12 &	51 &	9.6$^{+1.4}_{-1.6}$	& 123	&	19	&	93	&	14.6$^{+2.4}_{-2.6}$ \\
$< 45$~pc$^{-2}$ &	100~AU  & 73 &	9 &	58 &	6.3$^{+1.7}_{-1.3}$	& 74		&	9	&	75	&	5.5$^{+1.5}_{-1.5}$ \\
$>45$~pc$^{-2}$ &	$r_{25\%}$ & 56 &	12 &	31.5 &10.5$^{+1.5}_{-1.5}$& 123	&	19	&	52.2	&	16.5$^{+1.5}_{-1.5}$ \\
$<45$~pc$^{-2}$ &	$r_{25\%}$ & 73 &	9 &	33.6 &7.4$^{+1.6}_{-1.4}$	& 74		&	9	&	41.9	&	7.0$^{+1.0}_{-1.0}$ \\
$>45$~pc$^{-2}$ &	$r_{99\%}$ & 56 &	12 &	26.6 &10.7$^{+1.3}_{-0.7}$& 123	&	19	&	39.6	&	17.1$^{+1.9}_{-1.1}$ \\
$<45$~pc$^{-2}$ &	$r_{99\%}$ & 73 &	9 &23.7 &	7.9$^{+1.1}_{-0.9}$	& 74		&	9	&	31.2	&	7.5$^{+1.5}_{-1.5}$ \\
\enddata
\tablenotetext{1} {Number of sources between projected separations of 100 and 1000~AU.}
\tablenotetext{2} {Number of sources between  2000 and 5000~AU.}
\tablenotetext{3} {Number of companions calculated using $r_{inner}$.}
\tablenotetext{4} {Pre-main sequence stars identified by the presence of an IR-excess due to a dusty disk. }
\tablenotetext{5} {Includes both high and low YSO surface density regions.}

\end{deluxetable}

\begin{deluxetable}{lccc}
\tabletypesize{\scriptsize}
\tablewidth{0pt}
\tablecaption{CFs vs Sample \label{tab:csf_all}}
\tablehead{
\colhead{Sample} & \colhead{Protostars\tablenotemark{1}}  & \colhead{Pre-ms\tablenotemark{1,2}} & 
\colhead{Merged\tablenotemark{1,3}}}
\startdata
Combined 	&	10.4$^{+1.6}_{-1.4}$  & 10.1$^{+1.5}_{-1.6}$   &  10.3$^{+1.0}_{-1.0}$ \\
HST		 	&	11.4$^{+1.5}_{-1.3}$   & 10.2$^{+1.5}_{-1.6}$   &  10.8$^{+1.0}_{-0.9}$ \\ 
WFC3 		&	12.3$^{+1.7}_{-1.4}$   & 10.2$^{+1.5}_{-1.6}$   &  11.0$^{+1.2}_{-1.2}$ \\
\enddata
\tablenotetext{1} {Calculated using an $r_{inner} = 100$~AU}
\tablenotetext{2} {Pre-main sequence stars with disks. }
\tablenotetext{3} {Merged sample of dusty YSOs. }

\end{deluxetable}

\begin{deluxetable}{lcccc}
\tabletypesize{\scriptsize}
\tablewidth{0pt}
\tablecaption{WFC3 Sample: CF vs.  Surface Density \label{tab:csf_wfc3}}
\tablehead{
\colhead{$\Sigma_{YSO}$} & \colhead{$r_{inner}$} & \colhead{Protostars} & \colhead{Pre-ms stars.\tablenotemark{1}} & \colhead{Merged.\tablenotemark{2}}
}
\startdata
All \tablenotemark{3} & 100~AU & 12.3$^{+1.7}_{-1.4}\%$ & 10.2$^{+1.5}_{-1.6}\%$ & 11.0$^{+1.2}_{-1.2}\%$ \\
All \tablenotemark{3} & $r_{25\%}$ & 13.9$^{+1.6}_{-1.5}\%$ & 12.0$^{+1.2}_{-1.3}\%$ & 12.7$^{+0.8}_{-0.8}\%$ \\
All \tablenotemark{3} & $r_{99\%}$ & 14.4$^{+1.1}_{-1.3}\%$ & 12.5$^{+1.2}_{-0.8}\%$ & 13.3$^{+0.8}_{-0.7}\%$ \\
$ > 45$~pc$^{-2}$ & 100~AU & 17.1$^{+2.5}_{-2.8}\%$ & 11.9$^{+1.9}_{-1.3}\%$ & 13.5$^{+1.6}_{-1.2}\%$ \\
$ < 45$~pc$^{-2}$ & 100~AU & ~8.6$^{+2.4}_{-1.7}\%$ &  ~7.4$^{+2.1}_{-2.0}\%$ &  ~8.0$^{+1.5}_{-1.9}\%$ \\
$ > 45$~pc$^{-2}$ & $r_{25\%}$ & 18.8$^{+2.6}_{-2.7}\%$ & 13.4$^{+1.2}_{-1.3}\%$ & 15.1$^{+1.1}_{-1.1}\%$ \\
$ < 45$~pc$^{-2}$ & $r_{25\%}$&  10.2$^{+2.2}_{-1.9}\%$ &  ~9.5$^{+1.3}_{-1.4}\%$ &  ~9.8$^{+1.1}_{-1.0}\%$ \\
$ > 45$~pc$^{-2}$ & $r_{99\%}$ & 19.2$^{+2.2}_{-1.3}\%$ & 13.9$^{+0.7}_{-0.9}\%$ & 15.6$^{+1.2}_{-1.0}\%$ \\
$ < 45$~pc$^{-2}$ & $r_{99\%}$ & 10.8$^{+1.5}_{-1.2}\%$ &  10.2$^{+2.0}_{-2.1}\%$ &  10.5$^{+1.1}_{-1.0}\%$ \\
\enddata
\tablenotetext{1} {Pre-main sequence stars identified by the presence of an IR-excess due to a dusty disk. }
\tablenotetext{2} {Includes both protostars and pre-main sequence stars with disks.}
\tablenotetext{3} {Includes both high and low YSO surface density regions.}
\end{deluxetable}

\begin{deluxetable}{lcccccccccccc}
\tabletypesize{\scriptsize}
\tablewidth{0pt}
\tablecaption{WFC3 Sample: CF and CP Ratios$^1$ vs. Evolution and Completeness Limit for Threshold $\Sigma_{YSO}$ = 45~pc$^{-2}$ \label{tab:ratio_wfc3_limit}}
\tablehead{
 & \multicolumn{4}{c}{Protostars} & \multicolumn{4}{c}{Pre-Main Sequence Stars} & \multicolumn{4}{c}{Merged} \\
& \multicolumn{2}{c}{CF} & \multicolumn{2}{c}{CP} & \multicolumn{2}{c}{CF} & \multicolumn{2}{c}{CP} & \multicolumn{2}{c}{CF} & \multicolumn{2}{c}{CP} \\
\colhead{$r_{inner}$ \tablenotemark{2}} & \colhead{$P(R \le 1$)} & \colhead{R} & \colhead{$P(R \le 1)$} & \colhead{R} & \colhead{$P(R \le 1)$} & \colhead{R} & \colhead{$P(R \le 1)$} & \colhead{R} & \colhead{$P(R \le 1)$} & \colhead{R} & \colhead{$P(R \le 1)$} & \colhead{R} 
}
\startdata
100 AU &   0.0124 & 1.96  & 0.11 & 1.95 & 0.0792 & 1.50 &  0.22 & 1.53 & 0.0081 & 1.64 & 0.098 & 1.65 \\
$r_{25\%}$ &    0.0022 & 1.79  & 0.10 & 1.80 & 0.0292 & 1.35  &  0.25 & 1.34  & 0.0011 & 1.50 & 0.11  & 1.52 \\
$r_{99\%}$  &    0.0013 & 1.74  & 0.10 & 1.75 & 0.014 &  1.28 & 0.26 & 1.32  &  0.0004 &  1.45  & 0.10 & 1.47 \\
\enddata
\tablenotetext{1} {Determined using Bayesian parameter estimation as described in the Appendices B and C.}
\tablenotetext{2} {Inner radius used for determining line of sight contamination based on adopted completeness limit.}
%\tablenotetext{3} {Includes both high and low YSO surface density regions.}
\end{deluxetable}

\clearpage

\clearpage

\clearpage

\clearpage

\clearpage

\end{document}